\PassOptionsToPackage{table}{xcolor}
\documentclass{article}
\usepackage[OT1]{fontenc}
\usepackage{iclr2027_conference,times}

\usepackage{amsmath,amsfonts,bm}

\def\eqref#1{equation~\ref{#1}}

\def\1{\bm{1}}

\DeclareMathAlphabet{\mathsfit}{\encodingdefault}{\sfdefault}{m}{sl}
\SetMathAlphabet{\mathsfit}{bold}{\encodingdefault}{\sfdefault}{bx}{n}

\usepackage{hyperref}
\hypersetup{colorlinks=true,linkcolor=black,citecolor=black,filecolor=black,urlcolor=blue}
\usepackage{url}
\usepackage{graphicx}
\usepackage{wrapfig}
\usepackage{needspace}
\usepackage{booktabs}
\usepackage{xcolor}
\usepackage{enumitem}
\usepackage{siunitx}
\providecommand{\PaperResultSet}{cross}
\ExplSyntaxOn
\tl_new:N \l_paper_first_tl
\tl_new:N \l_paper_second_tl
\int_new:N \l_paper_largest_int
\NewDocumentCommand{\SetScore}{mmm}
  { \cs_gset:cpn { paper_score_#1_#2 } {#3} }

\cs_new_protected:Npn \paper_number:nn #1#2
  {
    \group_begin:
    \sisetup{round-mode=places,round-precision=#1,round-pad=true,
      round-half=up,group-digits=false,mode=text,detect-weight=true}
    \exp_args:Nx \num {#2}
    \group_end:
  }

\NewDocumentCommand{\WER}{O{3}mm}
  {
    \tl_set:Nx \l_paper_first_tl { \use:c { paper_score_#2_#3 } }
    \tl_if_eq:NnTF \l_paper_first_tl {XX}
      {XX}
      {\paper_number:nn {#1}{\l_paper_first_tl}}
  }

\cs_new_protected:Npn \paper_comparison:nnnnn #1#2#3#4#5
  {
    \tl_set:Nx \l_paper_first_tl { \use:c { paper_score_#3_#5 } }
    \tl_set:Nx \l_paper_second_tl { \use:c { paper_score_#4_#5 } }
    \tl_if_eq:NnTF \l_paper_first_tl {XX}
      {XX}
      {
        \tl_if_eq:NnTF \l_paper_second_tl {XX}
          {XX}
          {\paper_number:nn {#1}{\fp_eval:n {#2}}}
      }
  }

\NewDocumentCommand{\PP}{O{2}mmm}
  {\paper_comparison:nnnnn {#1}{\l_paper_first_tl-\l_paper_second_tl}{#2}{#3}{#4}}
\NewDocumentCommand{\RR}{O{1}mmm}
  {\paper_comparison:nnnnn {#1}
    {100*(\l_paper_first_tl-\l_paper_second_tl)/\l_paper_first_tl}{#2}{#3}{#4}}

\NewDocumentCommand{\LargestK}{m}
  {
    \tl_set:Nx \l_paper_first_tl {\use:c {paper_score_mas_#1}}
    \tl_if_eq:NnTF \l_paper_first_tl {XX}
      {XX}
      {
        \int_zero:N \l_paper_largest_int
        \clist_map_inline:nn {1,2,3,4,6,9,12,18,25,37,50,75,100}
          {
            \fp_compare:nNnT {\l_paper_first_tl} <
              {\use:c {paper_score_cascade##1_#1}}
              {\int_set:Nn \l_paper_largest_int {##1}}
          }
        \int_compare:nNnTF {\l_paper_largest_int} = {0}
          {\text{none}}{\int_use:N \l_paper_largest_int}
      }
  }
\ExplSyntaxOff

\SetScore{feng}{t12}{24.69}
\SetScore{rnn}{t12}{9.76}
\SetScore{rnn}{t15public}{6.67}
\SetScore{timemasked}{t12}{7.98}
\SetScore{modalone}{t12}{18.8}
\newcommand{\MoDAlOneCI}{0.5}
\SetScore{modal}{t12}{17.7}
\newcommand{\MoDAlCI}{0.4}
\SetScore{bit}{t12}{15.67}
\SetScore{bit}{t15public}{11.06}
\SetScore{bitensemble}{t12}{10.22}
\SetScore{bitensemble}{t15public}{7.76}
\SetScore{bitcascade}{t12}{6.35}
\SetScore{bitcascade}{t15public}{4.06}

\SetScore{cascade100}{t12}{7.876573284612262}
\SetScore{cascade100}{t15public}{5.969}
\SetScore{cascade100}{t15private}{6.026}
\SetScore{cascade75}{t12}{7.876573284612262}
\SetScore{cascade75}{t15public}{5.939}
\SetScore{cascade75}{t15private}{6.026}
\SetScore{cascade50}{t12}{7.9307078089051295}
\SetScore{cascade50}{t15public}{5.939}
\SetScore{cascade50}{t15private}{6.057}
\SetScore{cascade37}{t12}{7.9307078089051295}
\SetScore{cascade37}{t15public}{5.939}
\SetScore{cascade37}{t15private}{6.073}
\SetScore{cascade25}{t12}{7.917174177831912}
\SetScore{cascade25}{t15public}{5.939}
\SetScore{cascade25}{t15private}{6.12}
\SetScore{cascade18}{t12}{7.9307078089051295}
\SetScore{cascade18}{t15public}{5.969}
\SetScore{cascade18}{t15private}{6.089}
\SetScore{cascade12}{t12}{7.957775071051564}
\SetScore{cascade12}{t15public}{6.03}
\SetScore{cascade12}{t15private}{6.136}
\SetScore{cascade9}{t12}{7.97130870212478}
\SetScore{cascade9}{t15public}{6.03}
\SetScore{cascade9}{t15private}{6.183}
\SetScore{cascade6}{t12}{8.01190959534443}
\SetScore{cascade6}{t15public}{6.09}
\SetScore{cascade6}{t15private}{6.214}
\SetScore{cascade4}{t12}{8.120178643930167}
\SetScore{cascade4}{t15public}{6.181}
\SetScore{cascade4}{t15private}{6.293}
\SetScore{cascade3}{t12}{8.255514954662337}
\SetScore{cascade3}{t15public}{6.09}
\SetScore{cascade3}{t15private}{6.387}
\SetScore{cascade2}{t12}{8.458519420760592}
\SetScore{cascade2}{t15public}{6.424}
\SetScore{cascade2}{t15private}{6.45}
\SetScore{cascade1}{t12}{9.297604547300041}
\SetScore{cascade1}{t15public}{7.272}
\SetScore{cascade1}{t15private}{7.062}

\newcommand{\ReportedRnnEtoEWER}{19.9}
\newcommand{\ReportedBitTfsWER}{16.5}
\newcommand{\ReportedBitHumanWER}{15.8}
\newcommand{\ReportedRnnEtoEFifteenWER}{16.0}
\newcommand{\ReportedBitTfsFifteenWER}{12.2}
\newcommand{\ReportedBitHumanFifteenWER}{11.4}
\newcommand{\ReportedBrainWhispererWER}{10.2}
\newcommand{\ReportedBrainWhispererCrossWER}{8.7}

\SetScore{bestnolm}{t12}{12.61}
\newcommand{\ReportedBestFifteenWER}{5.17}

\newcommand{\ReportedFiedlerConformerWER}{26.67}
\newcommand{\ReportedFiedlerWavWER}{30.97}
\newcommand{\ReportedOlakBeamWER}{24.8}
\newcommand{\ReportedOlakBeamSD}{0.4}
\newcommand{\ReportedOlakGreedyWER}{25.6}
\newcommand{\ReportedOlakGreedySD}{0.2}
\newcommand{\ReportedOlakPhoneWER}{22.6}
\newcommand{\ReportedOlakPhoneSD}{0.2}
\newcommand{\ReportedOlakAgreementWER}{23.2}
\newcommand{\ReportedOlakAgreementSD}{0.4}
\newcommand{\ReportedBrainWhispererTestWER}{9.8}
\newcommand{\ReportedBrainWhispererCrossTestWER}{8.9}
\newcommand{\ReportedFengBpeTestWER}{28.5}
\newcommand{\ReportedFengPhoneTestWER}{27.3}
\newcommand{\ReportedFengBiGruTestWER}{26.3}
\newcommand{\ReportedFengVocalTestWER}{26.2}
\newcommand{\ReportedMoDAlOneTestWER}{22.4}
\newcommand{\ReportedMoDAlOneTestCI}{0.2}
\newcommand{\ReportedMoDAlFullTestWER}{21.6}
\newcommand{\ReportedMoDAlFullTestCI}{0.1}
\newcommand{\ReportedMoDAlTwoTestWER}{22.5}
\newcommand{\ReportedMoDAlTwoTestCI}{0.3}

\SetScore{mas}{t12}{6.92921910948708}
\SetScore{mas}{t15public}{3.606}
\SetScore{mas}{t15private}{3.892}
\SetScore{joint}{t12}{7.94424143997835}
\SetScore{joint}{t15public}{4.212}
\SetScore{joint}{t15private}{4.692}
\SetScore{nna}{t12}{7.05102178914603}
\SetScore{nna}{t15public}{3.909}
\SetScore{nna}{t15private}{3.939}
\SetScore{hard}{t12}{7.15929083773176}
\SetScore{hard}{t15public}{4.181}
\SetScore{hard}{t15private}{4.409}
\SetScore{softsg}{t12}{7.3487616727568}
\SetScore{softsg}{t15public}{3.666}
\SetScore{softsg}{t15private}{4.315}
\SetScore{single}{t12}{9.9878197320341}
\SetScore{single}{t15public}{6.121}
\SetScore{single}{t15private}{6.669}
\SetScore{separate}{t12}{7.97130870212478}
\SetScore{separate}{t15public}{4.12100}
\SetScore{separate}{t15private}{4.62900}

\newcommand{\SetCaseCount}[3]{\expandafter\def\csname casecount@#1@#2\endcsname{#3}}
\newcommand{\CaseCount}[2]{\csname casecount@#1@#2\endcsname}
\newcommand{\SetArbitrationCount}[3]{\expandafter\def\csname arbitration@#1@#2\endcsname{#3}}
\newcommand{\ArbCount}[2]{\csname arbitration@#1@#2\endcsname}

\SetArbitrationCount{t12}{rows}{880}
\SetArbitrationCount{t12}{reference_words}{5496}
\SetArbitrationCount{t12}{neural_edits}{778}
\SetArbitrationCount{t12}{phonetic_edits}{801}
\SetArbitrationCount{t12}{arbiter_edits}{557}
\SetArbitrationCount{t12}{oracle_edits}{553}
\SetArbitrationCount{t12}{oracle_correct}{575}
\SetArbitrationCount{t12}{arbiter_correct}{570}
\SetArbitrationCount{t12}{arbiter_wrong}{310}
\SetArbitrationCount{t12}{exact_gains}{31}
\SetArbitrationCount{t12}{exact_losses}{36}
\SetArbitrationCount{t12}{net_exact}{-5}
\SetArbitrationCount{t12}{edit_gains}{81}
\SetArbitrationCount{t12}{edit_losses}{85}
\SetArbitrationCount{t12}{net_edit_reduction}{-4}
\SetArbitrationCount{t12}{improved}{65}
\SetArbitrationCount{t12}{tied}{744}
\SetArbitrationCount{t12}{worsened}{71}
\SetArbitrationCount{t12}{rescued_identical_wrong_drafts}{8}
\SetArbitrationCount{t12}{both_correct_rows}{394}
\SetArbitrationCount{t12}{both_correct_arbiter_correct}{392}
\SetArbitrationCount{t12}{both_correct_arbiter_wrong}{2}
\SetArbitrationCount{t12}{neural_only_rows}{83}
\SetArbitrationCount{t12}{neural_only_arbiter_correct}{70}
\SetArbitrationCount{t12}{neural_only_arbiter_wrong}{13}
\SetArbitrationCount{t12}{phonetic_only_rows}{98}
\SetArbitrationCount{t12}{phonetic_only_arbiter_correct}{77}
\SetArbitrationCount{t12}{phonetic_only_arbiter_wrong}{21}
\SetArbitrationCount{t12}{both_wrong_rows}{305}
\SetArbitrationCount{t12}{both_wrong_arbiter_correct}{31}
\SetArbitrationCount{t12}{both_wrong_arbiter_wrong}{274}
\SetArbitrationCount{t12}{neural_correct}{477}
\SetArbitrationCount{t12}{phonetic_correct}{492}
\SetArbitrationCount{t12}{neural_relative_reduction}{28.4}
\SetArbitrationCount{t12}{phonetic_relative_reduction}{30.5}
\SetArbitrationCount{t12}{net_exact_signed}{-5}
\SetArbitrationCount{t12}{net_edit_reduction_signed}{-4}
\SetCaseCount{t12}{rows}{880}
\SetCaseCount{t12}{both_experts_wrong}{305}
\SetCaseCount{t12}{arbiter_recovers}{31}
\SetCaseCount{t12}{rescued_identical_wrong_drafts}{8}
\SetCaseCount{t12}{neural_correct_utterances}{83}
\SetCaseCount{t12}{neural_correct_arbiter_correct}{70}
\SetCaseCount{t12}{phonetic_correct_utterances}{98}
\SetCaseCount{t12}{phonetic_correct_arbiter_correct}{77}
\SetArbitrationCount{t15}{rows}{1426}
\SetArbitrationCount{t15}{reference_words}{9445}
\SetArbitrationCount{t15}{neural_edits}{627}
\SetArbitrationCount{t15}{phonetic_edits}{703}
\SetArbitrationCount{t15}{arbiter_edits}{420}
\SetArbitrationCount{t15}{oracle_edits}{446}
\SetArbitrationCount{t15}{oracle_correct}{1164}
\SetArbitrationCount{t15}{arbiter_correct}{1189}
\SetArbitrationCount{t15}{arbiter_wrong}{237}
\SetArbitrationCount{t15}{exact_gains}{49}
\SetArbitrationCount{t15}{exact_losses}{24}
\SetArbitrationCount{t15}{net_exact}{25}
\SetArbitrationCount{t15}{edit_gains}{83}
\SetArbitrationCount{t15}{edit_losses}{57}
\SetArbitrationCount{t15}{net_edit_reduction}{26}
\SetArbitrationCount{t15}{improved}{66}
\SetArbitrationCount{t15}{tied}{1309}
\SetArbitrationCount{t15}{worsened}{51}
\SetArbitrationCount{t15}{rescued_identical_wrong_drafts}{2}
\SetArbitrationCount{t15}{both_correct_rows}{973}
\SetArbitrationCount{t15}{both_correct_arbiter_correct}{973}
\SetArbitrationCount{t15}{both_correct_arbiter_wrong}{0}
\SetArbitrationCount{t15}{neural_only_rows}{105}
\SetArbitrationCount{t15}{neural_only_arbiter_correct}{93}
\SetArbitrationCount{t15}{neural_only_arbiter_wrong}{12}
\SetArbitrationCount{t15}{phonetic_only_rows}{86}
\SetArbitrationCount{t15}{phonetic_only_arbiter_correct}{74}
\SetArbitrationCount{t15}{phonetic_only_arbiter_wrong}{12}
\SetArbitrationCount{t15}{both_wrong_rows}{262}
\SetArbitrationCount{t15}{both_wrong_arbiter_correct}{49}
\SetArbitrationCount{t15}{both_wrong_arbiter_wrong}{213}
\SetArbitrationCount{t15}{neural_correct}{1078}
\SetArbitrationCount{t15}{phonetic_correct}{1059}
\SetArbitrationCount{t15}{neural_relative_reduction}{33.0}
\SetArbitrationCount{t15}{phonetic_relative_reduction}{40.3}
\SetArbitrationCount{t15}{net_exact_signed}{+25}
\SetArbitrationCount{t15}{net_edit_reduction_signed}{+26}
\SetCaseCount{t15}{rows}{1426}
\SetCaseCount{t15}{both_experts_wrong}{262}
\SetCaseCount{t15}{arbiter_recovers}{49}
\SetCaseCount{t15}{rescued_identical_wrong_drafts}{2}
\SetCaseCount{t15}{neural_correct_utterances}{105}
\SetCaseCount{t15}{neural_correct_arbiter_correct}{93}
\SetCaseCount{t15}{phonetic_correct_utterances}{86}
\SetCaseCount{t15}{phonetic_correct_arbiter_correct}{74}

\newcommand{\CascadeTwelveRAM}{321.8}
\newcommand{\MasTwelveRAM}{12.7}
\newcommand{\CascadeFifteenRAM}{334.8}
\newcommand{\MasFifteenRAM}{31.1}
\newcommand{\HostRAMTwelveRatio}{25.4}

\newcommand{\HostRAMFifteenRatio}{10.8}

\providecommand{\SetWordStat}[3]{\expandafter\def\csname wordstat@#1@#2\endcsname{#3}}
\providecommand{\WordStat}[2]{\csname wordstat@#1@#2\endcsname}
\SetWordStat{t12}{rows}{880}
\SetWordStat{t12}{reference_words}{5496}
\SetWordStat{t12}{neural_edits}{778}
\SetWordStat{t12}{phonetic_edits}{801}
\SetWordStat{t12}{neural_exact}{477}
\SetWordStat{t12}{phonetic_exact}{492}
\SetWordStat{t12}{whole_oracle_edits}{553}
\SetWordStat{t12}{whole_oracle_exact}{575}
\SetWordStat{t12}{graph_oracle_edits}{517}
\SetWordStat{t12}{graph_oracle_exact}{584}
\SetWordStat{t12}{fixed_alignment_edits}{518}
\SetWordStat{t12}{fixed_alignment_exact}{584}
\SetWordStat{t12}{arbiter_edits}{557}
\SetWordStat{t12}{arbiter_exact}{570}
\SetWordStat{t12}{better_than_graph}{52}
\SetWordStat{t12}{tied_with_graph}{745}
\SetWordStat{t12}{worse_than_graph}{83}
\SetWordStat{t12}{edits_gained_vs_graph}{62}
\SetWordStat{t12}{edits_lost_vs_graph}{102}
\SetWordStat{t12}{net_edits_vs_graph}{-40}
\SetWordStat{t12}{net_exact_vs_graph}{-14}
\SetWordStat{t12}{exact_draft_selection}{539}
\SetWordStat{t12}{exact_graph_recombination}{5}
\SetWordStat{t12}{exact_outside_graph}{26}
\SetWordStat{t12}{outside_graph_with_absent_word}{26}
\SetWordStat{t12}{optimal_alignment_sensitivity_rows}{1}
\SetWordStat{t12}{corrected_S}{75}
\SetWordStat{t12}{corrected_D}{0}
\SetWordStat{t12}{corrected_I}{13}
\SetWordStat{t12}{introduced_S}{79}
\SetWordStat{t12}{introduced_D}{4}
\SetWordStat{t12}{introduced_I}{9}
\SetWordStat{t12}{optional_deletion_edits}{496}
\SetWordStat{t12}{optional_deletion_exact}{584}
\SetWordStat{t12}{optional_better}{47}
\SetWordStat{t12}{optional_tied}{741}
\SetWordStat{t12}{optional_worse}{92}
\SetWordStat{t12}{optional_gained}{56}
\SetWordStat{t12}{optional_lost}{117}
\SetWordStat{t12}{neural_only_sites}{259}
\SetWordStat{t12}{neural_only_retained}{222}
\SetWordStat{t12}{neural_only_S}{254}
\SetWordStat{t12}{neural_only_D}{5}
\SetWordStat{t12}{neural_only_content}{204}
\SetWordStat{t12}{neural_only_function}{50}
\SetWordStat{t12}{neural_only_other}{5}
\SetWordStat{t12}{neural_only_unseen}{31}
\SetWordStat{t12}{neural_only_freq1_5}{71}
\SetWordStat{t12}{neural_only_freq6_20}{48}
\SetWordStat{t12}{neural_only_freqgt20}{109}
\SetWordStat{t12}{neural_only_phone_mean}{0.597}
\SetWordStat{t12}{neural_only_phone_covered}{245}
\SetWordStat{t12}{neural_only_phone_total}{254}
\SetWordStat{t12}{phonetic_only_sites}{241}
\SetWordStat{t12}{phonetic_only_retained}{190}
\SetWordStat{t12}{phonetic_only_S}{237}
\SetWordStat{t12}{phonetic_only_D}{4}
\SetWordStat{t12}{phonetic_only_content}{185}
\SetWordStat{t12}{phonetic_only_function}{53}
\SetWordStat{t12}{phonetic_only_other}{3}
\SetWordStat{t12}{phonetic_only_unseen}{22}
\SetWordStat{t12}{phonetic_only_freq1_5}{51}
\SetWordStat{t12}{phonetic_only_freq6_20}{44}
\SetWordStat{t12}{phonetic_only_freqgt20}{124}
\SetWordStat{t12}{phonetic_only_phone_mean}{0.534}
\SetWordStat{t12}{phonetic_only_phone_covered}{223}
\SetWordStat{t12}{phonetic_only_phone_total}{237}
\SetWordStat{t15}{rows}{1426}
\SetWordStat{t15}{reference_words}{9445}
\SetWordStat{t15}{neural_edits}{627}
\SetWordStat{t15}{phonetic_edits}{703}
\SetWordStat{t15}{neural_exact}{1078}
\SetWordStat{t15}{phonetic_exact}{1059}
\SetWordStat{t15}{whole_oracle_edits}{446}
\SetWordStat{t15}{whole_oracle_exact}{1164}
\SetWordStat{t15}{graph_oracle_edits}{410}
\SetWordStat{t15}{graph_oracle_exact}{1180}
\SetWordStat{t15}{fixed_alignment_edits}{412}
\SetWordStat{t15}{fixed_alignment_exact}{1179}
\SetWordStat{t15}{arbiter_edits}{420}
\SetWordStat{t15}{arbiter_exact}{1189}
\SetWordStat{t15}{better_than_graph}{54}
\SetWordStat{t15}{tied_with_graph}{1311}
\SetWordStat{t15}{worse_than_graph}{61}
\SetWordStat{t15}{edits_gained_vs_graph}{65}
\SetWordStat{t15}{edits_lost_vs_graph}{75}
\SetWordStat{t15}{net_edits_vs_graph}{-10}
\SetWordStat{t15}{net_exact_vs_graph}{9}
\SetWordStat{t15}{exact_draft_selection}{1140}
\SetWordStat{t15}{exact_graph_recombination}{11}
\SetWordStat{t15}{exact_outside_graph}{38}
\SetWordStat{t15}{outside_graph_with_absent_word}{36}
\SetWordStat{t15}{optimal_alignment_sensitivity_rows}{2}
\SetWordStat{t15}{corrected_S}{83}
\SetWordStat{t15}{corrected_D}{1}
\SetWordStat{t15}{corrected_I}{8}
\SetWordStat{t15}{introduced_S}{60}
\SetWordStat{t15}{introduced_D}{1}
\SetWordStat{t15}{introduced_I}{5}
\SetWordStat{t15}{optional_deletion_edits}{401}
\SetWordStat{t15}{optional_deletion_exact}{1184}
\SetWordStat{t15}{optional_better}{51}
\SetWordStat{t15}{optional_tied}{1310}
\SetWordStat{t15}{optional_worse}{65}
\SetWordStat{t15}{optional_gained}{60}
\SetWordStat{t15}{optional_lost}{79}
\SetWordStat{t15}{neural_only_sites}{264}
\SetWordStat{t15}{neural_only_retained}{233}
\SetWordStat{t15}{neural_only_S}{261}
\SetWordStat{t15}{neural_only_D}{3}
\SetWordStat{t15}{neural_only_content}{215}
\SetWordStat{t15}{neural_only_function}{42}
\SetWordStat{t15}{neural_only_other}{7}
\SetWordStat{t15}{neural_only_unseen}{31}
\SetWordStat{t15}{neural_only_freq1_5}{86}
\SetWordStat{t15}{neural_only_freq6_20}{45}
\SetWordStat{t15}{neural_only_freqgt20}{102}
\SetWordStat{t15}{neural_only_phone_mean}{0.594}
\SetWordStat{t15}{neural_only_phone_covered}{251}
\SetWordStat{t15}{neural_only_phone_total}{261}
\SetWordStat{t15}{phonetic_only_sites}{187}
\SetWordStat{t15}{phonetic_only_retained}{150}
\SetWordStat{t15}{phonetic_only_S}{184}
\SetWordStat{t15}{phonetic_only_D}{3}
\SetWordStat{t15}{phonetic_only_content}{153}
\SetWordStat{t15}{phonetic_only_function}{29}
\SetWordStat{t15}{phonetic_only_other}{5}
\SetWordStat{t15}{phonetic_only_unseen}{21}
\SetWordStat{t15}{phonetic_only_freq1_5}{60}
\SetWordStat{t15}{phonetic_only_freq6_20}{41}
\SetWordStat{t15}{phonetic_only_freqgt20}{65}
\SetWordStat{t15}{phonetic_only_phone_mean}{0.528}
\SetWordStat{t15}{phonetic_only_phone_covered}{173}
\SetWordStat{t15}{phonetic_only_phone_total}{184}

\newcommand{\WordComplementarityRows}{%
Word positions & \WordStat{t12}{neural_only_sites} & \WordStat{t12}{phonetic_only_sites} & \WordStat{t15}{neural_only_sites} & \WordStat{t15}{phonetic_only_sites} \\
Arbiter correct & \WordStat{t12}{neural_only_retained} & \WordStat{t12}{phonetic_only_retained} & \WordStat{t15}{neural_only_retained} & \WordStat{t15}{phonetic_only_retained} \\
Other expert: substitution & \WordStat{t12}{neural_only_S} & \WordStat{t12}{phonetic_only_S} & \WordStat{t15}{neural_only_S} & \WordStat{t15}{phonetic_only_S} \\
Other expert: deletion & \WordStat{t12}{neural_only_D} & \WordStat{t12}{phonetic_only_D} & \WordStat{t15}{neural_only_D} & \WordStat{t15}{phonetic_only_D} \\
Content words & \WordStat{t12}{neural_only_content} & \WordStat{t12}{phonetic_only_content} & \WordStat{t15}{neural_only_content} & \WordStat{t15}{phonetic_only_content} \\
Function words & \WordStat{t12}{neural_only_function} & \WordStat{t12}{phonetic_only_function} & \WordStat{t15}{neural_only_function} & \WordStat{t15}{phonetic_only_function} \\
Other tags & \WordStat{t12}{neural_only_other} & \WordStat{t12}{phonetic_only_other} & \WordStat{t15}{neural_only_other} & \WordStat{t15}{phonetic_only_other} \\
Unseen in training & \WordStat{t12}{neural_only_unseen} & \WordStat{t12}{phonetic_only_unseen} & \WordStat{t15}{neural_only_unseen} & \WordStat{t15}{phonetic_only_unseen} \\
Training count 1--5 & \WordStat{t12}{neural_only_freq1_5} & \WordStat{t12}{phonetic_only_freq1_5} & \WordStat{t15}{neural_only_freq1_5} & \WordStat{t15}{phonetic_only_freq1_5} \\
Training count 6--20 & \WordStat{t12}{neural_only_freq6_20} & \WordStat{t12}{phonetic_only_freq6_20} & \WordStat{t15}{neural_only_freq6_20} & \WordStat{t15}{phonetic_only_freq6_20} \\
Training count $>20$ & \WordStat{t12}{neural_only_freqgt20} & \WordStat{t12}{phonetic_only_freqgt20} & \WordStat{t15}{neural_only_freqgt20} & \WordStat{t15}{phonetic_only_freqgt20} \\
Mean phoneme distance & \WordStat{t12}{neural_only_phone_mean} & \WordStat{t12}{phonetic_only_phone_mean} & \WordStat{t15}{neural_only_phone_mean} & \WordStat{t15}{phonetic_only_phone_mean} \\
Dictionary coverage & \WordStat{t12}{neural_only_phone_covered}/\WordStat{t12}{neural_only_phone_total} & \WordStat{t12}{phonetic_only_phone_covered}/\WordStat{t12}{phonetic_only_phone_total} & \WordStat{t15}{neural_only_phone_covered}/\WordStat{t15}{neural_only_phone_total} & \WordStat{t15}{phonetic_only_phone_covered}/\WordStat{t15}{phonetic_only_phone_total} \\
}

\providecommand{\InputStat}[3]{\csname inputstat@#1@#2@#3\endcsname}
\expandafter\def\csname inputstat@t12@full@both_experts_wrong\endcsname{305}
\expandafter\def\csname inputstat@t12@full@both_wrong_recovered\endcsname{31}
\expandafter\def\csname inputstat@t12@full@delta_wer_pp\endcsname{0.000}
\expandafter\def\csname inputstat@t12@full@exact_sentences\endcsname{570}
\expandafter\def\csname inputstat@t12@full@full_correct_became_wrong\endcsname{0}
\expandafter\def\csname inputstat@t12@full@full_wrong_became_correct\endcsname{0}
\expandafter\def\csname inputstat@t12@full@reference_words\endcsname{5496}
\expandafter\def\csname inputstat@t12@full@rows\endcsname{880}
\expandafter\def\csname inputstat@t12@full@wer_percent\endcsname{10.135}
\expandafter\def\csname inputstat@t12@full@word_edits\endcsname{557}
\expandafter\def\csname inputstat@t12@without_neural_message@both_experts_wrong\endcsname{305}
\expandafter\def\csname inputstat@t12@without_neural_message@both_wrong_recovered\endcsname{33}
\expandafter\def\csname inputstat@t12@without_neural_message@delta_wer_pp\endcsname{0.873}
\expandafter\def\csname inputstat@t12@without_neural_message@exact_sentences\endcsname{559}
\expandafter\def\csname inputstat@t12@without_neural_message@full_correct_became_wrong\endcsname{35}
\expandafter\def\csname inputstat@t12@without_neural_message@full_wrong_became_correct\endcsname{24}
\expandafter\def\csname inputstat@t12@without_neural_message@reference_words\endcsname{5496}
\expandafter\def\csname inputstat@t12@without_neural_message@rows\endcsname{880}
\expandafter\def\csname inputstat@t12@without_neural_message@wer_percent\endcsname{11.008}
\expandafter\def\csname inputstat@t12@without_neural_message@word_edits\endcsname{605}
\expandafter\def\csname inputstat@t12@without_phonetic_message@both_experts_wrong\endcsname{305}
\expandafter\def\csname inputstat@t12@without_phonetic_message@both_wrong_recovered\endcsname{34}
\expandafter\def\csname inputstat@t12@without_phonetic_message@delta_wer_pp\endcsname{1.346}
\expandafter\def\csname inputstat@t12@without_phonetic_message@exact_sentences\endcsname{545}
\expandafter\def\csname inputstat@t12@without_phonetic_message@full_correct_became_wrong\endcsname{47}
\expandafter\def\csname inputstat@t12@without_phonetic_message@full_wrong_became_correct\endcsname{22}
\expandafter\def\csname inputstat@t12@without_phonetic_message@reference_words\endcsname{5496}
\expandafter\def\csname inputstat@t12@without_phonetic_message@rows\endcsname{880}
\expandafter\def\csname inputstat@t12@without_phonetic_message@wer_percent\endcsname{11.481}
\expandafter\def\csname inputstat@t12@without_phonetic_message@word_edits\endcsname{631}
\expandafter\def\csname inputstat@t12@without_both_messages@both_experts_wrong\endcsname{305}
\expandafter\def\csname inputstat@t12@without_both_messages@both_wrong_recovered\endcsname{24}
\expandafter\def\csname inputstat@t12@without_both_messages@delta_wer_pp\endcsname{4.130}
\expandafter\def\csname inputstat@t12@without_both_messages@exact_sentences\endcsname{465}
\expandafter\def\csname inputstat@t12@without_both_messages@full_correct_became_wrong\endcsname{126}
\expandafter\def\csname inputstat@t12@without_both_messages@full_wrong_became_correct\endcsname{21}
\expandafter\def\csname inputstat@t12@without_both_messages@reference_words\endcsname{5496}
\expandafter\def\csname inputstat@t12@without_both_messages@rows\endcsname{880}
\expandafter\def\csname inputstat@t12@without_both_messages@wer_percent\endcsname{14.265}
\expandafter\def\csname inputstat@t12@without_both_messages@word_edits\endcsname{784}
\expandafter\def\csname inputstat@t12@without_direct_neural@both_experts_wrong\endcsname{305}
\expandafter\def\csname inputstat@t12@without_direct_neural@both_wrong_recovered\endcsname{6}
\expandafter\def\csname inputstat@t12@without_direct_neural@delta_wer_pp\endcsname{4.567}
\expandafter\def\csname inputstat@t12@without_direct_neural@exact_sentences\endcsname{518}
\expandafter\def\csname inputstat@t12@without_direct_neural@full_correct_became_wrong\endcsname{70}
\expandafter\def\csname inputstat@t12@without_direct_neural@full_wrong_became_correct\endcsname{18}
\expandafter\def\csname inputstat@t12@without_direct_neural@reference_words\endcsname{5496}
\expandafter\def\csname inputstat@t12@without_direct_neural@rows\endcsname{880}
\expandafter\def\csname inputstat@t12@without_direct_neural@wer_percent\endcsname{14.702}
\expandafter\def\csname inputstat@t12@without_direct_neural@word_edits\endcsname{808}
\expandafter\def\csname inputstat@t15@full@both_experts_wrong\endcsname{262}
\expandafter\def\csname inputstat@t15@full@both_wrong_recovered\endcsname{49}
\expandafter\def\csname inputstat@t15@full@delta_wer_pp\endcsname{0.000}
\expandafter\def\csname inputstat@t15@full@exact_sentences\endcsname{1189}
\expandafter\def\csname inputstat@t15@full@full_correct_became_wrong\endcsname{0}
\expandafter\def\csname inputstat@t15@full@full_wrong_became_correct\endcsname{0}
\expandafter\def\csname inputstat@t15@full@reference_words\endcsname{9445}
\expandafter\def\csname inputstat@t15@full@rows\endcsname{1426}
\expandafter\def\csname inputstat@t15@full@wer_percent\endcsname{4.447}
\expandafter\def\csname inputstat@t15@full@word_edits\endcsname{420}
\expandafter\def\csname inputstat@t15@without_neural_message@both_experts_wrong\endcsname{262}
\expandafter\def\csname inputstat@t15@without_neural_message@both_wrong_recovered\endcsname{45}
\expandafter\def\csname inputstat@t15@without_neural_message@delta_wer_pp\endcsname{0.720}
\expandafter\def\csname inputstat@t15@without_neural_message@exact_sentences\endcsname{1158}
\expandafter\def\csname inputstat@t15@without_neural_message@full_correct_became_wrong\endcsname{52}
\expandafter\def\csname inputstat@t15@without_neural_message@full_wrong_became_correct\endcsname{21}
\expandafter\def\csname inputstat@t15@without_neural_message@reference_words\endcsname{9445}
\expandafter\def\csname inputstat@t15@without_neural_message@rows\endcsname{1426}
\expandafter\def\csname inputstat@t15@without_neural_message@wer_percent\endcsname{5.167}
\expandafter\def\csname inputstat@t15@without_neural_message@word_edits\endcsname{488}
\expandafter\def\csname inputstat@t15@without_phonetic_message@both_experts_wrong\endcsname{262}
\expandafter\def\csname inputstat@t15@without_phonetic_message@both_wrong_recovered\endcsname{42}
\expandafter\def\csname inputstat@t15@without_phonetic_message@delta_wer_pp\endcsname{0.720}
\expandafter\def\csname inputstat@t15@without_phonetic_message@exact_sentences\endcsname{1153}
\expandafter\def\csname inputstat@t15@without_phonetic_message@full_correct_became_wrong\endcsname{51}
\expandafter\def\csname inputstat@t15@without_phonetic_message@full_wrong_became_correct\endcsname{15}
\expandafter\def\csname inputstat@t15@without_phonetic_message@reference_words\endcsname{9445}
\expandafter\def\csname inputstat@t15@without_phonetic_message@rows\endcsname{1426}
\expandafter\def\csname inputstat@t15@without_phonetic_message@wer_percent\endcsname{5.167}
\expandafter\def\csname inputstat@t15@without_phonetic_message@word_edits\endcsname{488}
\expandafter\def\csname inputstat@t15@without_both_messages@both_experts_wrong\endcsname{262}
\expandafter\def\csname inputstat@t15@without_both_messages@both_wrong_recovered\endcsname{23}
\expandafter\def\csname inputstat@t15@without_both_messages@delta_wer_pp\endcsname{2.668}
\expandafter\def\csname inputstat@t15@without_both_messages@exact_sentences\endcsname{1043}
\expandafter\def\csname inputstat@t15@without_both_messages@full_correct_became_wrong\endcsname{167}
\expandafter\def\csname inputstat@t15@without_both_messages@full_wrong_became_correct\endcsname{21}
\expandafter\def\csname inputstat@t15@without_both_messages@reference_words\endcsname{9445}
\expandafter\def\csname inputstat@t15@without_both_messages@rows\endcsname{1426}
\expandafter\def\csname inputstat@t15@without_both_messages@wer_percent\endcsname{7.115}
\expandafter\def\csname inputstat@t15@without_both_messages@word_edits\endcsname{672}
\expandafter\def\csname inputstat@t15@without_direct_neural@both_experts_wrong\endcsname{262}
\expandafter\def\csname inputstat@t15@without_direct_neural@both_wrong_recovered\endcsname{5}
\expandafter\def\csname inputstat@t15@without_direct_neural@delta_wer_pp\endcsname{49.465}
\expandafter\def\csname inputstat@t15@without_direct_neural@exact_sentences\endcsname{544}
\expandafter\def\csname inputstat@t15@without_direct_neural@full_correct_became_wrong\endcsname{651}
\expandafter\def\csname inputstat@t15@without_direct_neural@full_wrong_became_correct\endcsname{6}
\expandafter\def\csname inputstat@t15@without_direct_neural@reference_words\endcsname{9445}
\expandafter\def\csname inputstat@t15@without_direct_neural@rows\endcsname{1426}
\expandafter\def\csname inputstat@t15@without_direct_neural@wer_percent\endcsname{53.912}
\expandafter\def\csname inputstat@t15@without_direct_neural@word_edits\endcsname{5092}
\newcommand{\InputMaskRows}{%
Complete input & \InputStat{t12}{full}{word_edits} & \InputStat{t12}{full}{wer_percent} & \InputStat{t12}{full}{exact_sentences} & \InputStat{t15}{full}{word_edits} & \InputStat{t15}{full}{wer_percent} & \InputStat{t15}{full}{exact_sentences} \\
Neural message masked & \InputStat{t12}{without_neural_message}{word_edits} & \InputStat{t12}{without_neural_message}{wer_percent} & \InputStat{t12}{without_neural_message}{exact_sentences} & \InputStat{t15}{without_neural_message}{word_edits} & \InputStat{t15}{without_neural_message}{wer_percent} & \InputStat{t15}{without_neural_message}{exact_sentences} \\
Phonetic message masked & \InputStat{t12}{without_phonetic_message}{word_edits} & \InputStat{t12}{without_phonetic_message}{wer_percent} & \InputStat{t12}{without_phonetic_message}{exact_sentences} & \InputStat{t15}{without_phonetic_message}{word_edits} & \InputStat{t15}{without_phonetic_message}{wer_percent} & \InputStat{t15}{without_phonetic_message}{exact_sentences} \\
Both messages masked & \InputStat{t12}{without_both_messages}{word_edits} & \InputStat{t12}{without_both_messages}{wer_percent} & \InputStat{t12}{without_both_messages}{exact_sentences} & \InputStat{t15}{without_both_messages}{word_edits} & \InputStat{t15}{without_both_messages}{wer_percent} & \InputStat{t15}{without_both_messages}{exact_sentences} \\
Direct neural input zeroed & \InputStat{t12}{without_direct_neural}{word_edits} & \InputStat{t12}{without_direct_neural}{wer_percent} & \InputStat{t12}{without_direct_neural}{exact_sentences} & \InputStat{t15}{without_direct_neural}{word_edits} & \InputStat{t15}{without_direct_neural}{wer_percent} & \InputStat{t15}{without_direct_neural}{exact_sentences} \\
}

\providecommand{\NeuralInputStat}[2]{\csname neuralinput@#1@#2\endcsname}
\expandafter\def\csname neuralinput@t12@zero_retained\endcsname{2}
\expandafter\def\csname neuralinput@t12@paired_outside\endcsname{25}
\expandafter\def\csname neuralinput@t12@paired_recovered\endcsname{30}
\expandafter\def\csname neuralinput@t12@paired_rows\endcsname{856}
\expandafter\def\csname neuralinput@t12@full_rows\endcsname{880}
\expandafter\def\csname neuralinput@t12@paired_words\endcsname{5,258}
\expandafter\def\csname neuralinput@t12@baseline_edits\endcsname{494}
\expandafter\def\csname neuralinput@t12@donor_mean_edits\endcsname{5,611}
\expandafter\def\csname neuralinput@t12@baseline_wer\endcsname{9.395}
\expandafter\def\csname neuralinput@t12@donor_mean_wer\endcsname{106.714}
\expandafter\def\csname neuralinput@t15@zero_retained\endcsname{4}
\expandafter\def\csname neuralinput@t15@paired_outside\endcsname{36}
\expandafter\def\csname neuralinput@t15@paired_recovered\endcsname{43}
\expandafter\def\csname neuralinput@t15@paired_rows\endcsname{1,363}
\expandafter\def\csname neuralinput@t15@full_rows\endcsname{1,426}
\expandafter\def\csname neuralinput@t15@paired_words\endcsname{8,902}
\expandafter\def\csname neuralinput@t15@baseline_edits\endcsname{398}
\expandafter\def\csname neuralinput@t15@donor_mean_edits\endcsname{9,589}
\expandafter\def\csname neuralinput@t15@baseline_wer\endcsname{4.471}
\expandafter\def\csname neuralinput@t15@donor_mean_wer\endcsname{107.717}
\ExplSyntaxOn
\str_if_eq:eeF {\PaperResultSet} {cross}
{\PackageError{masbit-neural-input}{Neural~input~analysis~requires~Cross}{Use~the~verified~cohort.}}
\ExplSyntaxOff

\providecommand{\CapacityStat}[1]{\csname capacitystat@#1\endcsname}
\SetScore{single_r24}{t12}{10.231425091352}
\SetScore{single_r24}{t15public}{5.787}
\SetScore{single_r24}{t15private}{6.889}
\SetScore{message_only}{t12}{8.444985789687372}
\SetScore{message_only}{t15public}{4.939}
\SetScore{message_only}{t15private}{5.492}
\expandafter\def\csname capacitystat@t12_N_unique\endcsname{259}
\expandafter\def\csname capacitystat@t12_N_retained\endcsname{222}
\expandafter\def\csname capacitystat@t12_N_unseen_positions\endcsname{216}
\expandafter\def\csname capacitystat@t12_N_unseen_unique\endcsname{31}
\expandafter\def\csname capacitystat@t12_N_unseen_retained\endcsname{30}
\expandafter\def\csname capacitystat@t12_N_dictionary_covered_substitutions\endcsname{245}
\expandafter\def\csname capacitystat@t12_N_retention_percent\endcsname{85.7}
\expandafter\def\csname capacitystat@t12_N_unseen_unique_percent\endcsname{14.4}
\expandafter\def\csname capacitystat@t12_N_unseen_retention_percent\endcsname{96.8}
\expandafter\def\csname capacitystat@t12_N_phone_mean\endcsname{0.597}
\expandafter\def\csname capacitystat@t12_P_unique\endcsname{241}
\expandafter\def\csname capacitystat@t12_P_retained\endcsname{190}
\expandafter\def\csname capacitystat@t12_P_unseen_positions\endcsname{216}
\expandafter\def\csname capacitystat@t12_P_unseen_unique\endcsname{22}
\expandafter\def\csname capacitystat@t12_P_unseen_retained\endcsname{17}
\expandafter\def\csname capacitystat@t12_P_dictionary_covered_substitutions\endcsname{223}
\expandafter\def\csname capacitystat@t12_P_retention_percent\endcsname{78.8}
\expandafter\def\csname capacitystat@t12_P_unseen_unique_percent\endcsname{10.2}
\expandafter\def\csname capacitystat@t12_P_unseen_retention_percent\endcsname{77.3}
\expandafter\def\csname capacitystat@t12_P_phone_mean\endcsname{0.534}
\expandafter\def\csname capacitystat@t15_N_unique\endcsname{264}
\expandafter\def\csname capacitystat@t15_N_retained\endcsname{233}
\expandafter\def\csname capacitystat@t15_N_unseen_positions\endcsname{229}
\expandafter\def\csname capacitystat@t15_N_unseen_unique\endcsname{31}
\expandafter\def\csname capacitystat@t15_N_unseen_retained\endcsname{28}
\expandafter\def\csname capacitystat@t15_N_dictionary_covered_substitutions\endcsname{251}
\expandafter\def\csname capacitystat@t15_N_retention_percent\endcsname{88.3}
\expandafter\def\csname capacitystat@t15_N_unseen_unique_percent\endcsname{13.5}
\expandafter\def\csname capacitystat@t15_N_unseen_retention_percent\endcsname{90.3}
\expandafter\def\csname capacitystat@t15_N_phone_mean\endcsname{0.594}
\expandafter\def\csname capacitystat@t15_P_unique\endcsname{187}
\expandafter\def\csname capacitystat@t15_P_retained\endcsname{150}
\expandafter\def\csname capacitystat@t15_P_unseen_positions\endcsname{229}
\expandafter\def\csname capacitystat@t15_P_unseen_unique\endcsname{21}
\expandafter\def\csname capacitystat@t15_P_unseen_retained\endcsname{19}
\expandafter\def\csname capacitystat@t15_P_dictionary_covered_substitutions\endcsname{173}
\expandafter\def\csname capacitystat@t15_P_retention_percent\endcsname{80.2}
\expandafter\def\csname capacitystat@t15_P_unseen_unique_percent\endcsname{9.2}
\expandafter\def\csname capacitystat@t15_P_unseen_retention_percent\endcsname{90.5}
\expandafter\def\csname capacitystat@t15_P_phone_mean\endcsname{0.528}
\expandafter\def\csname capacitystat@all_N_unique\endcsname{523}
\expandafter\def\csname capacitystat@all_N_retained\endcsname{455}
\expandafter\def\csname capacitystat@all_N_unseen_positions\endcsname{445}
\expandafter\def\csname capacitystat@all_N_unseen_unique\endcsname{62}
\expandafter\def\csname capacitystat@all_N_unseen_retained\endcsname{58}
\expandafter\def\csname capacitystat@all_N_dictionary_covered_substitutions\endcsname{496}
\expandafter\def\csname capacitystat@all_N_retention_percent\endcsname{87.0}
\expandafter\def\csname capacitystat@all_N_unseen_unique_percent\endcsname{13.9}
\expandafter\def\csname capacitystat@all_N_unseen_retention_percent\endcsname{93.5}
\expandafter\def\csname capacitystat@all_N_phone_mean\endcsname{0.596}
\expandafter\def\csname capacitystat@all_P_unique\endcsname{428}
\expandafter\def\csname capacitystat@all_P_retained\endcsname{340}
\expandafter\def\csname capacitystat@all_P_unseen_positions\endcsname{445}
\expandafter\def\csname capacitystat@all_P_unseen_unique\endcsname{43}
\expandafter\def\csname capacitystat@all_P_unseen_retained\endcsname{36}
\expandafter\def\csname capacitystat@all_P_dictionary_covered_substitutions\endcsname{396}
\expandafter\def\csname capacitystat@all_P_retention_percent\endcsname{79.4}
\expandafter\def\csname capacitystat@all_P_unseen_unique_percent\endcsname{9.7}
\expandafter\def\csname capacitystat@all_P_unseen_retention_percent\endcsname{83.7}
\expandafter\def\csname capacitystat@all_P_phone_mean\endcsname{0.532}
\expandafter\def\csname capacitystat@t12_speedup\endcsname{1.8}
\expandafter\def\csname capacitystat@t12_mas_minutes\endcsname{26}
\expandafter\def\csname capacitystat@t12_cascade_minutes\endcsname{46}
\expandafter\def\csname capacitystat@t15_speedup\endcsname{1.8}
\expandafter\def\csname capacitystat@t15_mas_minutes\endcsname{35}
\expandafter\def\csname capacitystat@t15_cascade_minutes\endcsname{63}

\title{MAS-BIT: A Multi-Agent System for End-to-End BraIn-to-Text Decoding}

\author{Xinjie Zhang$^{1,*}$\quad Linyang He$^{1,*,\dagger}$\quad Yizi Zhang$^{2}$\quad Yuesheng Ma$^{1}$\quad Nima Mesgarani$^{1}$\\
$^{1}$Columbia University \quad $^{2}$Stanford University \\
$^*$Equal Contribution \quad $^\dagger$Project Lead \\
\texttt{\{xinjie.zhang, linyang.he\}@columbia.edu \quad nima@ee.columbia.edu}
}

\newif\ifshowrevisions
\showrevisionsfalse
\newcommand{\red}[1]{\ifshowrevisions\textcolor{red}{#1}\else\textcolor{.}{#1}\fi}
\SetScore{cm1cascade100}{t12}{7.294627148463934}
\SetScore{cm1cascade100}{t15public}{4.696}
\SetScore{cm1cascade100}{t15private}{4.629}
\SetScore{cm1cascade75}{t12}{7.3081607795371495}
\SetScore{cm1cascade75}{t15public}{4.696}
\SetScore{cm1cascade75}{t15private}{4.629}
\SetScore{cm1cascade50}{t12}{7.362295303830018}
\SetScore{cm1cascade50}{t15public}{4.787}
\SetScore{cm1cascade50}{t15private}{4.645}
\SetScore{cm1cascade37}{t12}{7.375828934903235}
\SetScore{cm1cascade37}{t15public}{4.787}
\SetScore{cm1cascade37}{t15private}{4.645}
\SetScore{cm1cascade25}{t12}{7.389362565976451}
\SetScore{cm1cascade25}{t15public}{4.787}
\SetScore{cm1cascade25}{t15private}{4.708}
\SetScore{cm1cascade18}{t12}{7.402896197049669}
\SetScore{cm1cascade18}{t15public}{4.818}
\SetScore{cm1cascade18}{t15private}{4.676}
\SetScore{cm1cascade12}{t12}{7.443497090269319}
\SetScore{cm1cascade12}{t15public}{4.757}
\SetScore{cm1cascade12}{t15private}{4.692}
\SetScore{cm1cascade9}{t12}{7.538232507781838}
\SetScore{cm1cascade9}{t15public}{4.727}
\SetScore{cm1cascade9}{t15private}{4.739}
\SetScore{cm1cascade6}{t12}{7.524698876708621}
\SetScore{cm1cascade6}{t15public}{4.909}
\SetScore{cm1cascade6}{t15private}{4.802}
\SetScore{cm1cascade4}{t12}{7.660035187440791}
\SetScore{cm1cascade4}{t15public}{5.06}
\SetScore{cm1cascade4}{t15private}{4.959}
\SetScore{cm1cascade3}{t12}{7.808905129246177}
\SetScore{cm1cascade3}{t15public}{5.151}
\SetScore{cm1cascade3}{t15private}{5.006}
\SetScore{cm1cascade2}{t12}{8.01190959534443}
\SetScore{cm1cascade2}{t15public}{5.363}
\SetScore{cm1cascade2}{t15private}{5.037}
\SetScore{cm1cascade1}{t12}{8.891595615103533}
\SetScore{cm1cascade1}{t15public}{5.787}
\SetScore{cm1cascade1}{t15private}{5.555}
\SetScore{single_borealis}{t12}{12.478007849506023}
\SetScore{single_borealis}{t15public}{8.515}
\SetScore{single_borealis}{t15private}{8.553}
\SetScore{refinement}{t12}{7.1592908377317634}
\SetScore{refinement}{t15public}{3.939}
\SetScore{refinement}{t15private}{4.739}
\SetScore{lightbeam}{t12}{8.066044119637299}
\SetScore{lightbeam}{t15public}{6.151}
\SetScore{lightbeam}{t15private}{7.438}
\newcommand{\MasTwelveSeconds}{1538.8}
\newcommand{\CascadeTwelveSeconds}{2789.6}
\newcommand{\MasFifteenSeconds}{2091.4}
\newcommand{\CascadeFifteenSeconds}{3761.6}
\newcommand{\LightBeamTwelveRAM}{17.2}
\newcommand{\LightBeamFifteenRAM}{31.8}
\newcommand{\LightBeamTwelveSeconds}{3563.5}
\newcommand{\LightBeamFifteenSeconds}{6008.0}
\newcommand{\LightBeamTwelveRatio}{2.3}
\newcommand{\LightBeamFifteenRatio}{2.9}
\newcommand{\MasTwelveGPU}{8.559}
\newcommand{\MasFifteenGPU}{8.565}
\newcommand{\CascadeTwelveGPU}{25.211}
\newcommand{\CascadeFifteenGPU}{25.274}
\newcommand{\LightBeamTwelveGPU}{12.553}
\newcommand{\LightBeamFifteenGPU}{14.750}
\newcommand{\LaptopMasTwelveDecode}{1210.2}
\newcommand{\LaptopMasFifteenDecode}{1690.6}
\newcommand{\LaptopLightBeamTwelveDecode}{3204.6}
\newcommand{\LaptopLightBeamFifteenDecode}{6021.1}
\newcommand{\LaptopDecodeTwelveRatio}{2.6}
\newcommand{\LaptopDecodeFifteenRatio}{3.6}
\newcommand{\LaptopMasTwelveRAM}{12.6}
\newcommand{\LaptopMasFifteenRAM}{31.0}
\newcommand{\LaptopLightBeamTwelveRAM}{17.3}
\newcommand{\LaptopLightBeamFifteenRAM}{32.0}
\newcommand{\LaptopMasTwelveGPU}{8.662}
\newcommand{\LaptopMasFifteenGPU}{8.648}
\newcommand{\LaptopLightBeamTwelveGPU}{12.629}
\newcommand{\LaptopLightBeamFifteenGPU}{14.785}
\newcommand{\DesktopMasTwelveDecode}{550.5}
\newcommand{\DesktopMasFifteenDecode}{775.1}
\newcommand{\DesktopLightBeamTwelveDecode}{1483.5}
\newcommand{\DesktopLightBeamFifteenDecode}{2332.7}
\newcommand{\DesktopDecodeTwelveRatio}{2.7}
\newcommand{\DesktopDecodeFifteenRatio}{3.0}
\newcommand{\DesktopMasTwelveRAM}{12.8}
\newcommand{\DesktopMasFifteenRAM}{31.2}
\newcommand{\DesktopLightBeamTwelveRAM}{17.4}
\newcommand{\DesktopLightBeamFifteenRAM}{32.0}
\newcommand{\DesktopMasTwelveGPU}{8.695}
\newcommand{\DesktopMasFifteenGPU}{8.695}
\newcommand{\DesktopLightBeamTwelveGPU}{12.648}
\newcommand{\DesktopLightBeamFifteenGPU}{14.887}

\providecommand{\papermode}{2}
\ifcase\papermode\relax
\or
  \iclrfinalcopy
\or
  \iclrfinalcopy
\else
  \PackageError{masbit}{Invalid papermode: use 0, 1, or 2}
    {Set \string\papermode\space to 0 (review), 1 (published), or 2 (arXiv).}
\fi
\begin{document}

\maketitle
\ifnum\papermode=2\relax
  \lhead{Preprint}
\fi

\begin{abstract}
Speech brain-computer interfaces aim to restore communication by translating neural activity into text, yet end-to-end decoders have historically trailed cascaded systems in accuracy. We introduce a Multi-Agent System for BraIn-to-Text (MAS-BIT), the first multi-agent framework for end-to-end brain-to-text decoding from intracortical recordings. MAS-BIT coordinates three specialized agents: a neural agent that interprets continuous neural representations, a phonetic agent that decodes soft phoneme posteriors, and an arbitration agent that integrates their complementary evidence with direct access to neural representations to generate the final transcription. The expert agents communicate with the arbiter through token-embedding-based latent communication, constructing continuous messages as probability-weighted combinations of token embeddings. These messages retain alternative token hypotheses and provide a differentiable interface through which sentence-level supervision jointly trains the neural encoders and specialized agents across participants. MAS-BIT achieves state-of-the-art end-to-end decoding performance on the Brain-to-Text ’24 and ’25 benchmarks, with word error rates of 6.929\% on ’24 and 3.606\% on the ’25 public test set (3.892\% on the private test set). These results surpass the published BIT cascade on the ’25 public set and narrow the gap on ’24, without ensembling independently trained models or using external language-model rescoring.\ifnum\papermode>0\relax\footnote{Code is available \href{https://github.com/BaochaiXue/Multi_Agent_System_for_BraIn_to_Text}{here}, and model weights are available \href{https://huggingface.co/XinjieZhang/MAS-BIT}{here}.}\fi
\end{abstract}

\section{Introduction}
\label{sec:introduction}

Speech brain-computer interfaces (BCIs) aim to restore communication by translating neural activity associated with attempted speech into text~\citep{akbari2019towards,moses2021neuroprosthesis,metzger2023high,willett2023high,card2024accurate,wairagkar2025instantaneous}.
Accurate decoding requires recovering linguistic content from noisy recordings and limited paired training data.
Cascaded systems combine a neural-to-phoneme decoder with external $n$-gram/WFST decoding (and optional rescoring) to reconstruct sentences. In contrast, end-to-end systems connect neural representations directly to a pretrained language model and optimize for the target transcription
~\citep{feng2024towards,zhang2026bit}.
In the Brain-to-Text benchmark comparisons reported by \citet{zhang2026bit}, end-to-end decoding remains less accurate than cascaded decoding.

End-to-end decoding is nevertheless attractive on both modeling and deployment grounds. By mapping neural activity directly to text, it avoids the information bottleneck and error propagation of a separate phoneme-to-word stage, directly aligns neural representations with sentence-level objectives, and leverages pretrained language models without fixed pronunciation lexicons or heavily tuned search heuristics. Moreover, conventional WFST-based cascades rely on a separate decoding stack of large n-gram/WFST graphs and external language-model rescoring that can impose hundreds of gigabytes of host-memory overhead (Table~\ref{tab:inference_memory}), whereas an end-to-end decoder offers a single, substantially lighter inference pipeline. We therefore ask whether end-to-end decoding can approach or surpass cascaded accuracy while preserving these advantages.

A neural encoder provides two views of the same recording: continuous hidden representations and phoneme posteriors obtained from a supervised phoneme readout.
The hidden representations expose learned neural features to a language decoder, while the posteriors organize this evidence around phoneme identities and their relative probabilities.
Posterior-based interfaces in speech recognition illustrate how such structured distributions can condition a language model~\citep{ma2025legoslm}.
For brain-to-text decoding, these two views suggest a division of labor: one expert interprets continuous neural representations, and another interprets explicit phonetic evidence.
Their different inputs can lead to different transcription hypotheses, even though both originate from the same neural observation.
We investigate whether these specialists can learn to provide complementary evidence for a common final transcription.

The communication interface determines which parts of the experts' judgments reach the final decoder.
Passing a single generated text sequence conveys the selected tokens but discards their relative support against alternative tokens at each position.
The discrete selection also blocks ordinary backpropagation through the message into the expert probabilities that produced it.
Prior work on differentiable communication and continuous exchange between language-model agents provides alternatives to this text interface~\citep{foerster2016communicate,pham2024cipher,zou2026latentmas}.
For neural decoding, we seek an interface that conveys information about competing token hypotheses and allows the final transcription objective to shape the experts through their messages.

Here, we introduce an end-to-end \textbf{M}ulti-\textbf{A}gent \textbf{S}ystem for \textbf{B}ra\textbf{I}n-to-\textbf{T}ext (\textbf{MAS-BIT}) that separates specialist decoding from arbitration.
A \emph{neural expert} reads continuous encoder representations, a \emph{phonetic expert} reads soft phoneme posteriors, and an \emph{arbiter} integrates their messages with direct neural evidence.
The roles share one language-model backbone with separate adapters.
Rather than passing only selected text, the experts communicate through probability-weighted token embeddings that retain alternative hypotheses and allow the final transcription loss to supervise the experts through their messages.
\red{Participant-specific input projections allow two BIT-initialized encoders to process recordings from either participant, and staged specialization followed by joint optimization trains the encoders and agents together} in a single end-to-end system.

We evaluate \red{one jointly trained model} on the Brain-to-Text '24 and '25 benchmarks, obtaining \red{\WER{mas}{t12}}\% WER on the '24 holdout and \red{\WER{mas}{t15public}}\%/\red{\WER{mas}{t15private}}\% on the '25 public/private subsets.
On the '24 holdout and '25 public subset, MAS-BIT achieves state-of-the-art end-to-end performance, surpassing BIT's single-model decoder and end-to-end ensemble without ensembling independently trained models or external language-model rescoring~\citep{zhang2026bit}.
\red{It also outperforms the published BIT cascade on '25 public and narrows the gap on '24, while substantially reducing memory use and decoding time relative to the reproduced cascade and BIT + LightBeam~\citep{feghhi2026lightbeam}.}
Controlled comparisons show that specialized experts outperform a single decoder given both views \red{(including capacity-expanded variants) and sequential refinement controls matched in parameters and decoding passes; on the validation sets,} the arbiter recovers correct transcriptions when both drafts are wrong.
These results show that jointly trained specialists make better use of neural and phonetic evidence than feeding both views to one decoder or selecting between expert drafts.

\section{Related Work}
\label{sec:related_work}

\noindent\textbf{Speech neuroprostheses and end-to-end neural decoding.}
Intracortical speech neuroprostheses commonly decode intermediate phonetic units before language-model-based sentence reconstruction
\citep{willett2023high,card2024accurate}.
More recent work directly connects neural representations to pretrained language models
\citep{feng2024towards,zhang2026bit}.
In particular, BIT combines cross-task, cross-species neural pretraining with a phoneme-adapted encoder and an audio-LLM decoder
\citep{zhang2026bit}.
MoDAl combines parallel neural encoders through alignment and decorrelation losses~\citep{chen2026modal}, while BrainWhisperer adapts a pretrained Whisper encoder--decoder to neural inputs~\citep{boccato2026brainwhisperer}.
Other direct decoders use a BART word decoder~\citep{olak2026decoding}, transfer Wav2Vec2 representations to character CTC decoding~\citep{fiedler2025wav2vec2}, or train a character-CTC Conformer~\citep{khanday2026endtoend}.
\citet{Fogg2026.07.23.739430} train a shared neural phoneme decoder across six participants using participant-specific input projections.

\noindent\textbf{Posterior-based interfaces and signal-grounded refinement.}
CTC posteriors provide a probabilistic interface between low-level sequence representations and language models.
LegoSLM uses posterior-weighted token embeddings to connect a speech encoder to an LLM without committing to a one-best intermediate sequence
\citep{ma2025legoslm}.
Deliberation systems refine first-pass hypotheses while retaining access to the original signal
\citep{hu2020deliberation,dekel2026nle}.
MAS-BIT combines these ideas via a neural expert that reads continuous encoder states, a phonetic expert that reads CTC posteriors, and an arbiter that integrates their messages with direct neural evidence.

\noindent\textbf{Multi-agent and differentiable language-model communication.}
Multi-agent language-model systems improve generation by separating proposal generation from downstream aggregation
\citep{du2024debate,wang2025moa}.
Continuous communication further allows information to pass through differentiable representations rather than only discrete text
\citep{foerster2016communicate,pham2024cipher,ramesh2025activations,zou2026latentmas}.
MAS-BIT adopts a fixed two-expert--one-arbiter topology with a shared frozen language-model backbone and role-specific LoRA adapters
\citep{hu2022lora}.
Each expert specializes in its input evidence, and its generated draft is replayed as latent messages optimized via the downstream arbiter loss.

\begin{figure}[!t]
\centering
\includegraphics[width=\linewidth]{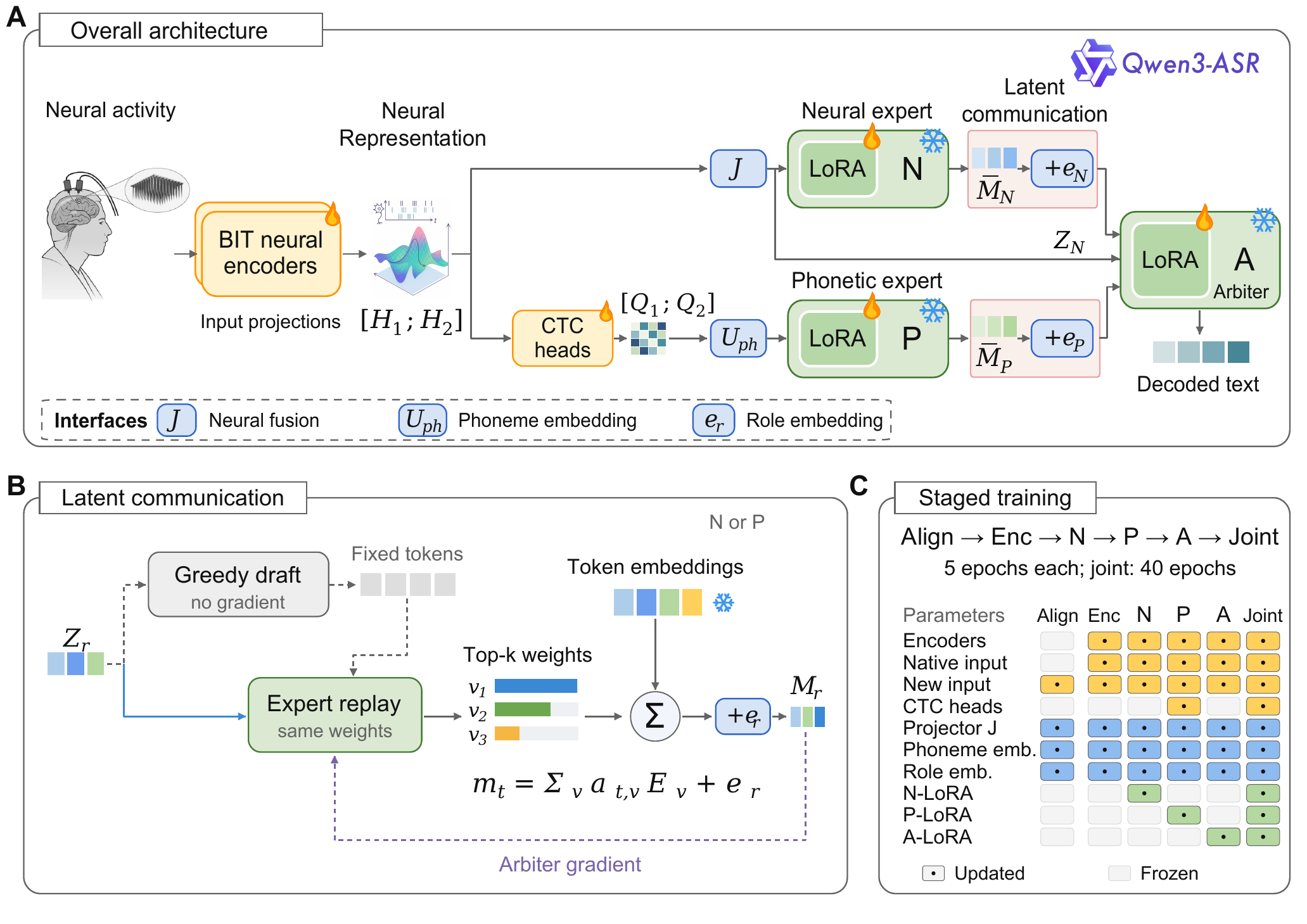}
\caption{\footnotesize \emph{MAS-BIT combines neural and phonetic evidence through specialized experts and an arbiter.} \textbf{(A)} Two BIT-initialized encoders process each recording through participant-specific input projections. Fused hidden states condition the neural expert and arbiter; both phoneme posteriors condition the phonetic expert. All three language agents share a frozen text backbone with separate adapters. \textbf{(B)} Each expert replays its greedy draft to construct latent messages from probability-weighted token embeddings. $Z_r$ denotes its input, $Z_N$ or $Z_P$, and $e_r$ identifies the message source. The arbiter loss trains active upstream components through these latent messages. \textbf{(C)} Five-epoch alignment\red{, encoder adaptation,} and specialization stages precede 40 epochs of joint optimization. Filled cells indicate parameter updates; frozen downstream weights retain input gradients. }
\label{fig:masbit_overview}
\end{figure}

\section{Methods}
\label{sec:methods}

MAS-BIT maps a neural recording to a sentence through three language agents: two specialized experts, the neural expert $N$ and phonetic expert $P$, and an arbiter $A$ (Figure~\ref{fig:masbit_overview}).
The experts read complementary representations of the same recording and send latent messages to the arbiter.
The arbiter combines these messages with direct neural evidence to produce the final transcription.
We describe the shared-backbone architecture and its staged-to-joint training procedure.

\subsection{Architecture}
\label{sec:neural_evidence}

\paragraph{Neural and phonetic evidence.}

Let $X\in\mathbb{R}^{T\times C}$ denote an utterance of intracortical features, $y=(y_1,\ldots,y_U)$ its text-token sequence including the end-of-sequence token, and $\pi$ its phoneme target.
The input combines threshold-crossing activity and spike-band power~\citep{nason2020low}; we describe preprocessing in Appendix~\ref{app:model}.
We process each recording with two neural Transformers initialized from the released T12 and T15 BIT encoders and their matching phoneme readouts~\citep{zhang2026bit}.
For input participant $s$, an input projection $R_{j,s}$ maps neural patches into encoder $j\in\{1,2\}$, producing $H_j=f_{\theta_j}(R_{j,s}(X))\in\mathbb{R}^{S\times d_n}$, where $S=\lceil T/5\rceil$ and $d_n=384$.
Each encoder retains its released input projection for its original participant and learns a new input projection for the other participant.
The two encoders process the same recording; the three language agents share their outputs.

A two-layer projector $J$ fuses the layer-normalized hidden states into the language model's embedding space.
Each encoder's phoneme readout produces a separate posterior, and the two expected phoneme embeddings are summed:
\begin{equation}
    \begin{aligned}
    Z_N &= J([\operatorname{LN}(H_1);\operatorname{LN}(H_2)]),\\
    Q_j &= \operatorname{softmax}(H_j W_{C,j}+b_{C,j}),\\
    Z_P &= Q_1 U_{\mathrm{ph},1}+Q_2 U_{\mathrm{ph},2}.
    \end{aligned}
    \label{eq:evidence_views}
\end{equation}
Here, $d=2048$ is the language-model width, and each $Q_j\in\mathbb{R}^{S\times 41}$ contains phoneme posteriors over 39 speech phones, silence, and the CTC blank.
The learned tables $U_{\mathrm{ph},j}\in\mathbb{R}^{41\times d}$ convert the two independently normalized posteriors into expected phoneme embeddings.
Thus, $Z_N$ retains continuous neural features, while $Z_P$ expresses the evidence in terms of the supervised phoneme inventory.
The phonetic interface preserves the full distribution, including blank probability, at every valid patch position.

\paragraph{Language agents and roles.}

All three agents use the Qwen3-ASR-1.7B~\citep{qwen2026asr} text decoder.
We share and freeze its weights and token embedding table, while each role has an independent LoRA adapter~\citep{hu2022lora}.
Appendix~\ref{app:model} lists the adapter settings.
We initialize the neural projector, phoneme embedding \red{tables}, message-role embeddings, and role adapters for MAS-BIT.

The neural expert $N$ conditions on $Z_N$ and predicts an English transcription directly from neural representations.
The phonetic expert $P$ conditions on $Z_P$ and also predicts English text tokens; its phonetic specialization comes from the posterior-based input.
The arbiter $A$ conditions on $Z_N$ together with the two expert messages defined below.
Each role conditions on its continuous input sequence and autoregressively generates text.
Role-specific prompts describe the input evidence and request an English transcription (Appendix~\ref{app:model}; Table~\ref{tab:role_prompts}).
Together, the input evidence, prompt, and separate adapter define each decoding role.
Padding positions are masked.

\subsection{Latent communication}
\label{sec:communication}

Latent messages are sequences of probability-weighted token embeddings, with learned role embeddings identifying their source.
Each expert $r\in\{N, P\}$ receives a continuous input $Z_r$: projected neural states $Z_N$ for the neural expert, or posterior-weighted phoneme embeddings $Z_P$ for the phonetic expert.
Conditioned on this input, the expert generates a greedy draft $\hat y^r$.
Draft generation uses the expert's previously generated tokens and stops at an end-of-sequence token or a fixed length cap.
During training, this rollout runs without gradient tracking and with language-model dropout disabled.
We then replay the generated draft through the same expert to compute next-token logits
\begin{equation}
    \ell^r_t = g_r(Z_r,\hat y^r_{<t}).
    \label{eq:replay}
\end{equation}
The replay is differentiable and uses the expert's own draft as its text context.
Let $\mathcal{K}^r_t$ be the indices of the 16 largest logits at position $t$.
With temperature $\tau=1$, the message content is a probability-weighted token embedding~\citep{pham2024cipher}:
\begin{equation}
    a^r_{t,v} =
    \frac{\exp(\ell^r_{t,v}/\tau)}
         {\sum_{u\in\mathcal{K}^r_t}\exp(\ell^r_{t,u}/\tau)},
    \qquad
    \bar m^r_t = \sum_{v\in\mathcal{K}^r_t}a^r_{t,v}E_v,
    \label{eq:soft_message}
\end{equation}
where $E_v\in\mathbb{R}^{d}$ is the shared frozen embedding of text token $v$.
The weights are renormalized over the selected support, allowing several candidate tokens to contribute to each message position.
We add a learned message-role embedding $e_r$ to each message token to indicate whether it originates from the neural or phonetic expert.
At inference, this gives $m^r_t=\bar m^r_t+e_r$.
Training applies dropout to the message content before adding the role embedding (Appendix~\ref{app:messages}).

The arbiter's continuous prefix is the concatenation
\begin{equation}
    Z_A = [Z_N;M_N;M_P],
    \qquad
    p_A(y\mid X)=\prod_{t=1}^{U}p_A(y_t\mid Z_A,y_{<t}).
    \label{eq:arbiter}
\end{equation}
Message masks retain valid draft positions, including the first end-of-sequence token.
This defines a single round of expert-to-arbiter communication with direct neural conditioning.
The arbiter loss backpropagates through the selected softmax weights and replay computation into active expert adapters, input interfaces, and the neural encoder.
The discrete draft tokens and top-$k$ membership are held fixed during this backward pass.

\paragraph{Inference.}
\label{sec:inference}
At inference, the model encodes the recording once, generates and replays the two expert drafts, and greedily decodes the arbiter's final sentence.
Neural augmentation and dropout are disabled.
The arbiter's transcription is the final system output.

\subsection{Training objectives and staged training}
\label{sec:objective}

We supervise each agent with teacher-forced cross-entropy on the target sentence, while constructing the expert messages from generated drafts.
Let $\mathcal{L}_{\mathrm{CE}}^r$ denote the summed token loss for role $r$, normalized by the number of target text tokens, and $\mathcal{L}_{\mathrm{CTC}}^{(l)}$ the mean phoneme CTC loss~\citep{graves2006connectionist} across the two encoders at layer $l$, normalized by the number of target phones.
We apply auxiliary CTC losses to intermediate encoder layers~\citep{lee2021interctc}.
The training objective is
\begin{equation}
\begin{split}
    \mathcal{L} ={}&
    \mathcal{L}_{\mathrm{CE}}^N+
    \mathcal{L}_{\mathrm{CE}}^P+
    \mathcal{L}_{\mathrm{CE}}^A+
    0.3\,\mathcal{L}_{\mathrm{CTC}}^{(7)}\\
    &+0.05\,\rho\,
    \frac{\mathcal{L}_{\mathrm{CTC}}^{(3)}+
          \mathcal{L}_{\mathrm{CTC}}^{(5)}}{2}.
\end{split}
\label{eq:objective}
\end{equation}
The coefficient $\rho$ increases linearly from zero to one over the first epoch in which the encoder is updated and remains one thereafter.
All three sentence losses are computed in every training stage; the stage determines which parameters are updated.
Text and phoneme losses retain their separate denominators, with the distributed normalization specified in Appendix~\ref{app:optimization}.

Inspired by the neural augmentation approach of \citet{Fogg2026.07.23.739430}, all three agents share the same augmented recording during expert generation, replay, and sentence supervision.
Augmentation includes temporal and electrode perturbations, as well as patch masking (Appendix~\ref{app:augmentation}).

\paragraph{Staged training.}
\label{sec:staged_optimization}

Training begins with five alignment epochs that update the input interfaces and newly initialized participant-specific input projections.
Five epochs of encoder adaptation are followed by five epochs each of neural-expert, phonetic-expert, and arbiter specialization.
All adaptation parameters are subsequently optimized together for 40 epochs:
\begin{equation}
    \red{\mathrm{Align}_5}\;\rightarrow\;\red{\mathrm{Enc}_5}\;\rightarrow\;N_5\;\rightarrow\;P_5\;\rightarrow\;A_5\;\rightarrow\;\mathrm{Joint}_{40}.
    \label{eq:stage_schedule}
\end{equation}
Subscripts denote epoch counts, giving a maximum of 65 epochs.
Figure~\ref{fig:masbit_overview}C summarizes the stage-specific updates; Table~\ref{tab:ownership} in Appendix~\ref{app:optimization} lists the active parameter groups.
The input interfaces remain active throughout training; both encoder trunks and participant-specific input projections remain active after alignment.
The $N$, $P$, and $A$ stages also update their role adapters, with both phoneme readouts updated in the $P$ stage.
The final joint stage updates all adaptation parameters.

Frozen downstream modules transmit input gradients, allowing the arbiter objective to supervise active upstream components, but receive no parameter or optimizer-state updates.

We optimize encoder weight matrices with Muon and the rest with AdamW~\citep{jordan2024muon,loshchilov2017decoupled}, using block-local warmup--cosine schedules (Appendix~\ref{app:optimization}).

\section{Evaluation}
\label{sec:evaluation}

\subsection{Tasks}
\label{sec:selection}

We evaluate MAS-BIT on attempted speech decoding using the Brain-to-Text '24 and '25 datasets from participants T12 and T15, respectively~\citep{willett2023high,willett2024brain,card2024accurate}.
In this task, we record neural activity while participants attempt to articulate a sentence, and the decoder predicts the sentence's transcription.
We measure performance by corpus word error rate (WER): the total number of word substitutions, deletions, and insertions divided by the number of reference words; lower WER indicates more accurate decoding.

We train one model jointly on 8,800 T12 and 8,072 T15 utterances, using participant-specific input projections and shared encoders, interfaces, and language agents.
We use all 880 and 1,426 provided labeled validation utterances for checkpoint selection.
Our benchmark comparisons, including all five controls, report performance on the hidden holdouts of 1,200 and 1,450 utterances.
T15 public and private holdout scores are reported separately.
Checkpoint selection minimizes the pooled word-edit count across both provided labeled validation sets, with decoding settings fixed before holdout evaluation.
All predictions are generated without access to reference transcriptions, and every utterance in the declared split contributes to evaluation.
Appendix~\ref{app:evaluation} gives details.

\subsection{Baselines}
\label{sec:controls}

We benchmark MAS-BIT against published end-to-end and cascaded decoders, examine candidate-sentence counts in our cascaded reproduction, and evaluate five core ablations of the multi-agent framework and \red{adapter-capacity and larger-backbone variants} of the single-agent control \red{along with sequential dual-view refinement, detached-latent-message, and message-only arbiter controls}.

\paragraph{Published baselines.}
Our primary comparison uses reported corpus WER on matched participant and hidden-holdout subsets (Table~\ref{tab:published_comparison}).
End-to-end decoding without ensembling generates text from neural input without an external WFST, $n$-gram model, pronunciation lexicon, or independent language-model reranker; phoneme pretraining, auxiliary CTC losses, and staged training are compatible with this definition.
BIT End-to-End + Ensemble is listed separately, and published cascaded references appear in Appendix~\ref{app:benchmark_tables}.
Appendix~\ref{app:other_evaluations} separately summarizes published results on other evaluation sets.

\paragraph{Candidate-sentence analysis.}
\label{sec:cascade_protocol}
We reproduce cascaded decoding for each participant using the released BIT encoder.
We initialize MAS-BIT from the same encoder.
Following \citet{zhang2026bit}, we pass CTC phoneme logits to an OpenWebText2 5-gram language model and rescore the candidate sentences with OPT-6.7B~\citep{zhang2022opt}.
We vary the number of retained candidates, $K\in\{1,2,3,4,6,9,12,18,25,37,50,75,100\}$, while keeping the encoder, language models, and search settings fixed.
\red{To separate encoder adaptation from multi-agent decoding, we also evaluate this cascade using each participant's fine-tuned encoder branch and phoneme readout from MAS-BIT.}
Appendix~\ref{app:cascade} describes lattice rescoring and candidate selection.

\paragraph{Multi-agent ablations.}
We compare MAS-BIT with \emph{Single agent, two views}, \emph{Two neural experts}, \emph{Hard-text communication}, \emph{Separate participant training}, and \emph{No staged training}.
All controls share the two-encoder frontend, backbone, initialization, augmentation, and checkpoint-selection protocol; corresponding components start from the same weights.
Appendix~\ref{app:controls} defines each control.

\section{Experiments and Results}
\label{sec:results}

\subsection{MAS-BIT establishes state-of-the-art end-to-end decoding}
\label{sec:bit_results}
\label{sec:decoding_results}

\begin{figure}[!b]
\centering
\includegraphics[width=\linewidth]{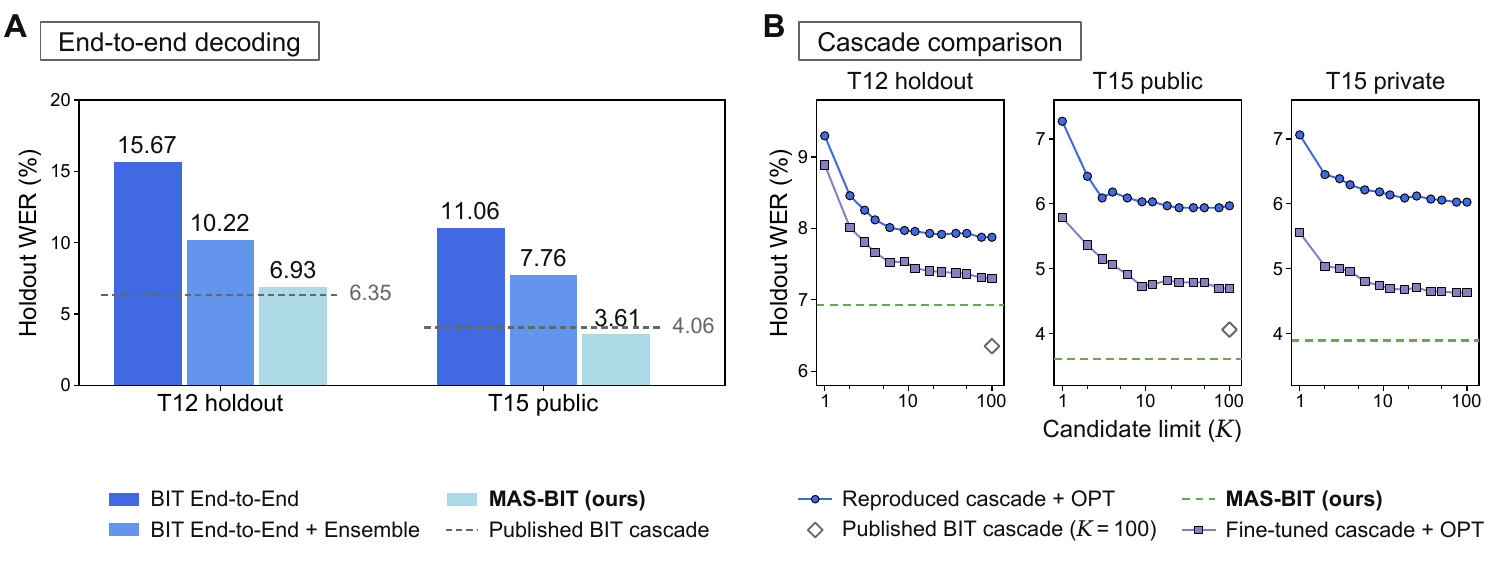}
\caption{\footnotesize \emph{Jointly trained multi-agent decoding achieves state-of-the-art end-to-end performance.} \textbf{(A)} Published BIT end-to-end results~\citep{zhang2026bit} and MAS-BIT on T12 holdout and T15 public. \textbf{(B)} Blue curves use the released participant-specific BIT encoders; purple curves use the corresponding fine-tuned encoder branch and final phoneme readout from MAS-BIT. Both cascades use the same WFST, unpruned 5-gram, and OPT rescoring settings at the 13 measured candidate limits; green lines show MAS-BIT. Gray diamonds at $K=100$ mark the published BIT cascade results on T12 and T15 public. All panels report corpus holdout WER (\%, lower is better); candidate limits use a logarithmic axis.}
\label{fig:benchmark}
\label{fig:cascade_candidates}
\end{figure}

MAS-BIT establishes a new end-to-end state of the art on the Brain-to-Text '24 holdout and '25 public benchmark, achieving \WER{mas}{t12}\% and \WER{mas}{t15public}\% WER, respectively (Figure~\ref{fig:benchmark}A; Table~\ref{tab:published_comparison}).
The same model achieves \WER{mas}{t15private}\% WER on the '25 private subset.
\red{Achieved with one unified cross-participant model and greedy decoding, these results outperform every published end-to-end speech BCI, with or without ensembling, on matched subsets of both benchmarks.}

\begin{table}[!t]
\centering
\caption{\footnotesize \emph{Published end-to-end benchmark results.} Values are corpus WER (\%); \red{\citet{feng2024towards}} and BIT scores follow \citet{zhang2026bit}. MoDAl reports mean $\pm$ 95\% CI over five seeds; MoDAl-Full also uses area 44. BrainWhisperer-cross uses additional neural datasets. \red{$^{*}$T15 public/private subset unspecified.} -- denotes an unreported result.}
\label{tab:published_comparison}
\label{tab:bit_comparison}
\small
\setlength{\tabcolsep}{6pt}
\begin{tabular}{lccc}
\toprule
& T12 & \multicolumn{2}{c}{T15} \\
\cmidrule(lr){2-2}\cmidrule(lr){3-4}
\textbf{Method} & \textbf{Holdout} & \textbf{Public} & \textbf{Private} \\
\midrule
\multicolumn{4}{l}{\textit{End-to-end decoding without ensembling}} \\
\citet{feng2024towards} & \WER[2]{feng}{t12} & -- & -- \\
RNN encoder + E2E LLM decoder~\citep{zhang2026bit} & \ReportedRnnEtoEWER & \ReportedRnnEtoEFifteenWER & -- \\
MoDAl-1~\citep{chen2026modal} & \WER[1]{modalone}{t12}\,$\pm$\,\MoDAlOneCI & -- & -- \\
MoDAl-Full~\citep{chen2026modal} & \WER[1]{modal}{t12}\,$\pm$\,\MoDAlCI & -- & -- \\
BIT-TFS End-to-End~\citep{zhang2026bit} & \ReportedBitTfsWER & \ReportedBitTfsFifteenWER & -- \\
BIT-Human End-to-End~\citep{zhang2026bit} & \ReportedBitHumanWER & \ReportedBitHumanFifteenWER & -- \\
BIT End-to-End~\citep{zhang2026bit} & \WER[2]{bit}{t12} & \WER[2]{bit}{t15public} & -- \\
BrainWhisperer~\citep{boccato2026brainwhisperer} & -- & \multicolumn{2}{c}{\ReportedBrainWhispererWER$^{*}$} \\
BrainWhisperer-cross~\citep{boccato2026brainwhisperer} & -- & \multicolumn{2}{c}{\ReportedBrainWhispererCrossWER$^{*}$} \\
\red{BEST (no external LM)~\citep{feng2026best}} & \red{\WER[2]{bestnolm}{t12}} & \multicolumn{2}{c}{\red{\ReportedBestFifteenWER$^{*}$}} \\
\rowcolor{blue!10} \textbf{MAS-BIT (ours)} & \textbf{\WER{mas}{t12}} & \textbf{\WER{mas}{t15public}} & \textbf{\WER{mas}{t15private}} \\
\midrule
\multicolumn{4}{l}{\textit{End-to-end decoding with ensembling}} \\
BIT End-to-End + Ensemble~\citep{zhang2026bit} & \WER[2]{bitensemble}{t12} & \WER[2]{bitensemble}{t15public} & -- \\
\bottomrule
\end{tabular}
\end{table}

Relative to single-model BIT, WER falls by \PP[2]{bit}{mas}{t12} \red{and} \PP[2]{bit}{mas}{t15public} points (\RR[1]{bit}{mas}{t12}\% \red{and} \RR[1]{bit}{mas}{t15public}\% relative) on T12 \red{and} T15 public\red{, and MAS-BIT further surpasses} BIT's end-to-end ensembles \red{by} \PP[2]{bitensemble}{mas}{t12} \red{and} \PP[2]{bitensemble}{mas}{t15public} points (\RR[1]{bitensemble}{mas}{t12}\% \red{and} \RR[1]{bitensemble}{mas}{t15public}\% \red{relative})~\citep{zhang2026bit}.
\red{On T12, MAS-BIT outperforms MoDAl-Full's reported mean by \PP[1]{modal}{mas}{t12} percentage points~\citep{chen2026modal} and the concurrent no-external-LM BEST decoder by \PP[2]{bestnolm}{mas}{t12} points~\citep{feng2026best}.}
\red{On T15,} BrainWhisperer and BrainWhisperer-cross report competition WERs of \ReportedBrainWhispererWER\% and \ReportedBrainWhispererCrossWER\%~\citep{boccato2026brainwhisperer}\red{, and BEST reports \ReportedBestFifteenWER\%~\citep{feng2026best}, all with the public/private subset unspecified. MAS-BIT's WER on either T15 subset is below each of them, and less than half of both BrainWhisperer results.}

\subsection{MAS-BIT surpasses the reproduced cascade on both benchmarks}
\label{sec:cascade_results}

We compare MAS-BIT with participant-specific cascaded decoding from the released BIT encoders, \red{which remain fixed in the baseline cascade while} MAS-BIT jointly adapts both encoders with the language agents across participants.
MAS-BIT achieves lower holdout WER on T12 and both T15 subsets across all tested candidate limits up to 100 sentences (Figure~\ref{fig:benchmark}B)\red{, improving upon} the lowest reproduced WER \red{by \PP[3]{cascade100}{mas}{t12} percentage points on T12 (from \WER{cascade100}{t12}\% to \WER{mas}{t12}\%) and by} \PP{cascade25}{mas}{t15public} and \PP{cascade100}{mas}{t15private} points on T15 public and private.
\red{To separate encoder adaptation from multi-agent decoding, running the same cascade with each participant's fine-tuned encoder branch and phoneme readout from MAS-BIT lowers $K=100$ WER to \WER{cm1cascade100}{t12}\% on T12 and \WER{cm1cascade100}{t15public}\%/\WER{cm1cascade100}{t15private}\% on T15 public/private (Table~\ref{tab:cascade_results}). MAS-BIT still outperforms this adapted-encoder cascade at every candidate limit (by \PP{cm1cascade100}{mas}{t12} points on T12 and \PP{cm1cascade100}{mas}{t15public}/\PP{cm1cascade100}{mas}{t15private} points on T15 at $K=100$), showing that encoder adaptation and multi-agent decoding both contribute.}
Published BIT cascaded decoding provides a separate reference at $K=100$: \WER[2]{bitcascade}{t12}\% on T12 and \WER[2]{bitcascade}{t15public}\% on T15 public~\citep{zhang2026bit}.
\red{The published cascade used phoneme-level encoders that are not publicly released; our reproduction starts from the released sentence-level checkpoints (Appendix~\ref{app:cascade}).}
MAS-BIT improves upon this published T15 public result by \PP[3]{bitcascade}{mas}{t15public} percentage points \red{and comes within \PP[2]{mas}{bitcascade}{t12} points on T12, where BIT's end-to-end decoder trailed by \PP[2]{bit}{bitcascade}{t12} points,} without WFST decoding or external candidate rescoring.

\subsection{MAS-BIT lowers inference memory and decoding time}
\label{sec:efficiency_results}

\begin{table}[t]
\centering
\caption{\footnotesize \emph{Peak host RAM\red{, GPU memory,} and decoding time on one NVIDIA A40 GPU.} \red{Memory} includes loading and audits; time excludes both (Appendix~\ref{app:memory}).}
\label{tab:inference_memory}
\small
\setlength{\tabcolsep}{2pt}
\begin{tabular}{@{}lcccccc@{}}
\toprule
& \multicolumn{2}{c}{\red{\textbf{Host RAM (GiB)}}} & \multicolumn{2}{c}{\red{\textbf{Peak GPU (GiB)}}} & \multicolumn{2}{c}{\textbf{Decode time (s)}} \\
\cmidrule(lr){2-3}\cmidrule(lr){4-5}\cmidrule(lr){6-7}
\textbf{Method} & \textbf{T12} & \textbf{T15} & \red{\textbf{T12}} & \red{\textbf{T15}} & \textbf{T12} & \textbf{T15} \\
\midrule
Reproduced cascade ($K=100$)~\citep{zhang2026bit} & \CascadeTwelveRAM & \CascadeFifteenRAM & \red{\CascadeTwelveGPU} & \red{\CascadeFifteenGPU} & \CascadeTwelveSeconds{} & \CascadeFifteenSeconds{} \\
\red{BIT + LightBeam~\citep{feghhi2026lightbeam}} & \red{\LightBeamTwelveRAM} & \red{\LightBeamFifteenRAM} & \red{\LightBeamTwelveGPU} & \red{\LightBeamFifteenGPU} & \red{\LightBeamTwelveSeconds} & \red{\LightBeamFifteenSeconds} \\
\rowcolor{blue!10}\textbf{MAS-BIT (ours)} & \textbf{\MasTwelveRAM} & \textbf{\MasFifteenRAM} & \red{\textbf{\MasTwelveGPU}} & \red{\textbf{\MasFifteenGPU}} & \textbf{\MasTwelveSeconds{}} & \textbf{\MasFifteenSeconds{}} \\
\bottomrule
\end{tabular}
\end{table}

\red{Sharing one frozen 1.7B backbone without external graph search or candidate rescoring also lowers inference cost (Table~\ref{tab:inference_memory}).}
\red{On an A40 GPU}, the reproduced $K=100$ cascade \red{requires \HostRAMFifteenRatio--\HostRAMTwelveRatio$\times$ MAS-BIT's peak host RAM and $2.95\times$ its peak GPU memory, while MAS-BIT decodes} approximately \CapacityStat{t12_speedup}$\times$ faster.
\red{Lexicon-constrained LightBeam decoding~\citep{feghhi2026lightbeam} reduces memory use but still needs 1.47--1.72$\times$ MAS-BIT's GPU memory and decodes \LightBeamTwelveRatio--\LightBeamFifteenRatio$\times$ slower. On consumer RTX 5090 laptop and RTX 4090 desktop GPUs, where the $K=100$ cascade exceeds host memory, MAS-BIT reproduces identical FP32 predictions under 8.7~GiB of GPU memory while decoding \LaptopDecodeTwelveRatio--\LaptopDecodeFifteenRatio$\times$ and \DesktopDecodeTwelveRatio--\DesktopDecodeFifteenRatio$\times$ faster than BIT + LightBeam (Tables~\ref{tab:laptop_efficiency}--\ref{tab:desktop_efficiency}).}

\subsection{Specialization, communication, and staged training improve decoding}
\label{sec:ablation_results}
\label{sec:cooperation_results}

\begin{figure}[!t]
\centering
\includegraphics[width=\linewidth]{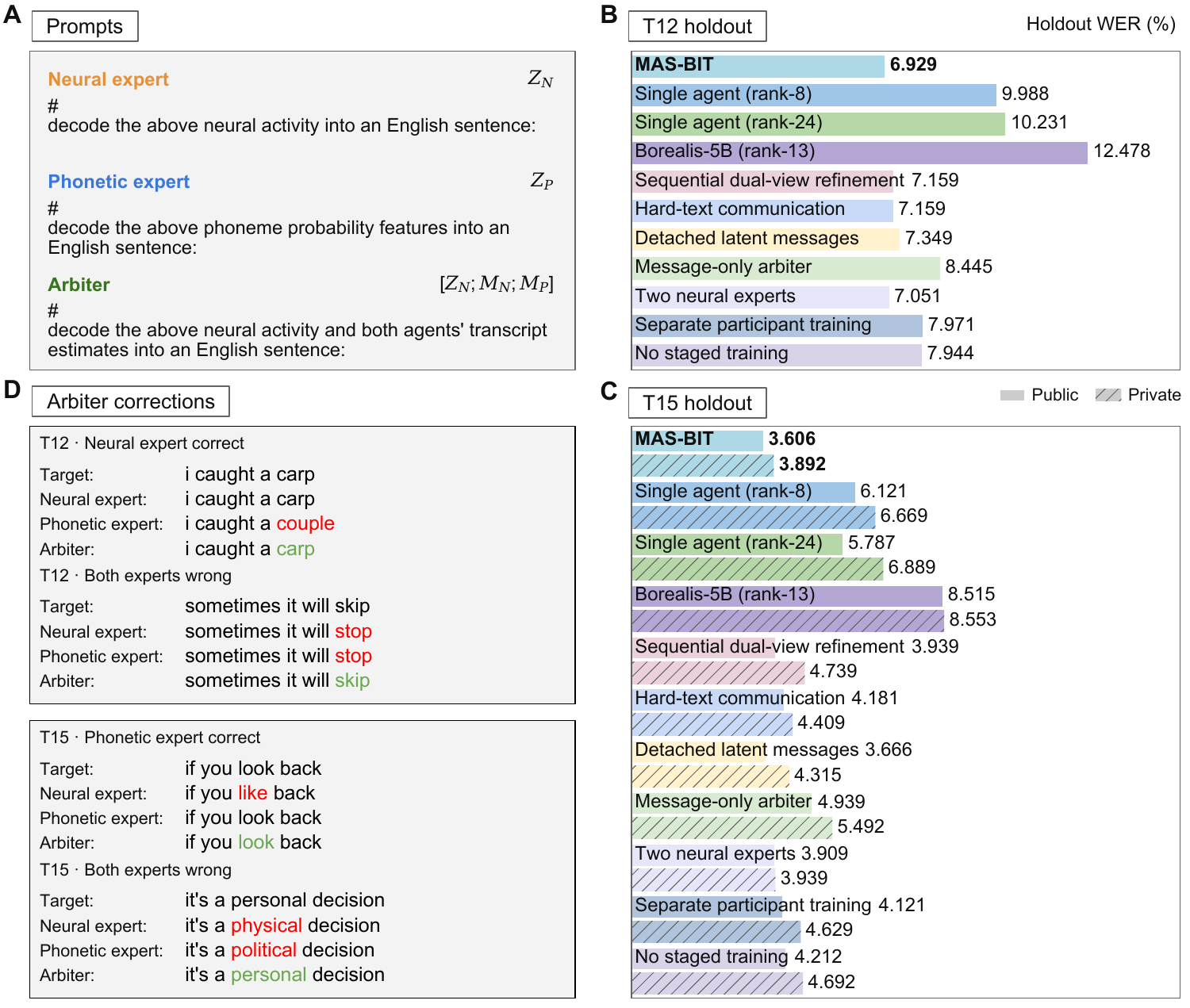}
\caption{\footnotesize \emph{Expert specialization and latent communication improve end-to-end decoding.} \textbf{(A)} Role-specific prompts; $\#$ marks the continuous-input slot. \textbf{(B--C)} T12 and T15 holdout WER; solid and hatched bars show T15 public and private. The \red{eight controls and rank-24 single-agent variant} share the backbone and two-encoder initialization. \red{The Borealis-5B single-agent variant uses a 4B decoder with rank-13 LoRA; all single-agent variants receive both input views.} Bars have the same zero-origin scale. \textbf{(D)} Selected validation transcripts show the arbiter matching a correct expert and recovering sentences when both drafts are wrong. Expert errors are red; corresponding correct arbiter words are green.}
\label{fig:ablations}
\end{figure}

\red{Each of five core controls degrades accuracy across all three holdout splits, isolating the roles of} expert specialization, phonetic evidence, latent messages, cross-participant joint training, and staged training (Figure~\ref{fig:ablations}B--C\red{; Table~\ref{tab:ablation_results}}).

\paragraph{Specialization makes better use of the same input evidence.}
Specialized experts and arbitration reduce WER relative to a single agent given both neural representations and phoneme posteriors\red{, whose} holdout WER is higher than MAS-BIT by \PP{single}{mas}{t12} percentage points on T12 and \PP{single}{mas}{t15public}/\PP{single}{mas}{t15private} points on T15 public/private.
\red{To test whether this gap reflects parameter capacity, expanding} the single-agent adapter to rank 24, matching the combined LoRA parameter count of the three rank-8 role adapters, gives \WER{single_r24}{t12}\% T12 and \WER{single_r24}{t15public}\%/\WER{single_r24}{t15private}\% T15 public/private WER\red{, while scaling the single-agent backbone to the 4B audio-trained decoder of Borealis-5B~\citep{borealis2025} (rank-13 LoRA) yields \WER{single_borealis}{t12}\% and \WER{single_borealis}{t15public}\%/\WER{single_borealis}{t15private}\% WER (Figure~\ref{fig:ablations}B--C; Appendix~\ref{app:controls}).
Even when matching MAS-BIT's three rank-8 adapters, intermediate supervision, and multi-pass latent-message decoding via sequential dual-view refinement, holdout WER remains \PP{refinement}{mas}{t12} points worse on T12 and \PP{refinement}{mas}{t15public}/\PP{refinement}{mas}{t15private} points worse on T15 (Table~\ref{tab:ablation_results}), so dedicated neural and phonetic experts improve further on matched capacity and decoding passes.}

\paragraph{Latent messages improve upon discrete expert drafts.}
Replacing latent messages with hard text increases holdout WER by \PP{hard}{mas}{t12} points on T12 and \PP{hard}{mas}{t15public}/\PP{hard}{mas}{t15private} points on T15 public/private.
The latent interface retains alternative token hypotheses and allows the arbiter loss to update expert probabilities through replay\red{, whereas} the hard-text control passes \red{only} the selected draft tokens and blocks this gradient path.
\red{To separate gradient transmission from alternative-token content, a detached-latent-message control preserves the top-16 probability-weighted token embeddings in the forward pass while stopping arbiter gradients at the messages (Appendix~\ref{app:controls}); this raises holdout WER by \PP{softsg}{mas}{t12} points on T12 and \PP{softsg}{mas}{t15public}/\PP{softsg}{mas}{t15private} points on T15, confirming that end-to-end message gradients improve expert coordination.}
\red{Conversely, a message-only arbiter that decodes solely from $[M_N; M_P]$ without direct neural input $Z_N$ degrades WER by \PP{message_only}{mas}{t12} points on T12 and \PP{message_only}{mas}{t15public}/\PP{message_only}{mas}{t15private} points on T15, exceeding the hard-text penalty on every split and showing that arbitration relies on direct neural grounding rather than refining expert drafts alone.}

\paragraph{Phonetic evidence complements neural interpretation.}
Replacing the phonetic expert with a second neural expert increases holdout WER by \PP{nna}{mas}{t12} points on T12 and \PP{nna}{mas}{t15public}/\PP{nna}{mas}{t15private} points on T15 public/private.
\red{The two-neural-expert system still improves on the single agent by \PP{single}{nna}{t12} points on T12 and \PP{single}{nna}{t15public}/\PP{single}{nna}{t15private} points on T15, so most of the gain comes from expert drafting and neurally grounded arbitration, with a smaller, consistent gain from the phoneme-posterior view.}
The two interfaces also expose more distinct sentence-level geometry than the two-neural-expert control (Appendix~\ref{app:rsa}).

\paragraph{Cross-participant joint training improves holdout generalization.}
Training separate MAS-BIT models on each participant's data gives \WER{separate}{t12}\% holdout WER on T12 and \WER{separate}{t15public}\%/\WER{separate}{t15private}\% on T15 public/private.
Under the same staged epoch schedule, joint training reduces WER by \PP[3]{separate}{mas}{t12} percentage points on T12 and \PP[3]{separate}{mas}{t15public}/\PP[3]{separate}{mas}{t15private} points on T15 public/private, corresponding to relative reductions of \RR[1]{separate}{mas}{t12}\% and \RR[1]{separate}{mas}{t15public}\%/\RR[1]{separate}{mas}{t15private}\%.
The jointly trained encoders, interfaces, and language agents thus improve holdout accuracy while serving both participants with a single model.

\paragraph{Specialization prepares the system for joint optimization.}
Training all adaptation groups from the first epoch increases holdout WER by \PP[3]{joint}{mas}{t12} points on T12 and \PP[3]{joint}{mas}{t15public}/\PP[3]{joint}{mas}{t15private} points on T15 public/private.
The staged prefix establishes the input interfaces and expert roles before the final joint phase.

\Needspace{26\baselineskip}
\suppressfloats[t]
\subsection{Arbitration integrates complementary expert judgments}
\label{sec:arbiter_cases}
\begingroup
\setlength{\intextsep}{3pt}
\setlength{\columnsep}{10pt}
\begin{wrapfigure}{r}{0.46\textwidth}
\centering
\includegraphics[width=\linewidth]{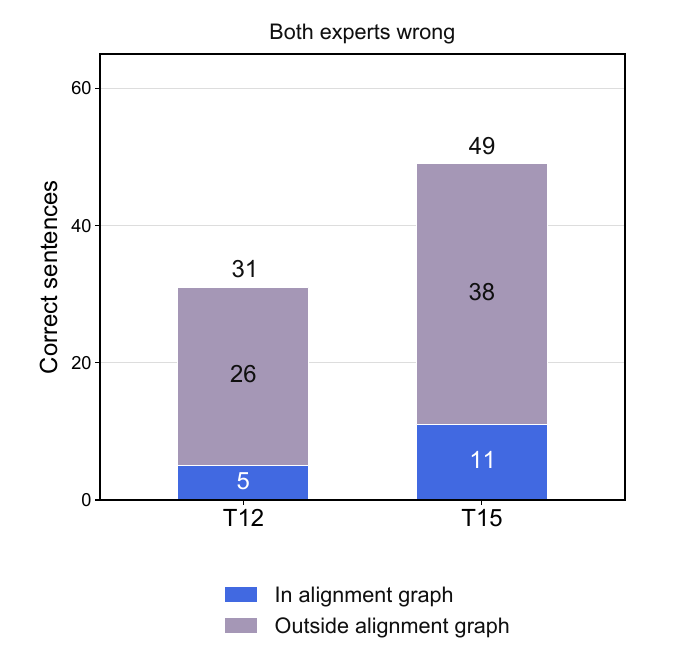}
\caption{\footnotesize \emph{Arbitration recovers sentences beyond draft recombination.} Exact recoveries when both experts are wrong, on the full validation sets.}
\label{fig:sentence_recovery}
\end{wrapfigure}
On the full validation sets, arbitration reduces word edits relative to the neural \red{and} phonetic experts by \ArbCount{t12}{neural_relative_reduction}\% \red{and} \ArbCount{t12}{phonetic_relative_reduction}\% on T12, and \red{by} \ArbCount{t15}{neural_relative_reduction}\% \red{and} \ArbCount{t15}{phonetic_relative_reduction}\% on T15 (Appendix~\ref{app:expert_cases}).
\red{Crucially, when both experts err, the} arbiter recovers \CaseCount{t12}{arbiter_recovers} \red{exact sentences on T12 and} \CaseCount{t15}{arbiter_recovers} \red{on T15: \WordStat{t12}{exact_outside_graph} and \WordStat{t15}{exact_outside_graph} cannot be formed by recombining words along any minimum-edit draft alignment, including \WordStat{t12}{outside_graph_with_absent_word} and \WordStat{t15}{outside_graph_with_absent_word}} with a correct word absent from both drafts (Figure~\ref{fig:sentence_recovery}\red{; examples in Figure~\ref{fig:ablations}D}).
The model therefore combines expert messages with neural evidence to produce corrections beyond selecting \red{or stitching expert drafts (Appendix~\ref{app:expert_cases})}.

At fixed weights and expert drafts, masking both messages increases validation WER by \InputStat{t12}{without_both_messages}{delta_wer_pp} \red{and} \InputStat{t15}{without_both_messages}{delta_wer_pp} percentage points on T12 \red{and} T15, while Top-16 messages remove 41 \red{and} 29 word edits relative to Top-1 (Appendices~\ref{app:arbiter_inputs}--\ref{app:soft_message_interventions}).
Zeroing the direct neural input $Z_N$ removes all \WordStat{t12}{exact_outside_graph} \red{and} \WordStat{t15}{exact_outside_graph} \red{recoveries beyond draft recombination, and} replacing it with a length-matched trial from the same participant removes all \NeuralInputStat{t12}{paired_outside} \red{and} \NeuralInputStat{t15}{paired_outside} such recoveries in the paired subsets (Appendix~\ref{app:arbiter_inputs}).
\par
\ifnum\value{WF@wrappedlines}>1
  \addtocounter{WF@wrappedlines}{-1}
  \vspace{\value{WF@wrappedlines}\baselineskip}
\fi
\WFclear
\endgroup

\section{Discussion}
\label{sec:discussion}

\red{\textbf{Summary.}} We introduce MAS-BIT, an end-to-end speech decoder that combines neural and phonetic experts through latent communication \red{and signal-grounded arbitration.
Rather than leaving interpretation and generation to one decoder, dedicated experts interpret continuous neural states and soft phoneme posteriors while an arbiter resolves their disagreements with direct access to the recording, recovering sentences missed by both experts.
Jointly trained across participants after staged specialization, one model sets a new end-to-end state of the art on Brain-to-Text '24 and '25 (\WER{mas}{t12}\% and \WER{mas}{t15public}\%/\WER{mas}{t15private}\% WER), surpasses BIT's published cascade on '25 public and comes within \PP[2]{mas}{bitcascade}{t12} points of it on T12, outperforms reproduced cascades at every candidate limit, and aligns cross-participant interfaces with speech and language geometry while preserving distinct views (Appendix~\ref{app:rsa}).}

\red{\textbf{Limitations and future directions.} Our evaluation is offline and covers two participants.
For online streaming BCIs, our encoders use bidirectional attention over complete utterances and sequential drafting precedes arbitration; adapting multi-agent coordination to causal encoders and lower-latency message passing is an important next step, alongside scaling shared interfaces to larger cohorts, inner speech, and continual drift adaptation.
Extending decoding to inner speech will also require explicit user consent and privacy safeguards~\citep{kunz2025inner}.}
\clearpage

\section*{LLM Usage Statement}
We used generative AI tools, including ChatGPT, Codex, and Claude, to assist with grammar checking, text polishing, and software implementation (such as coding, debugging, and visualization scripts). All research ideas, hypotheses, experimental designs, and scientific interpretations originated from the authors, who reviewed, verified, and take full responsibility for all code, experimental results, and text in this paper.

\section*{Reproducibility Statement}
Detailed architecture, optimization, preprocessing, evaluation, and
analysis procedures are provided in the Methods, Evaluation, and supplementary sections.
\ifnum\papermode>0\relax
Code and model weights are publicly available on \href{https://github.com/BaochaiXue/Multi_Agent_System_for_BraIn_to_Text}{GitHub} and \href{https://huggingface.co/XinjieZhang/MAS-BIT}{Hugging Face}.
\else
Upon publication, we will release the source code, configurations, and model checkpoints to facilitate reproduction of the reported results.
\fi
\bibliography{iclr2027_conference}

\begin{thebibliography}{43}
\providecommand{\natexlab}[1]{#1}
\providecommand{\url}[1]{\texttt{#1}}
\expandafter\ifx\csname urlstyle\endcsname\relax
  \providecommand{\doi}[1]{doi: #1}\else
  \providecommand{\doi}{doi: \begingroup \urlstyle{rm}\Url}\fi

\bibitem[Akbari et~al.(2019)Akbari, Khalighinejad, Herrero, Mehta, and Mesgarani]{akbari2019towards}
Hassan Akbari, Bahar Khalighinejad, Jose~L Herrero, Ashesh~D Mehta, and Nima Mesgarani.
\newblock Towards reconstructing intelligible speech from the human auditory cortex.
\newblock \emph{Scientific reports}, 9\penalty0 (1):\penalty0 874, 2019.

\bibitem[Benster et~al.(2024)Benster, Wilson, Elisha, Willett, and Druckmann]{benster2024cross}
Tyler Benster, Guy Wilson, Reshef Elisha, Francis~R Willett, and Shaul Druckmann.
\newblock A cross-modal approach to silent speech with {LLM}-enhanced recognition, 2024.
\newblock URL \url{https://arxiv.org/abs/2403.05583}.

\bibitem[Boccato et~al.(2026)Boccato, Olak, and Ferrante]{boccato2026brainwhisperer}
Tommaso Boccato, Michal Olak, and Matteo Ferrante.
\newblock {B}rain{W}hisperer: {L}everaging {L}arge-{S}cale {ASR} {M}odels for {N}eural {S}peech {D}ecoding, 2026.
\newblock arXiv:2603.13321v1.

\bibitem[Card et~al.(2024)Card, Wairagkar, Iacobacci, Hou, Singer-Clark, Willett, Kunz, Fan, Vahdati~Nia, Deo, Srinivasan, Choi, Glasser, Hochberg, Henderson, Shahlaie, Stavisky, and Brandman]{card2024accurate}
Nicholas~S. Card, Maitreyee Wairagkar, Carrina Iacobacci, Xianda Hou, Tyler Singer-Clark, Francis~R. Willett, Erin~M. Kunz, Chaofei Fan, Maryam Vahdati~Nia, Darrel~R. Deo, Aparna Srinivasan, Eun~Young Choi, Matthew~F. Glasser, Leigh~R. Hochberg, Jaimie~M. Henderson, Kiarash Shahlaie, Sergey~D. Stavisky, and David~M. Brandman.
\newblock {An Accurate and Rapidly Calibrating Speech Neuroprosthesis}.
\newblock \emph{New England Journal of Medicine}, 391\penalty0 (7):\penalty0 609--618, 2024.
\newblock ISSN 1533-4406.
\newblock \doi{10.1056/nejmoa2314132}.
\newblock URL \url{https://doi.org/10.1056/NEJMoa2314132}.

\bibitem[Chen \& Chin(2026)Chen and Chin]{chen2026modal}
Yuanhao Chen and Peter Chin.
\newblock {MoDAl}: Self-supervised neural modality discovery via decorrelation for speech neuroprosthesis, 2026.
\newblock URL \url{https://arxiv.org/abs/2605.00025}.

\bibitem[Dekel et~al.(2026)Dekel, Thomas, Fukada, and Saon]{dekel2026nle}
Avihu Dekel, Samuel Thomas, Takashi Fukada, and George Saon.
\newblock {NLE: Non-autoregressive LLM-based ASR by Transcript Editing}, 2026.
\newblock URL \url{https://arxiv.org/abs/2603.08397}.

\bibitem[Du et~al.(2024)Du, Li, Torralba, Tenenbaum, and Mordatch]{du2024debate}
Yilun Du, Shuang Li, Antonio Torralba, Joshua~B. Tenenbaum, and Igor Mordatch.
\newblock Improving factuality and reasoning in language models through multiagent debate.
\newblock In Ruslan Salakhutdinov, Zico Kolter, Katherine Heller, Adrian Weller, Nuria Oliver, Jonathan Scarlett, and Felix Berkenkamp (eds.), \emph{Proceedings of the 41st International Conference on Machine Learning}, volume 235 of \emph{Proceedings of Machine Learning Research}, pp.\  11733--11763. PMLR, 21--27 Jul 2024.
\newblock URL \url{https://proceedings.mlr.press/v235/du24e.html}.

\bibitem[Feghhi et~al.(2025)Feghhi, Kaasyap, Hadidi, and Kao]{feghhi2025time}
Ebrahim Feghhi, Shreyas Kaasyap, Nima Hadidi, and Jonathan~C. Kao.
\newblock Time-masked transformers with lightweight test-time adaptation for neural speech decoding.
\newblock In \emph{Advances in Neural Information Processing Systems}, volume~38, 2025.
\newblock URL \url{https://papers.nips.cc/paper_files/paper/2025/hash/85e351d89eee3f33d3d1356f38590bb6-Abstract-Conference.html}.

\bibitem[Feghhi et~al.(2026)Feghhi, Hu, Hadidi, and Kao]{feghhi2026lightbeam}
Ebrahim Feghhi, Junlin Hu, Nima Hadidi, and Jonathan~C. Kao.
\newblock {LightBeam}: An accurate and memory-efficient {CTC} decoder for speech neuroprostheses, 2026.
\newblock URL \url{https://arxiv.org/abs/2603.14002}.

\bibitem[Feng et~al.(2024)Feng, Liu, Wang, and Wang]{feng2024towards}
Sheng Feng, Heyang Liu, Yu~Wang, and Yanfeng Wang.
\newblock {Towards an End-to-End Framework for Invasive Brain Signal Decoding with Large Language Models}.
\newblock In \emph{{Interspeech 2024}}, pp.\  1495--1499, 2024.
\newblock \doi{10.21437/Interspeech.2024-382}.

\bibitem[Feng et~al.(2026)Feng, Wang, Chen, Guan, and Han]{feng2026best}
Zehui Feng, Weichuan Wang, Xiaohan Chen, Cuntai Guan, and Ting Han.
\newblock Do speech {BCIs} need larger models? {R}ethinking neural decoding beyond scaling.
\newblock In \emph{Advances in Neural Information Processing Systems}, 2026.
\newblock URL \url{https://arxiv.org/abs/2610.23175}.

\bibitem[Fiedler et~al.(2025)Fiedler, Hermann, M{\"u}ller, Cohen, Chin, Friedrich, and Vaadia]{fiedler2025wav2vec2}
Tobias Fiedler, Leon Hermann, Florian M{\"u}ller, Sarel Cohen, Peter Chin, Tobias Friedrich, and Eilon Vaadia.
\newblock Teaching {Wav2Vec2} the language of the brain, 2025.
\newblock URL \url{https://arxiv.org/abs/2501.09459v1}.

\bibitem[Foerster et~al.(2016)Foerster, Assael, de~Freitas, and Whiteson]{foerster2016communicate}
Jakob Foerster, Ioannis~Alexandros Assael, Nando de~Freitas, and Shimon Whiteson.
\newblock Learning to communicate with deep multi-agent reinforcement learning.
\newblock In D.~Lee, M.~Sugiyama, U.~Luxburg, I.~Guyon, and R.~Garnett (eds.), \emph{Advances in Neural Information Processing Systems}, volume~29. Curran Associates, Inc., 2016.
\newblock URL \url{https://proceedings.neurips.cc/paper_files/paper/2016/file/c7635bfd99248a2cdef8249ef7bfbef4-Paper.pdf}.

\bibitem[Fogg et~al.(2026)Fogg, Card, Wairagkar, Srinivasan, Singer-Clark, Hou, Okorokova, Peracha, Iacobacci, Brailow, Jude, Levi-Aharoni, Le, Mifsud, Deevi, Nason-Tomaszewski, Pritchard, Zhang, Richards, Bechefsky, Hochberg, Williams, Shahlaie, AuYong, Rubin, Pandarinath, Brandman, and Stavisky]{Fogg2026.07.23.739430}
Zachery~M. Fogg, Nicholas~S. Card, Maitreyee Wairagkar, Aparna Srinivasan, Tyler Singer-Clark, Xianda Hou, Elizaveta Okorokova, Hamza Peracha, Carrina Iacobacci, Tiffany Brailow, Justin Jude, Hadar Levi-Aharoni, Trung Le, Domenick Mifsud, Pranav~I. Deevi, Samuel~R. Nason-Tomaszewski, Anna~L. Pritchard, Yunnuo Zhang, Brice Richards, Payton Bechefsky, Leigh~R. Hochberg, Ziv Williams, Kiarash Shahlaie, Nicholas AuYong, Daniel~B. Rubin, Chethan Pandarinath, David~M. Brandman, and Sergey~D. Stavisky.
\newblock A generalizable speech neuroprosthesis.
\newblock \emph{bioRxiv}, 2026.
\newblock \doi{10.64898/2026.07.23.739430}.
\newblock URL \url{https://www.biorxiv.org/content/early/2026/07/27/2026.07.23.739430}.

\bibitem[Graves et~al.(2006)Graves, Fern{\'a}ndez, Gomez, and Schmidhuber]{graves2006connectionist}
Alex Graves, Santiago Fern{\'a}ndez, Faustino Gomez, and J{\"u}rgen Schmidhuber.
\newblock Connectionist temporal classification: labelling unsegmented sequence data with recurrent neural networks.
\newblock In \emph{Proceedings of the 23rd international conference on Machine learning - ICML '06}, ICML '06, pp.\  369--376. ACM Press, 2006.
\newblock \doi{10.1145/1143844.1143891}.
\newblock URL \url{http://dx.doi.org/10.1145/1143844.1143891}.

\bibitem[Hu et~al.(2022)Hu, Shen, Wallis, Allen-Zhu, Li, Wang, Wang, and Chen]{hu2022lora}
Edward~J. Hu, Yelong Shen, Phillip Wallis, Zeyuan Allen-Zhu, Yuanzhi Li, Shean Wang, Lu~Wang, and Weizhu Chen.
\newblock {LoRA}: Low-rank adaptation of large language models.
\newblock In \emph{ICLR 2022}, April 2022.
\newblock URL \url{https://www.microsoft.com/en-us/research/publication/lora-low-rank-adaptation-of-large-language-models/}.

\bibitem[Hu et~al.(2020)Hu, Sainath, Pang, and Prabhavalkar]{hu2020deliberation}
Ke~Hu, Tara~N. Sainath, Ruoming Pang, and Rohit Prabhavalkar.
\newblock {Deliberation Model Based Two-Pass End-To-End Speech Recognition}.
\newblock In \emph{ICASSP 2020 - 2020 IEEE International Conference on Acoustics, Speech and Signal Processing (ICASSP)}, pp.\  7799--7803. IEEE, 2020.
\newblock \doi{10.1109/icassp40776.2020.9053606}.
\newblock URL \url{https://doi.org/10.1109/ICASSP40776.2020.9053606}.

\bibitem[Jordan et~al.(2024)Jordan, Jin, Boza, You, Cesista, Newhouse, and Bernstein]{jordan2024muon}
Keller Jordan, Yuchen Jin, Vlado Boza, Jiacheng You, Franz Cesista, Laker Newhouse, and Jeremy Bernstein.
\newblock Muon: An optimizer for hidden layers in neural networks, 2024.
\newblock URL \url{https://kellerjordan.github.io/posts/muon/}.

\bibitem[Khanday et~al.(2026)Khanday, Gonzalez-Lopez, Ouellet, Gald{\'o}n, and Granado]{khanday2026endtoend}
Owais~M. Khanday, Jose~A. Gonzalez-Lopez, Marc Ouellet, Alberto Gald{\'o}n, and Gonzalo~Olivares Granado.
\newblock End-to-end intracortical speech decoding from neural activity.
\newblock In \emph{Odyssey 2026}, pp.\  165--172, 2026.
\newblock \doi{10.21437/Odyssey.2026-24}.
\newblock URL \url{https://www.isca-archive.org/odyssey_2026/khanday26_odyssey.html}.

\bibitem[Kriegeskorte et~al.(2008)Kriegeskorte, Mur, and Bandettini]{kriegeskorte2008representational}
Nikolaus Kriegeskorte, Marieke Mur, and Peter~A. Bandettini.
\newblock Representational similarity analysis - connecting the branches of systems neuroscience.
\newblock \emph{Frontiers in Systems Neuroscience}, Volume 2 - 2008, 2008.
\newblock ISSN 1662-5137.
\newblock \doi{10.3389/neuro.06.004.2008}.
\newblock URL \url{https://www.frontiersin.org/journals/systems-neuroscience/articles/10.3389/neuro.06.004.2008}.

\bibitem[Kunz et~al.(2025)Kunz, Abramovich~Krasa, Kamdar, Avansino, Hahn, Yoon, Singh, Nason-Tomaszewski, Card, Jude, et~al.]{kunz2025inner}
Erin~M Kunz, Benyamin Abramovich~Krasa, Foram Kamdar, Donald~T Avansino, Nick Hahn, Seonghyun Yoon, Akansha Singh, Samuel~R Nason-Tomaszewski, Nicholas~S Card, Justin~J Jude, et~al.
\newblock Inner speech in motor cortex and implications for speech neuroprostheses.
\newblock \emph{Cell}, 188\penalty0 (17):\penalty0 4658--4673.e17, 2025.
\newblock \doi{10.1016/j.cell.2025.06.015}.

\bibitem[Lee \& Watanabe(2021)Lee and Watanabe]{lee2021interctc}
Jaesong Lee and Shinji Watanabe.
\newblock Intermediate loss regularization for {CTC}-based speech recognition.
\newblock In \emph{ICASSP 2021 - 2021 IEEE International Conference on Acoustics, Speech and Signal Processing (ICASSP)}, pp.\  6224--6228. IEEE, June 2021.
\newblock \doi{10.1109/ICASSP39728.2021.9414594}.
\newblock URL \url{https://doi.org/10.1109/ICASSP39728.2021.9414594}.

\bibitem[Loshchilov \& Hutter(2019)Loshchilov and Hutter]{loshchilov2017decoupled}
Ilya Loshchilov and Frank Hutter.
\newblock {Decoupled Weight Decay Regularization}.
\newblock In \emph{International Conference on Learning Representations}, 2019.
\newblock URL \url{https://mlanthology.org/iclr/2019/loshchilov2019iclr-decoupled/}.

\bibitem[Ma et~al.(2025)Ma, Chen, Audhkhasi, and Ramabhadran]{ma2025legoslm}
Rao Ma, Tongzhou Chen, Kartik Audhkhasi, and Bhuvana Ramabhadran.
\newblock {L}ego{SLM}: Connecting {LLM} with speech encoder using {CTC} posteriors.
\newblock In Christos Christodoulopoulos, Tanmoy Chakraborty, Carolyn Rose, and Violet Peng (eds.), \emph{Findings of the Association for Computational Linguistics: EMNLP 2025}, pp.\  18171--18186, Suzhou, China, November 2025. Association for Computational Linguistics.
\newblock ISBN 979-8-89176-335-7.
\newblock \doi{10.18653/v1/2025.findings-emnlp.985}.
\newblock URL \url{https://aclanthology.org/2025.findings-emnlp.985/}.

\bibitem[Metzger et~al.(2023)Metzger, Littlejohn, Silva, Moses, Seaton, Wang, Dougherty, Liu, Wu, Berger, et~al.]{metzger2023high}
Sean~L Metzger, Kaylo~T Littlejohn, Alexander~B Silva, David~A Moses, Margaret~P Seaton, Ran Wang, Maximilian~E Dougherty, Jessie~R Liu, Peter Wu, Michael~A Berger, et~al.
\newblock A high-performance neuroprosthesis for speech decoding and avatar control.
\newblock \emph{Nature}, 620\penalty0 (7976):\penalty0 1037--1046, 2023.

\bibitem[Moses et~al.(2021)Moses, Metzger, Liu, Anumanchipalli, Makin, Sun, Chartier, Dougherty, Liu, Abrams, et~al.]{moses2021neuroprosthesis}
David~A Moses, Sean~L Metzger, Jessie~R Liu, Gopala~K Anumanchipalli, Joseph~G Makin, Pengfei~F Sun, Josh Chartier, Maximilian~E Dougherty, Patricia~M Liu, Gary~M Abrams, et~al.
\newblock Neuroprosthesis for decoding speech in a paralyzed person with anarthria.
\newblock \emph{New England Journal of Medicine}, 385\penalty0 (3):\penalty0 217--227, 2021.

\bibitem[Nason et~al.(2020)Nason, Vaskov, Willsey, Welle, An, Vu, Bullard, Nu, Kao, Shenoy, Jang, Kim, Blaauw, Patil, and Chestek]{nason2020low}
Samuel~R. Nason, Alex~K. Vaskov, Matthew~S. Willsey, Elissa~J. Welle, Hyochan An, Philip~P. Vu, Autumn~J. Bullard, Chrono~S. Nu, Jonathan~C. Kao, Krishna~V. Shenoy, Taekwang Jang, Hun-Seok Kim, David Blaauw, Parag~G. Patil, and Cynthia~A. Chestek.
\newblock A low-power band of neuronal spiking activity dominated by local single units improves the performance of brain--machine interfaces.
\newblock \emph{Nature biomedical engineering}, 4\penalty0 (10):\penalty0 973--983, 2020.
\newblock \doi{10.1038/s41551-020-0591-0}.
\newblock URL \url{https://doi.org/10.1038/s41551-020-0591-0}.

\bibitem[Olak et~al.(2026)Olak, Boccato, and Ferrante]{olak2026decoding}
Micha{\l} Olak, Tommaso Boccato, and Matteo Ferrante.
\newblock {B}rain {T}ranslation: {M}ultitask sequence-to-sequence modeling for intracortical attempted speech decoding.
\newblock In \emph{Findings of the Association for Computational Linguistics: EMNLP 2026}, pp.\  43901--43916, 2026.
\newblock URL \url{https://aclanthology.org/2026.findings-emnlp.2062/}.

\bibitem[Park et~al.(2019)Park, Chan, Zhang, Chiu, Zoph, Cubuk, and Le]{park2019specaugment}
Daniel~S. Park, William Chan, Yu~Zhang, Chung-Cheng Chiu, Barret Zoph, Ekin~D. Cubuk, and Quoc~V. Le.
\newblock {SpecAugment: A Simple Data Augmentation Method for Automatic Speech Recognition}.
\newblock In \emph{{Interspeech 2019}}, pp.\  2613--2617, 2019.
\newblock \doi{10.21437/Interspeech.2019-2680}.

\bibitem[Park \& Kim(2019)Park and Kim]{park2019g2p}
Kyubyong Park and Jongseok Kim.
\newblock {g2pE}.
\newblock \url{https://github.com/Kyubyong/g2p}, 2019.

\bibitem[Pham et~al.(2024)Pham, Liu, Yang, Chen, Liu, Yuan, Plummer, Wang, and Yang]{pham2024cipher}
Chau Pham, Boyi Liu, Yingxiang Yang, Zhengyu Chen, Tianyi Liu, Jianbo Yuan, Bryan Plummer, Zhaoran Wang, and Hongxia Yang.
\newblock Let models speak ciphers: Multiagent debate through embeddings.
\newblock In B.~Kim, Y.~Yue, S.~Chaudhuri, K.~Fragkiadaki, M.~Khan, and Y.~Sun (eds.), \emph{International Conference on Learning Representations}, volume 2024, pp.\  51899--51928, 2024.
\newblock URL \url{https://proceedings.iclr.cc/paper_files/paper/2024/file/e444859b2a22df6b56af9381ad1e9480-Paper-Conference.pdf}.

\bibitem[Ramesh \& Li(2025)Ramesh and Li]{ramesh2025activations}
Vignav Ramesh and Kenneth Li.
\newblock Communicating activations between language model agents.
\newblock In Aarti Singh, Maryam Fazel, Daniel Hsu, Simon Lacoste-Julien, Felix Berkenkamp, Tegan Maharaj, Kiri Wagstaff, and Jerry Zhu (eds.), \emph{Proceedings of the 42nd International Conference on Machine Learning}, volume 267 of \emph{Proceedings of Machine Learning Research}, pp.\  51094--51116. PMLR, 13--19 Jul 2025.
\newblock URL \url{https://proceedings.mlr.press/v267/ramesh25a.html}.

\bibitem[Shi et~al.(2026)Shi, Wang, Guo, Wang, Zhang, Zhang, Guo, Hao, Xi, Yang, Xu, Zhou, and Lin]{qwen2026asr}
Xian Shi, Xiong Wang, Zhifang Guo, Yongqi Wang, Pei Zhang, Xinyu Zhang, Zishan Guo, Hongkun Hao, Yu~Xi, Baosong Yang, Jin Xu, Jingren Zhou, and Junyang Lin.
\newblock {Qwen3-ASR} technical report, 2026.
\newblock URL \url{https://arxiv.org/abs/2601.21337}.

\bibitem[Su et~al.(2024)Su, Ahmed, Lu, Pan, Bo, and Liu]{su2024roformer}
Jianlin Su, Murtadha H.~M. Ahmed, Yu~Lu, Shengfeng Pan, Wen Bo, and Yunfeng Liu.
\newblock {RoFormer}: Enhanced transformer with rotary position embedding.
\newblock \emph{Neurocomputing}, 568:\penalty0 127063, 2024.
\newblock \doi{10.1016/J.NEUCOM.2023.127063}.
\newblock URL \url{https://doi.org/10.1016/j.neucom.2023.127063}.

\bibitem[{VikhrModels}(2025)]{borealis2025}
{VikhrModels}.
\newblock {Borealis}: Audio-language model for speech understanding, 2025.
\newblock URL \url{https://huggingface.co/Vikhrmodels/Borealis-5b-it}.

\bibitem[Wairagkar et~al.(2025)Wairagkar, Card, Singer-Clark, Hou, Iacobacci, Miller, Hochberg, Brandman, and Stavisky]{wairagkar2025instantaneous}
Maitreyee Wairagkar, Nicholas~S Card, Tyler Singer-Clark, Xianda Hou, Carrina Iacobacci, Lee~M Miller, Leigh~R Hochberg, David~M Brandman, and Sergey~D Stavisky.
\newblock An instantaneous voice-synthesis neuroprosthesis.
\newblock \emph{Nature}, 644\penalty0 (8075):\penalty0 145--152, 2025.

\bibitem[Wang et~al.(2025)Wang, Wang, Athiwaratkun, Zhang, and Zou]{wang2025moa}
Junlin Wang, Jue Wang, Ben Athiwaratkun, Ce~Zhang, and James~Y Zou.
\newblock Mixture-of-agents enhances large language model capabilities.
\newblock In Y.~Yue, A.~Garg, N.~Peng, F.~Sha, and R.~Yu (eds.), \emph{International Conference on Learning Representations}, volume 2025, pp.\  33944--33963, 2025.
\newblock URL \url{https://proceedings.iclr.cc/paper_files/paper/2025/file/5434be94e82c54327bb9dcaf7fca52b6-Paper-Conference.pdf}.

\bibitem[Willett et~al.(2023)Willett, Kunz, Fan, Avansino, Wilson, Choi, Kamdar, Glasser, Hochberg, Druckmann, Shenoy, and Henderson]{willett2023high}
Francis~R. Willett, Erin~M. Kunz, Chaofei Fan, Donald~T. Avansino, Guy~H. Wilson, Eun~Young Choi, Foram Kamdar, Matthew~F. Glasser, Leigh~R. Hochberg, Shaul Druckmann, Krishna~V. Shenoy, and Jaimie~M. Henderson.
\newblock {A high-performance speech neuroprosthesis}.
\newblock \emph{Nature}, 620\penalty0 (7976):\penalty0 1031--1036, 2023.
\newblock ISSN 1476-4687.
\newblock \doi{10.1038/s41586-023-06377-x}.
\newblock URL \url{https://doi.org/10.1038/s41586-023-06377-x}.

\bibitem[Willett et~al.(2024)Willett, Li, Le, Fan, Chen, Shlizerman, Chen, Zheng, Okubo, Benster, Lee, Kounga, Buchanan, Zoltowski, Linderman, and Henderson]{willett2024brain}
Francis~R. Willett, Jingyuan Li, Trung Le, Chaofei Fan, Mingfei Chen, Eli Shlizerman, Yue Chen, Xin Zheng, Tatsuo~S. Okubo, Tyler Benster, Hyun~Dong Lee, Maxwell Kounga, E.~Kelly Buchanan, David Zoltowski, Scott~W. Linderman, and Jaimie~M. Henderson.
\newblock Brain-to-text benchmark '24: Lessons learned, 2024.
\newblock URL \url{https://arxiv.org/abs/2412.17227}.

\bibitem[Yoon et~al.(2026)Yoon, Avansino, Madugula, Levin, Fan, Abramovich~Krasa, Singh, Vo, Hahn, Card, Fogg, Wairagkar, Nason-Tomaszewski, Jacques, Bechefsky, Iacobacci, Deo, Hochberg, Brandman, Stavisky, Au~Yong, Pandarinath, Henderson, and Willett]{yoon2026ensembles}
Seonghyun Yoon, Donald~T. Avansino, Sasidhar Madugula, Alisa~D. Levin, Chaofei Fan, Benyamin Abramovich~Krasa, Akansha Singh, Christina Vo, Nick~V. Hahn, Nicholas~S. Card, Zachery Fogg, Maitreyee Wairagkar, Samuel~R. Nason-Tomaszewski, Brandon~G. Jacques, Payton~H. Bechefsky, Carrina Iacobacci, Darrel~R. Deo, Leigh~R. Hochberg, David~M. Brandman, Sergey~D. Stavisky, Nicholas Au~Yong, Chethan Pandarinath, Jaimie~M. Henderson, and Francis~R. Willett.
\newblock Neural decoding of speech using deep neural ensembles.
\newblock \emph{bioRxiv}, 2026.
\newblock \doi{10.64898/2026.06.02.729705}.
\newblock URL \url{https://www.biorxiv.org/content/10.64898/2026.06.02.729705v1}.
\newblock Preprint.

\bibitem[Zhang et~al.(2022)Zhang, Roller, Goyal, Artetxe, Chen, Chen, Dewan, Diab, Li, Lin, Mihaylov, Ott, Shleifer, Shuster, Simig, Koura, Sridhar, Wang, and Zettlemoyer]{zhang2022opt}
Susan Zhang, Stephen Roller, Naman Goyal, Mikel Artetxe, Moya Chen, Shuohui Chen, Christopher Dewan, Mona Diab, Xian Li, Xi~Victoria Lin, Todor Mihaylov, Myle Ott, Sam Shleifer, Kurt Shuster, Daniel Simig, Punit~Singh Koura, Anjali Sridhar, Tianlu Wang, and Luke Zettlemoyer.
\newblock {OPT}: Open pre-trained transformer language models, 2022.
\newblock URL \url{https://arxiv.org/abs/2205.01068}.

\bibitem[Zhang et~al.(2026)Zhang, He, Fan, Liu, Yu, Le, Li, Linderman, Duncker, Willett, Mesgarani, and Paninski]{zhang2026bit}
Yizi Zhang, Linyang He, Chaofei Fan, Tingkai Liu, Han Yu, Trung Le, Jingyuan Li, Scott Linderman, Lea Duncker, Francis Willett, Nima Mesgarani, and Liam Paninski.
\newblock A cross-species neural foundation model for end-to-end speech decoding.
\newblock In C.~Vondrick, B.~Hariharan, C.~Raffel, L.~Pinto, D.~Yang, and A.~Faust (eds.), \emph{International Conference on Learning Representations}, volume 2026, pp.\  30773--30797, 2026.
\newblock URL \url{https://proceedings.iclr.cc/paper_files/paper/2026/file/34307cd20546fa1c6de90f303973ccf7-Paper-Conference.pdf}.

\bibitem[Zou et~al.(2026)Zou, Qiu, Li, Yang, Tieu, Lu, Shen, Tong, Choi, He, Zou, Wang, and Yang]{zou2026latentmas}
Jiaru Zou, Ruizhong Qiu, Gaotang Li, Xiyuan Yang, Katherine Tieu, Pan Lu, Ke~Shen, Hanghang Tong, Yejin Choi, Jingrui He, James Zou, Mengdi Wang, and Ling Yang.
\newblock Latent collaboration in multi-agent systems.
\newblock In \emph{Forty-third International Conference on Machine Learning}, 2026.

\end{thebibliography}
\bibliographystyle{iclr2027_conference}

\clearpage
\appendix
\section{Implementation Details}
\label{app:implementation}

\subsection{Data and initialization}
\label{app:model}

Section~\ref{sec:selection} describes dataset splits and checkpoint selection.

Input features are concatenated as threshold crossings followed by spike-band power.
Each input bin spans 20\,ms, giving $C=256$ channels for T12 and $C=512$ for T15.
The encoder uses Gaussian smoothing, non-overlapping five-frame patches, and seven bidirectional Transformer blocks.
T12 uses 128 electrodes, block-wise z-scoring, and subsequent day-level normalization estimated from that day's training trials.
T15 uses all 256 electrodes and the normalized features supplied in the released HDF5 files.
Participant-specific preprocessing retains the released 1,250/2,500-frame limits for T12/T15; both encoder trunks support the 2,500-frame horizon used for T15 input.
Transcriptions are lowercased, and punctuation other than apostrophes is removed.
Phoneme targets are generated from the transcriptions with \texttt{g2p\_en}~\citep{park2019g2p}, with stress markers removed, spaces mapped to silence, and a final silence appended.
The 41-class inventory contains 39 ARPAbet phones, silence, and blank.

The two released BIT models supply their neural encoders, native input projections, and matching phoneme readouts.
MAS-BIT introduces two cross-participant input projections, a neural fusion projector, two phoneme embedding tables, two message-role vectors, and three LoRA adapters on top of the pretrained Qwen text decoder.
The fusion projector independently layer-normalizes both 384-dimensional encoder states, concatenates them, and applies a $768\rightarrow1024\rightarrow2048$ MLP with GELU between the linear layers.
The two 41-row phoneme embedding tables are initialized with standard deviation scaled by $1/\sqrt{2}$ before their expected embeddings are summed.
These are trained in a single run, beginning with the alignment stage.
The neural encoder uses a width of 384, seven layers, six attention heads, a feedforward width of 512, GELU activations, and rotary positional embeddings~\citep{su2024roformer}.
Its participant-specific patch input layer applies LayerNorm, a linear projection from $5C$ to 384, and another LayerNorm.
Gaussian smoothing has a standard deviation of two input bins (40\,ms); embedding and Transformer dropout are $0.2$ and $0.4$.
The text decoder has width 2048 and 28 layers.
Each role's LoRA is applied to the query, key, value, output, gate, up, and down projections, with rank 8, scaling parameter 32 (a factor of $32/8$), and dropout $0.2$.

\newcommand{\RolePromptRows}{%
Neural expert & $Z_N$ & \texttt{\detokenize{<|neural_activity|>}}\#\allowbreak\texttt{\detokenize{<|/neural_activity|>}}\newline\emph{decode the above neural activity into an English sentence:} \\
\addlinespace[3pt]
Phonetic expert & $Z_P$ & \texttt{\detokenize{<|phoneme_probability_features|>}}\#\allowbreak\texttt{\detokenize{<|/phoneme_probability_features|>}}\newline\emph{decode the above phoneme probability features into an English sentence:} \\
\addlinespace[3pt]
Arbiter & $[Z_N;M_N;M_P]$ & \texttt{\detokenize{<|neural_and_expert_messages|>}}\#\allowbreak\texttt{\detokenize{<|/neural_and_expert_messages|>}}\newline\emph{decode the above neural activity and both agents' transcript estimates into an English sentence:} \\
\addlinespace[3pt]
Single agent, two views & $[Z_N+v_N;\;Z_P+v_P]$ & \texttt{\detokenize{<|neural_and_phoneme_features|>}}\#\allowbreak\texttt{\detokenize{<|/neural_and_phoneme_features|>}}\newline\emph{decode the above neural activity and phoneme probability features into an English sentence:} \\
}

\newcommand{\AdditionalRolePromptRows}{%
\addlinespace[3pt]
\red{Message-only arbiter} & \red{$[M_N;M_P]$} & \red{\texttt{\detokenize{<|expert_messages|>}}\#\allowbreak\texttt{\detokenize{<|/expert_messages|>}}\newline\emph{decode both agents' transcript estimates above into an English sentence:}} \\
\addlinespace[3pt]
\red{Refinement agent} & \red{$[Z_N;Z_P;M_1]$} & \red{\texttt{\detokenize{<|neural_phoneme_and_draft|>}}\#\allowbreak\texttt{\detokenize{<|/neural_phoneme_and_draft|>}}\newline\emph{refine the first draft using the above neural activity and phoneme probability features into an English sentence:}} \\
}

\paragraph{Role-specific prompts.}
Text delimiters identify each role's continuous input, followed by an instruction to produce an English sentence.
The neural expert's prompt names neural activity, the phonetic expert's prompt names phoneme probability features, and the arbiter's prompt names neural activity and both experts' transcript estimates.
Figure~\ref{fig:ablations}A shows the complete prompts, with $\#$ marking the continuous-input insertion.
Table~\ref{tab:role_prompts} gives the exact prompt text \red{for MAS-BIT and all ablation controls} used during training and inference.
Hard-text communication, \red{detached latent messages,} separate participant training, and no staged training retain the same three role prompts; the single-agent prompt names both input views.
\red{In the two-neural-expert control, both experts use the neural-expert template and the arbiter prompt is unchanged. The rank-24 and Borealis-5B single-agent variants use the same single-agent template.}
\red{The message-only control retains both expert prompts and uses the message-only arbiter template. Sequential dual-view refinement uses the single-agent text with $[Z_N;Z_P]$ for the first agent, the refinement-agent template with $[Z_N;Z_P;M_1]$ for the second, and the arbiter template with $[Z_N;M_1;M_2]$ for the final transcription.}
The delimiters and instructions use the pretrained tokenizer's existing vocabulary.
\red{We tokenize the literal text on each side of $\#$ with the corresponding decoder tokenizer, with automatic special-token insertion disabled and no additional system or chat messages.}
For the single-agent control, $v_N$ and $v_P$ are learned view embeddings for the neural and phonetic inputs, respectively.
\begin{table}[ht]
\centering
\caption{\footnotesize \emph{Prompts for MAS-BIT and \red{all ablation controls.}} The placeholder $\#$ is replaced by the continuous input in the middle column. Paired text delimiters and instruction text are shown verbatim. \red{Each template is a single line; line breaks are typographic.}}
\label{tab:role_prompts}
\small
\begin{tabular}{@{}p{0.22\linewidth}p{0.20\linewidth}p{\dimexpr0.58\linewidth-4\tabcolsep\relax}@{}}
\toprule
\textbf{Role} & \textbf{Input} & \textbf{Prompt} \\
\midrule
\RolePromptRows
\AdditionalRolePromptRows
\bottomrule
\end{tabular}
\end{table}

\subsection{Message construction}
\label{app:messages}

The continuous inputs to the neural expert, phonetic expert, and arbiter are $Z_N$, $Z_P$, and $[Z_N;M_N;M_P]$, respectively.
The hard-text control constructs $M_N$ and $M_P$ from the discrete draft-token embeddings.
The single-agent control concatenates $Z_N$ and $Z_P$ after adding learned view identifiers.
Cross-entropy supervises answer tokens, including EOS; conditioning tokens and padding receive no labels.
EOS token IDs are 151643 and 151645.

\begin{samepage}
For latent-message replay, the retained draft includes the first EOS token; generated padding is excluded by its validity mask.
Each expert's whole message is zeroed independently with probability $0.15$, without rescaling the retained messages; subsequent elementwise embedding dropout has probability $0.1$ and uses standard inverse-retention scaling.
Role vectors are added after both operations.
\end{samepage}

\subsection{Neural input augmentation}
\label{app:augmentation}

Every training example enters the augmentation pipeline.
Time warping and masking follow principles used in SpecAugment~\citep{park2019specaugment}.
First, we draw two interior cut points and remove duplicates to define temporal segments; we then resample each segment linearly by a factor drawn from $\mathcal{N}(1,0.05^2)$ and clip it to $[0.9,1.1]$.
Trials shorter than nine frames retain their original time grid.
We reject a proposed warp if it exceeds the input limit or leaves fewer encoder patches than required by the target-phone count plus adjacent repeated phones.
The remaining transformations still apply when warping is rejected.
Second, we mask each electrode independently with probability $0.05$, zeroing its threshold-crossing and spike-power features together across the trial.
Third, raw temporal spans of 10--20 frames are sampled with a cumulative span-length budget of $\lfloor0.1T_{\mathrm{aug}}\rfloor$, where $T_{\mathrm{aug}}$ is the post-warp sequence length; the last span is shortened to the remaining budget, and overlaps are allowed.
Fourth, we add independent Gaussian noise with standard deviation $0.2$ in normalized feature units.
These operations precede the encoder's Gaussian smoothing.

Let $S_{\mathrm{aug}}=\lceil T_{\mathrm{aug}}/5\rceil$ be the number of valid augmented patches.
After the patch input projection, sample a contiguous span length uniformly from $1$ to $\lfloor0.2S_{\mathrm{aug}}\rfloor$ and place it uniformly among valid starting positions.
Replace its patches with the frozen pretrained mask embedding, retaining sequence length and attention validity.
No span is applied when $\lfloor0.2S_{\mathrm{aug}}\rfloor=0$.
One augmentation realization is shared across the three agents for each example, including during the alignment stage.
Evaluation retains deterministic smoothing and disables these stochastic corruptions.

\subsection{Optimization and reproducibility}
\label{app:optimization}

\begin{table}[ht]
\centering
\caption{\footnotesize \emph{Parameter groups updated in each stage.} The pretrained language-model backbone remains frozen throughout.}
\label{tab:ownership}
\small
\begin{tabular}{@{}p{0.29\linewidth}p{0.67\linewidth}@{}}
\toprule
\textbf{Stage} & \textbf{Updated parameters} \\
\midrule
Alignment (1--5) & New participant-specific input projections, neural fusion projector, phoneme embedding tables, message-role embeddings \\
Encoder adaptation (6--10) & Both encoders and all input projections; all input interfaces \\
Neural specialization (11--15) & Encoder-adaptation groups and N adapter \\
Phonetic specialization (16--20) & Encoder-adaptation groups, both phoneme readouts, and P adapter \\
Arbiter specialization (21--25) & Encoder-adaptation groups and A adapter \\
Joint optimization (26--65) & All adaptation parameters, including all three role adapters and both phoneme readouts \\
\bottomrule
\end{tabular}
\end{table}

The schedule in Equation~\ref{eq:stage_schedule} comprises five five-epoch alignment, adaptation, and specialization blocks, followed by one 40-epoch joint block.
The intermediate-CTC ramp advances during encoder-active training, beginning at epoch 6.
All three cross-entropy terms have unit weight and use no label smoothing.
Intermediate CTC uses the same output normalization and phoneme readout as the final state, with total weight $0.05$ across layers 3 and 5 in addition to final-layer weight $0.3$.
The CTC implementation uses blank index zero and suppresses infinite losses for infeasible alignments.

We use Muon for the linear weight matrices in both encoder trunks, with momentum $0.95$, Nesterov updates, five Newton--Schulz iterations, weight decay $0.01$, and peak learning rate $4\times10^{-5}$~\citep{jordan2024muon}.
The update uses the matrix-shape scaling $0.2\sqrt{\max(m,n)}$ for a weight of shape $m\times n$.
The input projections, interfaces, phoneme readouts, role adapters, and encoder biases and normalization parameters use AdamW, with $(\beta_1,\beta_2)=(0.9,0.999)$, $\epsilon=10^{-8}$, and weight decay $10^{-5}$ for matrix parameters and zero for biases and normalization parameters.
AdamW peak learning rates are $2\times10^{-5}$ for encoder biases and normalization, all input projections, and phoneme readouts, and $10^{-4}$ for the interfaces and adapters; new input projections use $10^{-4}$ during alignment.
We apply a single gradient-norm clip of $1$ to the union of active parameters after accumulation and distributed reduction.
Learning rates follow a separate warmup--cosine curve within each block.
For a block containing $K$ optimizer updates, let $w=\max(1,\operatorname{round}(\omega K))$ and let $t$ count completed updates in that block.
An active group's learning rate is
\begin{equation}
    \eta_g(t)=\eta_g^{\mathrm{peak}}\,\mu_c\,s(t),
    \qquad
    s(t)=\begin{cases}
      0.1+0.9t/w, & t<w,\\
      0.1+0.45\left[1+\cos\!\left(\pi\frac{t-w}{K-w}\right)\right], & t\geq w,
    \end{cases}
    \label{eq:block_lr}
\end{equation}
The five-epoch blocks use peak multiplier $1$; the final 40-epoch joint block uses $0.125$.
AdamW uses warmup fraction $\omega=0.05$ throughout; Muon uses $0.10$ in each encoder-active five-epoch block and $0.05$ in the joint block.
The block-local warmup restarts at each stage transition, while optimizer moments are retained throughout the run.
The learning-rate schedule is fixed and is not reduced in response to validation plateaus.

We train on eight data-parallel ranks with an effective batch size of 32.
Each rank accumulates two microbatches of two utterances; each microbatch contains recordings from one participant.
Within each rank, cross-entropy sums are divided by that window's actual target-token count, including EOS, and CTC sums by its target-phone count.
The update averages the eight resulting rank-local objectives.
The last partial window uses its actual counts, and padding or inactive duplicate slots contribute zero loss.
Biases and one-dimensional parameters are excluded from weight decay.
Training uses BF16 autocast and random seed 42.
Validation generation and loss evaluation both use a batch size of 1.

\subsection{Checkpoint selection and evaluation}
\label{app:evaluation}

Each epoch is evaluated by greedily generating sentences on all provided labeled validation utterances.
Checkpoint selection begins at epoch 25, after the complete alignment and expert--arbiter prefix.
Earlier epochs are evaluated but are ineligible for selection.
We retain the eligible checkpoint with the smallest pooled word-edit count across 2,306 validation utterances (14,941 reference words) and stop after 15 epochs without a strict improvement, or at the maximum of 65 epochs.
The reference-word denominator is fixed, so minimizing edit count is equivalent to minimizing corpus validation-set WER.
Ties retain the earlier checkpoint; stage transitions do not reset patience, and training may stop within a block.

The T15 training set includes the three released ``do nothing'' control trials.
All declared rows are retained in training and evaluation.
The labeled validation sets support checkpoint selection and analysis of expert predictions.
Benchmark comparisons and all five trained controls report holdout WER; T15 public and private scores are kept separate.
For final holdout decoding, the selected checkpoint uses FP32 and a batch size of one.
Expert and arbiter generation are greedy, with fixed caps of 24 new tokens for T12 and 29 for T15, and terminate earlier at an end-of-sequence token.
Reference transcriptions are excluded from generation, stopping, and candidate selection.

\subsection{Implementation of the ablation controls}
\label{app:controls}

All controls share the data splits, augmentation, effective batch size 32, and selection rule.
Corresponding components use common initial weights; subsequent independent training can change their representations.

\paragraph{Single agent, two views.}
The input concatenates $Z_N$ and $Z_P$ and includes learned view identifiers.
Its prompt requests an English transcription from neural activity and phoneme probability features.
The single adapter starts from the untrained arbiter adapter and is active in all 55 scheduled post-alignment-and-encoder-adaptation epochs.
Its three five-epoch specialization blocks update the decoder each time, with phoneme-readout updates in the middle block, before the final 40-epoch joint block.
The decoder uses one role adapter and a single sentence loss.
An additional capacity control increases this adapter to rank 24 with scaling parameter 96, matching the three rank-8 role adapters on the same target modules while retaining the scaling factor of 4.
Its initial LoRA factors concatenate the three untrained role adapters along the rank dimension; the initial effective updates remain zero, and the remaining single-agent training and evaluation settings are unchanged.
\red{We also evaluate a larger-backbone single-agent control that replaces the frozen Qwen3-ASR-1.7B text decoder (width 2048) with the 4B audio-trained text decoder and tokenizer from Borealis-5B~\citep{borealis2025} (width 2560, 36 layers) and a rank-13 LoRA adapter with scaling parameter 52, using 26,836,992 LoRA parameters, 2.63\% more than the three rank-8 role adapters combined while projecting the neural and phonetic views to width 2560 under the same staged training schedule.}

\paragraph{\red{Sequential dual-view refinement.}}
\red{In this control, we retain the exact trainable parameter inventory, three rank-8 role adapters, three unit-weight sentence cross-entropy losses, and final and intermediate CTC losses of MAS-BIT, while replacing view-specific expert specialization with sequential dual-view drafting and refinement.
The first agent receives the concatenated continuous views $[Z_N; Z_P]$ with the single-agent dual-view prompt, generates a greedy draft, and replays that draft through its adapter to construct the top-16 probability-weighted latent message $M_1$.
The second agent receives $[Z_N; Z_P; M_1]$ with paired \texttt{\detokenize{<|neural_phoneme_and_draft|>}} delimiters and the instruction ``refine the first draft using the above neural activity and phoneme probability features into an English sentence:'', generates a refined greedy draft, and replays it to construct $M_2$; the arbiter receives $[Z_N; M_1; M_2]$ with MAS-BIT's arbiter prompt, reusing $M_1$ and its computation graph without regenerating or replaying it a second time.
This matches MAS-BIT's three greedy autoregressive generations and two message replays at inference (with message-role embeddings and message dropout retained during training), under the same model initialization, six-stage 65-epoch training schedule (updating the first-draft, refinement, and arbiter adapters in the three respective five-epoch specialization blocks before 40 epochs of joint optimization), optimizer hyperparameters, and pooled validation checkpoint selection.}

\paragraph{Hard-text communication.}
Each message uses the frozen token embeddings from the expert's greedy draft instead of probability-weighted embeddings.
The direct neural input to the arbiter and message-role embeddings are retained.
The discrete message blocks arbitration gradients into the sending expert, while the experts retain their own sentence supervision.
Consequently, this control changes both the retained token information and the gradient path.

\paragraph{\red{Detached latent messages.}}
\red{In this control, each expert replays its greedy draft to construct the same top-16 probability-weighted token embeddings, while detaching the replay logits before message construction so that the arbiter cross-entropy loss does not backpropagate through the messages into the sending experts, projectors, or neural encoders.
This preserves alternative token hypotheses in the forward pass while removing the message gradient path.
Direct neural input to the arbiter, message dropout, message-role embeddings, and the experts' own sentence supervision are retained under the same model initialization, staged training schedule, optimizer hyperparameters, and validation checkpoint selection as in MAS-BIT.}

\paragraph{\red{Message-only arbiter.}}
\red{In this control, the arbiter receives only the concatenated latent messages $[M_N; M_P]$, omitting the direct neural input $Z_N$ during training, validation, and holdout decoding.
To guide decoding, the arbiter prompt instructs the model to translate both agents' transcript estimates into an English sentence.
Both messages remain differentiable top-16 probability-weighted token embeddings with message-role embeddings and message dropout, allowing the arbiter cross-entropy loss to backpropagate through the messages into the sending experts, projectors, and neural encoders.
We retain the same model initialization, staged training schedule, optimizer hyperparameters, and validation checkpoint selection as in MAS-BIT.
Its holdout WER exceeds the hard-text control by \PP{message_only}{hard}{t12} points on T12 and \PP{message_only}{hard}{t15public}/\PP{message_only}{hard}{t15private} points on T15 public/private (Table~\ref{tab:ablation_results}).}

\paragraph{Two neural experts.}
The second expert receives the same fused neural view as the first, replacing the phoneme-posterior view with a second independently adapted neural expert.
Both retain their own adapters and message-role identifiers; the corresponding expert block updates the second neural adapter.
The two encoder trunks, direct neural input, staged-to-joint schedule, and auxiliary phoneme supervision are retained.

\paragraph{Separate participant training.}
We train two independent MAS-BIT models on the 8,800 T12 and 8,072 T15 training utterances, respectively, retaining the two-encoder frontend, common initialization, 65-epoch staged-to-joint schedule, effective batch size 32, and optimizer settings.
Block learning-rate schedules follow each participant's epoch length, giving 275/253 optimizer updates per epoch on T12/T15, compared with 528 for joint training.
Checkpoints are selected from epoch 25 on the corresponding participant's complete labeled validation set (880 T12 or 1,426 T15 utterances), with the same integer word-edit criterion, earlier-checkpoint tie rule, and 15-epoch patience before FP32 holdout evaluation.

\paragraph{No staged training.}
All adaptation groups, including both encoders, input projections, readouts, interfaces, and role adapters, update from epoch 1.
We use a single 65-epoch warmup--cosine schedule, with 10\% warmup for Muon and 5\% for AdamW, retaining the same optimizer, maximum epoch budget, and pooled checkpoint-selection rule.

\section{Analysis of expert predictions}
\label{app:expert_cases}
We analyze the selected joint checkpoint's complete FP32, batch-one validation predictions: 880 T12 and 1,426 T15 utterances.
These labeled validation sets support analysis of expert errors.
All three agents generate greedily from the same recording and checkpoint, and references are used after generation.
Arbitration, draft recombination, message interventions, and input masking share these expert outputs and the same complete-input arbiter baseline.
We first compare the three reference-blind outputs from the same model and recordings.

\begin{table}[ht]
\centering
\caption{\footnotesize \emph{Arbitration improves upon both expert outputs.} Word-edit and exact-match counts cover every validation utterance. All three outputs are generated without the reference.}
\label{tab:arbitration_gains}
\small
\begin{tabular}{lrrrr}
\toprule
& \multicolumn{2}{c}{T12} & \multicolumn{2}{c}{T15}\\
\cmidrule(lr){2-3}\cmidrule(lr){4-5}
\textbf{Output} & \textbf{Word edits} & \textbf{Exact matches} & \textbf{Word edits} & \textbf{Exact matches}\\
\midrule
Neural expert & \ArbCount{t12}{neural_edits} & \ArbCount{t12}{neural_correct} & \ArbCount{t15}{neural_edits} & \ArbCount{t15}{neural_correct}\\
Phonetic expert & \ArbCount{t12}{phonetic_edits} & \ArbCount{t12}{phonetic_correct} & \ArbCount{t15}{phonetic_edits} & \ArbCount{t15}{phonetic_correct}\\
Arbiter & \ArbCount{t12}{arbiter_edits} & \ArbCount{t12}{arbiter_correct} & \ArbCount{t15}{arbiter_edits} & \ArbCount{t15}{arbiter_correct}\\
\bottomrule
\end{tabular}
\end{table}
On T12, arbitration reduces the neural/phonetic experts' \ArbCount{t12}{neural_edits}/\ArbCount{t12}{phonetic_edits} word edits to \ArbCount{t12}{arbiter_edits}, reductions of \ArbCount{t12}{neural_relative_reduction}\%/\ArbCount{t12}{phonetic_relative_reduction}\%, while increasing exact matches from \ArbCount{t12}{neural_correct}/\ArbCount{t12}{phonetic_correct} to \ArbCount{t12}{arbiter_correct}.
On T15, the corresponding edit counts fall from \ArbCount{t15}{neural_edits}/\ArbCount{t15}{phonetic_edits} to \ArbCount{t15}{arbiter_edits} (\ArbCount{t15}{neural_relative_reduction}\%/\ArbCount{t15}{phonetic_relative_reduction}\% reductions), and exact matches increase from \ArbCount{t15}{neural_correct}/\ArbCount{t15}{phonetic_correct} to \ArbCount{t15}{arbiter_correct}.

\subsection{Arbitration beyond aligned draft recombination}
\label{app:word_oracle}

We construct a word-alignment graph from the neural and phonetic drafts before reading the reference.
We retain all minimum-cost alignments using unit costs for substitutions, deletions, and insertions.
At each alignment edge, the graph offers either expert's word, with an empty choice where the alignment has a gap.
The graph preserves word order and contains both original drafts.
For its set of candidate sentences $\mathcal{L}_i$, the minimum edit distance from the reference is
\[
e_i^{\mathrm{comb}}=\min_{z\in\mathcal{L}_i}\operatorname{ED}(z,y_i).
\]
We compute this distance by dynamic programming over the graph and reference positions after graph construction.
An exact arbiter recovery lies beyond aligned draft recombination when $e_i^{\mathrm{comb}}>0$.

Among the \CaseCount{t12}{arbiter_recovers}/\CaseCount{t15}{arbiter_recovers} sentences recovered when both drafts are wrong, \WordStat{t12}{exact_graph_recombination}/\WordStat{t15}{exact_graph_recombination} are available through the alignment graph and \WordStat{t12}{exact_outside_graph}/\WordStat{t15}{exact_outside_graph} lie outside it (Figure~\ref{fig:sentence_recovery}).
Of the latter, \WordStat{t12}{outside_graph_with_absent_word}/\WordStat{t15}{outside_graph_with_absent_word} contain a correct word absent from both drafts.
These recoveries show that the arbiter generates correct answers beyond aligned combinations of the expert drafts.

\paragraph{Word-level complementarity.}
We align each draft and the arbiter output to the reference with the same unit edit costs, resolving ties by diagonal, deletion, then insertion.
At each reference-word position, we record which experts are correct and whether the arbiter retains the correct word.
The neural expert is uniquely correct at \WordStat{t12}{neural_only_sites}/\WordStat{t15}{neural_only_sites} positions on T12/T15; the arbiter is correct at \WordStat{t12}{neural_only_retained}/\WordStat{t15}{neural_only_retained} of these positions.
The phonetic expert is uniquely correct at \WordStat{t12}{phonetic_only_sites}/\WordStat{t15}{phonetic_only_sites} positions, with \WordStat{t12}{phonetic_only_retained}/\WordStat{t15}{phonetic_only_retained} retained by the arbiter.

\begin{table}[ht]
\centering
\caption{\footnotesize \emph{Word positions correct in exactly one expert draft.} Columns identify the correct expert; substitution/deletion counts describe the other expert's error. Training frequency is measured in the pooled neural fine-tuning transcripts. Phoneme distance is averaged over dictionary-covered substitutions, with coverage shown explicitly.}
\label{tab:word_complementarity}
\small
\begin{tabular}{lrrrr}
\toprule
& \multicolumn{2}{c}{T12} & \multicolumn{2}{c}{T15} \\
\cmidrule(lr){2-3}\cmidrule(lr){4-5}
\textbf{Quantity} & \textbf{Neural} & \textbf{Phonetic} & \textbf{Neural} & \textbf{Phonetic} \\
\midrule
\WordComplementarityRows
\bottomrule
\end{tabular}
\end{table}

Substitutions account for most expert disagreements (Table~\ref{tab:word_complementarity}).
We count target-word occurrences in the pooled 16,872 neural fine-tuning transcripts (8,800 T12 and 8,072 T15) and group them as unseen, 1--5, 6--20, or more than 20 occurrences.
We use NLTK's contextual part-of-speech tags to group words: closed-class tags and a fixed pronoun/auxiliary list define function words, while the remaining noun, verb, adjective, and adverb tags define content words.
For substitutions, we use stress-stripped CMU pronunciations and take the minimum, over pronunciation pairs, of edit distance divided by the longer pronunciation length.
Words absent from the dictionary retain missing phoneme-distance values.

The arbiter retains \CapacityStat{t12_N_retention_percent}\%/\CapacityStat{t15_N_retention_percent}\% of the neural expert's uniquely correct positions and \CapacityStat{t12_P_retention_percent}\%/\CapacityStat{t15_P_retention_percent}\% of the phonetic expert's on T12/T15, giving pooled retention of \CapacityStat{all_N_retention_percent}\% (\CapacityStat{all_N_retained}/\CapacityStat{all_N_unique}) and \CapacityStat{all_P_retention_percent}\% (\CapacityStat{all_P_retained}/\CapacityStat{all_P_unique}), respectively.
Among \CapacityStat{all_N_unseen_positions} reference-word positions unseen in the pooled neural fine-tuning transcripts, the neural and phonetic experts are uniquely correct at \CapacityStat{all_N_unseen_unique} (\CapacityStat{all_N_unseen_unique_percent}\%) and \CapacityStat{all_P_unseen_unique} (\CapacityStat{all_P_unseen_unique_percent}\%) positions; the arbiter retains \CapacityStat{all_N_unseen_retained}/\CapacityStat{all_N_unseen_unique} (\CapacityStat{all_N_unseen_retention_percent}\%) and \CapacityStat{all_P_unseen_retained}/\CapacityStat{all_P_unseen_unique} (\CapacityStat{all_P_unseen_retention_percent}\%) of these correct words.
At positions uniquely correct in the phonetic expert, the other expert's dictionary-covered substitutions have smaller mean normalized phoneme distances (\CapacityStat{t12_P_phone_mean}/\CapacityStat{t15_P_phone_mean} on T12/T15) than at positions uniquely correct in the neural expert (\CapacityStat{t12_N_phone_mean}/\CapacityStat{t15_N_phone_mean}); the pooled means are \CapacityStat{all_P_phone_mean} and \CapacityStat{all_N_phone_mean}.
These complementary word-level contributions are retained by arbitration, including on words absent from neural fine-tuning.

\subsection{Expert messages complement direct neural conditioning}
\label{app:arbiter_inputs}
We freeze the selected checkpoint and decode the full validation sets under five input conditions (Table~\ref{tab:arbiter_inputs}).
For each utterance, the experts generate their drafts and latent messages once; the arbiter then decodes with the complete input, each message masked separately, both messages masked, or the direct neural input zeroed.
Message masking follows training dropout: we zero the message content before adding its role embedding, preserving sequence lengths, positions, and validity masks.
Zeroing the projected neural input leaves both expert messages unchanged.

\begin{table}[ht]
\centering
\caption{\footnotesize \emph{Arbitration is most accurate with both expert messages and direct neural conditioning.} Word edits, corpus validation WER (\%), and exact sentence matches cover all 880/1,426 T12/T15 utterances and 5,496/9,445 reference words. All conditions use fixed weights, FP32 batch-one decoding, and the same cached expert outputs. Complete-input predictions match the preceding arbitration analysis and the Top-16 message condition.}
\label{tab:arbiter_inputs}
\small
\setlength{\tabcolsep}{4pt}
\begin{tabular}{@{}lrrrrrr@{}}
\toprule
& \multicolumn{3}{c}{T12} & \multicolumn{3}{c}{T15} \\
\cmidrule(lr){2-4}\cmidrule(lr){5-7}
\textbf{Arbiter input} & \textbf{Edits} & \textbf{WER} & \textbf{Exact matches} & \textbf{Edits} & \textbf{WER} & \textbf{Exact matches} \\
\midrule
\InputMaskRows
\bottomrule
\end{tabular}
\end{table}

Complete input yields the lowest validation WER and the most exact matches on both participants.
Masking either message, both messages, or the direct neural input increases word errors and reduces exact matches.
Masking both messages adds \InputStat{t12}{without_both_messages}{delta_wer_pp}/\InputStat{t15}{without_both_messages}{delta_wer_pp} percentage points to WER and reduces exact matches from \InputStat{t12}{full}{exact_sentences}/\InputStat{t15}{full}{exact_sentences} to \InputStat{t12}{without_both_messages}{exact_sentences}/\InputStat{t15}{without_both_messages}{exact_sentences} on T12/T15.
With weights and expert drafts fixed, these interventions show that the trained arbiter uses both expert messages and direct neural evidence to improve full-set decoding accuracy.

\paragraph{Trial-specific neural input supports recovery beyond the drafts.}
We track retention of the full-input arbiter's exact recoveries on utterances where both expert drafts are wrong.
Zeroing $Z_N$ while preserving both complete latent messages retains \NeuralInputStat{t12}{zero_retained}/\NeuralInputStat{t15}{zero_retained} of the original \CaseCount{t12}{arbiter_recovers}/\CaseCount{t15}{arbiter_recovers} recoveries on T12/T15.
None of the original \WordStat{t12}{exact_outside_graph}/\WordStat{t15}{exact_outside_graph} outside-graph recoveries survives, including the \WordStat{t12}{outside_graph_with_absent_word}/\WordStat{t15}{outside_graph_with_absent_word} containing a correct word absent from both drafts.

To test neural content using nonzero inputs, we replace only $Z_N$ with another validation trial's representation from the same participant and with the same valid neural-token count, keeping the recipient's latent messages and masks fixed.
Three deterministic donor mappings select without reference or hypothesis text, prioritizing the same recording day and then the nearest day; donors are reused cyclically when fewer than three are available.
The paired subsets contain \NeuralInputStat{t12}{paired_rows}/\NeuralInputStat{t12}{full_rows} T12 and \NeuralInputStat{t15}{paired_rows}/\NeuralInputStat{t15}{full_rows} T15 utterances, with \NeuralInputStat{t12}{paired_words}/\NeuralInputStat{t15}{paired_words} reference words; other rows have no length-matched donor.
Self-replacement reproduces the baseline token sequences on every utterance.
Across the three donor mappings, mean word edits rise from \NeuralInputStat{t12}{baseline_edits}/\NeuralInputStat{t15}{baseline_edits} to \NeuralInputStat{t12}{donor_mean_edits}/\NeuralInputStat{t15}{donor_mean_edits}, increasing paired-set WER from \NeuralInputStat{t12}{baseline_wer}\%/\NeuralInputStat{t15}{baseline_wer}\% to \NeuralInputStat{t12}{donor_mean_wer}\%/\NeuralInputStat{t15}{donor_mean_wer}\%.
Each mapping removes all \NeuralInputStat{t12}{paired_recovered}/\NeuralInputStat{t15}{paired_recovered} original both-wrong recoveries in these subsets, including all \NeuralInputStat{t12}{paired_outside}/\NeuralInputStat{t15}{paired_outside} outside-graph recoveries.
These fixed-weight interventions show that the arbiter uses trial-specific neural content together with the expert messages.

\subsection{Token alternatives in frozen expert messages}
\label{app:soft_message_interventions}

We vary expert messages while fixing the selected model weights, greedy expert drafts, and direct neural input to the arbiter.
For each utterance, we cache the neural representations and the top-16 token logits from replay along each expert's own draft.
We retain the top $k\in\{1,2,4,8,16\}$ candidates at each message position, renormalize their probabilities at $\tau=1$, and construct the weighted embeddings in Equation~\ref{eq:soft_message}.
Both experts use the same $k$; message lengths, masks, and learned role embeddings remain fixed.

The arbiter uses the same greedy decoding rule throughout.

\begin{figure}[!t]
\centering
\includegraphics[width=\linewidth]{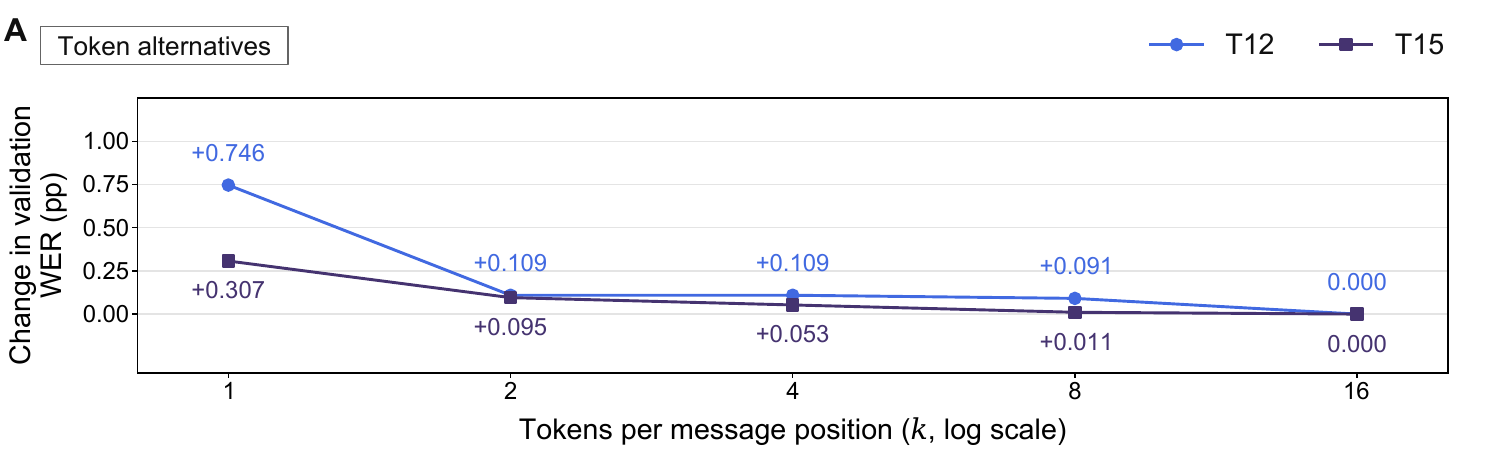}
\caption{\footnotesize \emph{Token alternatives improve decoding in both participants.} Changes in corpus validation WER are relative to top-16 messages at $\tau=1$ on all 880 T12 and 1,426 T15 validation utterances. Positive values indicate higher WER. Token support varies from 1 to 16 on a logarithmic axis. Values are labeled to three decimals. Model weights, expert drafts, direct neural input, and greedy arbiter decoding remain fixed.}
\label{fig:soft_messages}
\end{figure}

Figure~\ref{fig:soft_messages} shows the changes in corpus validation-set WER relative to the original top-16 messages.
Tables~\ref{tab:soft_message_results} and~\ref{tab:soft_message_groups} report the absolute WER on the complete provided validation sets and within each expert-draft group.
Expert agreement, disagreement, and correctness are determined from the original fixed drafts under the common text normalization.
These comparisons measure token-support sensitivity at fixed weights; the hard-text control in Appendix~\ref{app:controls} instead trains with a discrete message interface.

\begin{table}[ht]
\centering
\caption{\footnotesize \emph{Token support with fixed model weights and expert drafts.} At $\tau=1$, corpus validation-set WER (\%) and word edits are reported on all 880 T12 and 1,426 T15 provided validation utterances. The reference-word denominators are 5,496 and 9,445, respectively. Top-1 changes the messages of the trained MAS-BIT model; it is distinct from the separately trained hard-text control.}
\label{tab:soft_message_results}
\small
\begin{tabular}{@{}lrrrr@{}}
\toprule
& \multicolumn{2}{c}{T12} & \multicolumn{2}{c}{T15} \\
\cmidrule(lr){2-3}\cmidrule(lr){4-5}
\textbf{Messages} & \textbf{Edits} & \textbf{Validation WER} & \textbf{Edits} & \textbf{Validation WER} \\
\midrule
Top-1 & 598 & 10.881 & 449 & 4.754 \\
Top-2 & 563 & 10.244 & 429 & 4.542 \\
Top-4 & 563 & 10.244 & 425 & 4.500 \\
Top-8 & 562 & 10.226 & 421 & 4.457 \\
Top-16 & 557 & 10.135 & 420 & 4.447 \\
\bottomrule
\end{tabular}
\end{table}

\begin{table}[ht]
\centering
\caption{\footnotesize \emph{Token-support sensitivity across fixed expert-draft groups.} Values are corpus validation-set WER (\%). Groups are defined before intervention. Agreement and disagreement partition each validation set; both-wrong utterances overlap these groups.}
\label{tab:soft_message_groups}
\footnotesize
\setlength{\tabcolsep}{4pt}
\begin{tabular}{@{}llrrrr@{}}
\toprule
\textbf{Participant} & \textbf{Expert drafts} & \textbf{Rows} & \textbf{Words} & \textbf{Top-16} & \textbf{Top-1} \\
\midrule
T12 & Agree & 460 & 2761 & 2.789 & 2.970 \\
T12 & Disagree & 420 & 2735 & 17.550 & 18.867 \\
T12 & Both wrong & 305 & 1975 & 25.924 & 28.101 \\
T15 & Agree & 1015 & 6565 & 0.716 & 0.716 \\
T15 & Disagree & 411 & 2880 & 12.951 & 13.958 \\
T15 & Both wrong & 262 & 1829 & 21.542 & 23.127 \\
\bottomrule
\end{tabular}
\end{table}

Increasing token support reduces word edits from 598 to 557 on T12 and from 449 to 420 on T15.
Retaining 16 token alternatives instead of one removes 41/29 word edits, reducing validation WER by 0.746/0.307 percentage points.
These results show that multiple token hypotheses contribute beyond the experts' top-ranked tokens.

\section{Reproduced cascade}
\label{app:cascade}

\paragraph{Neural encoder and language models.}
We reproduce cascaded decoding with the released BIT encoder for each participant~\citep{zhang2026bit}.
These participant-specific checkpoints also initialize the two encoders jointly adapted in MAS-BIT.
\red{They are the checkpoints fine-tuned for sentence-level decoding; the encoders fine-tuned for phoneme-level decoding that produced the published cascade results are not part of the public release, so the reproduced and published cascades are not expected to coincide (Table~\ref{tab:cascade_results}).}

We decode phoneme logits using the benchmark's OpenWebText2 5-gram weighted finite-state transducer (WFST).
We rescored the lattice by replacing the pruned 5-gram contribution with the unpruned model before extracting up to 100 hypotheses.
This candidate limit is fixed across all experiments; $K$ controls only the number of ranked hypotheses passed to OPT-6.7B~\citep{zhang2022opt}.
If fewer than $K$ hypotheses are available, we retain all of them.
Table~\ref{tab:cascade_parameters} summarizes the decoder parameters.

\begin{table}[ht]
\centering
\caption{\footnotesize \emph{Hyperparameters used for cascaded decoding.}}
\label{tab:cascade_parameters}
\footnotesize
\begin{tabular}{@{}lc@{}}
\toprule
\textbf{Parameter} & \textbf{Value} \\
\midrule
Maximum 5-gram hypotheses & 100 \\
Pruning beam / lattice beam & 17 / 8 \\
Maximum / minimum active states & 7,000 / 200 \\
Acoustic score scale & 0.325 \\
Blank penalty / blank-skip threshold & 90 / 1 \\
OPT rescoring weight ($\alpha$) & 0.55 \\
Length penalty & 0 \\
\bottomrule
\end{tabular}
\end{table}

For each $K$, the output maximizes $s(h)=0.325\,s_{\mathrm{acoustic}}(h)+0.45\,s_{\mathrm{5gram}}(h)+0.55\,s_{\mathrm{OPT}}(h)$, where the OPT score is the sum of token log-probabilities without length normalization.
In a tie, the first candidate is retained. OPT rescores every retained candidate, including the single candidate at $K=1$.
We do not use reference labels to generate or rank hypotheses.

We retain the candidate sentences and their OPT scores and apply the same selection rule at each limit.
The cached scores reproduce the transcriptions from separate decoding at the original seven limits.
We specify all additional limits before obtaining their holdout scores.
For each limit, we evaluate WER on every holdout sentence; the public and private T15 scores refer to distinct hidden subsets.
Neural decoding and OPT scoring use FP32 without TF32 and process one utterance or candidate sentence at a time.
Figure~\ref{fig:benchmark}B shows the effect of the candidate limit on WER.
Straight segments connect the measured settings; no smoothed curve is fitted.
\red{To evaluate cascaded decoding after encoder adaptation, we also run the same FP32 pipeline across the 13 candidate limits using each participant's fine-tuned encoder branch, input projection, and final CTC readout from MAS-BIT (Table~\ref{tab:cascade_results}).
At $K=100$, this lowers cascade holdout WER by \PP{cascade100}{cm1cascade100}{t12} points on T12 and \PP{cascade100}{cm1cascade100}{t15public}/\PP{cascade100}{cm1cascade100}{t15private} points on T15 public/private relative to the released encoders.}

\paragraph{Peak host \red{and GPU} memory.}
\label{app:memory}
We compare peak host RAM \red{and GPU memory} in complete inference runs, including model and data loading, decoding, and audits.
The measurements use a single NVIDIA A40 GPU and 16 cores of an AMD EPYC 7662 CPU, with FP32 computation and batch size 1.
Each run loads the decoder and completes all 1,200 T12 or 1,450 T15 holdout utterances from one participant.
MAS-BIT uses the same jointly trained model for both participants; the reproduced cascade uses the released participant-specific checkpoints with $K=100$, including WFST decoding, unpruned 5-gram rescoring, and OPT rescoring.
\red{We also evaluate BIT + LightBeam~\citep{feghhi2026lightbeam} under the same FP32 setup (beam 1,000/900, 100k lexicon, pruned 4-gram KenLM, and Llama-3.2-1B with a reproduced next-word-prediction adapter), obtaining \WER{lightbeam}{t12}\% holdout WER on T12 and \WER{lightbeam}{t15public}\%/\WER{lightbeam}{t15private}\% on T15 public/private (Table~\ref{tab:cascade_results}).}
We sample \red{process-tree host resident memory and device-used GPU memory} once per second, and report the \red{respective peaks}.
Decoding time excludes loading and audits and is rounded to 0.1~s; GPU values come from matched reruns.

\paragraph{Offline decoding time.}
The one-second phase timestamps estimate complete-holdout decoding at about \MasTwelveSeconds{}/\MasFifteenSeconds{}~s for MAS-BIT versus \CascadeTwelveSeconds{}/\CascadeFifteenSeconds{}~s for the reproduced $K=100$ cascade on T12/T15, approximately \CapacityStat{t12_speedup}$\times$ faster on both participants.
These durations cover decoding all 1,200/1,450 utterances with the single-A40, 16-core, FP32 batch-one setup above; model/data loading and audit phases are excluded.
\red{Under the same setup, BIT + LightBeam takes \LightBeamTwelveSeconds/\LightBeamFifteenSeconds{}~s on T12/T15 (\LightBeamTwelveRatio$\times$/\LightBeamFifteenRatio$\times$ and 1.3$\times$/1.6$\times$ the decoding time of MAS-BIT and the $K=100$ cascade).}

\paragraph{\red{Consumer laptop verification.}}
\red{We also compare MAS-BIT with BIT + LightBeam~\citep{feghhi2026lightbeam} on a consumer laptop equipped with an NVIDIA GeForce RTX 5090 Laptop GPU, an Intel Core Ultra 9 275HX processor, and 62.3~GiB host RAM, running Ubuntu under WSL2.}
\red{Both systems use the same software environment, FP32 computation, batch size 1, and 16 logical CPUs. Each run processes all 1,200 T12 or 1,450 T15 holdout utterances, and all predictions match the corresponding A40 results.}
\red{We sample host resident memory and device-used GPU memory once per second and report the peaks across loading, decoding, and audits; decoding time excludes loading and audits.}
\red{The reproduced $K=100$ cascade is omitted because its measured \CascadeTwelveRAM/\CascadeFifteenRAM{}~GiB host memory footprint exceeds the laptop's available RAM.}
\red{MAS-BIT completes holdout decoding in \LaptopMasTwelveDecode/\LaptopMasFifteenDecode{}~s on T12/T15, compared with \LaptopLightBeamTwelveDecode/\LaptopLightBeamFifteenDecode{}~s for BIT + LightBeam, with lower peak host and GPU memory (Table~\ref{tab:laptop_efficiency}).}

\begin{table}[ht]
\centering
\caption{\footnotesize \red{\emph{Peak host RAM, GPU memory, and decoding time on a consumer RTX 5090 Laptop GPU.} Each value summarizes one complete holdout run under WSL2. Memory includes loading and audits; decode time excludes both and is rounded to 0.1~s. GPU values are one-second device-used peaks.}}
\label{tab:laptop_efficiency}
\small
\setlength{\tabcolsep}{2pt}
\begin{tabular}{@{}lcccccc@{}}
\toprule
& \multicolumn{2}{c}{\red{\textbf{Host RAM (GiB)}}} & \multicolumn{2}{c}{\red{\textbf{Peak GPU (GiB)}}} & \multicolumn{2}{c}{\red{\textbf{Decode time (s)}}} \\
\cmidrule(lr){2-3}\cmidrule(lr){4-5}\cmidrule(lr){6-7}
\red{\textbf{Method}} & \red{\textbf{T12}} & \red{\textbf{T15}} & \red{\textbf{T12}} & \red{\textbf{T15}} & \red{\textbf{T12}} & \red{\textbf{T15}} \\
\midrule
\red{BIT + LightBeam~\citep{feghhi2026lightbeam}} & \red{\LaptopLightBeamTwelveRAM} & \red{\LaptopLightBeamFifteenRAM} & \red{\LaptopLightBeamTwelveGPU} & \red{\LaptopLightBeamFifteenGPU} & \red{\LaptopLightBeamTwelveDecode} & \red{\LaptopLightBeamFifteenDecode} \\
\rowcolor{blue!10}\red{\textbf{MAS-BIT (ours)}} & \red{\textbf{\LaptopMasTwelveRAM}} & \red{\textbf{\LaptopMasFifteenRAM}} & \red{\textbf{\LaptopMasTwelveGPU}} & \red{\textbf{\LaptopMasFifteenGPU}} & \red{\textbf{\LaptopMasTwelveDecode}} & \red{\textbf{\LaptopMasFifteenDecode}} \\
\bottomrule
\end{tabular}
\end{table}

\paragraph{\red{Consumer desktop verification.}}
\red{We repeat this comparison on a consumer desktop equipped with an NVIDIA GeForce RTX 4090 GPU, an Intel Core i9-14900KF processor (5.6~GHz P-core frequency ceiling), and 61.5~GiB host RAM, running native Ubuntu.}
\red{Both systems use the same CUDA~13.2 software environment, FP32 computation, batch size 1, and 16 bound physical cores (eight P-cores and eight E-cores). Each run processes all 1,200 T12 or 1,450 T15 holdout utterances, and all predictions match the corresponding A40 results.}
\red{We sample host resident memory and per-process GPU memory once per second and report the peaks across loading, decoding, and audits; decoding time excludes loading and audits.}
\red{The reproduced $K=100$ cascade is omitted because its measured \CascadeTwelveRAM/\CascadeFifteenRAM{}~GiB host memory footprint exceeds the desktop's available RAM.}
\red{MAS-BIT completes holdout decoding in \DesktopMasTwelveDecode/\DesktopMasFifteenDecode{}~s on T12/T15, compared with \DesktopLightBeamTwelveDecode/\DesktopLightBeamFifteenDecode{}~s for BIT + LightBeam, with lower peak host and GPU memory (Table~\ref{tab:desktop_efficiency}).}

\begin{table}[ht]
\centering
\caption{\footnotesize \red{\emph{Peak host RAM, GPU memory, and decoding time on a consumer RTX 4090 desktop GPU.} Each value summarizes one complete holdout run on native Ubuntu. Memory includes loading and audits; decode time excludes both and is rounded to 0.1~s. GPU values are one-second per-process peaks.}}
\label{tab:desktop_efficiency}
\small
\setlength{\tabcolsep}{2pt}
\begin{tabular}{@{}lcccccc@{}}
\toprule
& \multicolumn{2}{c}{\red{\textbf{Host RAM (GiB)}}} & \multicolumn{2}{c}{\red{\textbf{Peak GPU (GiB)}}} & \multicolumn{2}{c}{\red{\textbf{Decode time (s)}}} \\
\cmidrule(lr){2-3}\cmidrule(lr){4-5}\cmidrule(lr){6-7}
\red{\textbf{Method}} & \red{\textbf{T12}} & \red{\textbf{T15}} & \red{\textbf{T12}} & \red{\textbf{T15}} & \red{\textbf{T12}} & \red{\textbf{T15}} \\
\midrule
\red{BIT + LightBeam~\citep{feghhi2026lightbeam}} & \red{\DesktopLightBeamTwelveRAM} & \red{\DesktopLightBeamFifteenRAM} & \red{\DesktopLightBeamTwelveGPU} & \red{\DesktopLightBeamFifteenGPU} & \red{\DesktopLightBeamTwelveDecode} & \red{\DesktopLightBeamFifteenDecode} \\
\rowcolor{blue!10}\red{\textbf{MAS-BIT (ours)}} & \red{\textbf{\DesktopMasTwelveRAM}} & \red{\textbf{\DesktopMasFifteenRAM}} & \red{\textbf{\DesktopMasTwelveGPU}} & \red{\textbf{\DesktopMasFifteenGPU}} & \red{\textbf{\DesktopMasTwelveDecode}} & \red{\textbf{\DesktopMasFifteenDecode}} \\
\bottomrule
\end{tabular}
\end{table}

\clearpage
\section{Benchmark and control results}
\label{app:benchmark_tables}

Tables~\ref{tab:cascade_results} and~\ref{tab:ablation_results} give the exact holdout results summarized in Figures~\ref{fig:benchmark}B and~\ref{fig:ablations}B--C, respectively.
\red{Table~\ref{tab:cascade_results} additionally reports the fine-tuned-encoder cascade control across all 13 candidate limits and BIT + LightBeam~\citep{feghhi2026lightbeam}.}
Table~\ref{tab:ablation_results} additionally reports the rank-24 single-agent capacity control\red{, the Borealis-5B single-agent backbone control, the sequential dual-view refinement control, the detached-latent-message control, and the message-only arbiter control}.
Table~\ref{tab:published_comparison} reports published benchmark comparisons.

\paragraph{Published end-to-end protocols.}
The \red{\citeauthor{feng2024towards}} and BIT scores follow the benchmark tables of \citet{zhang2026bit}; BIT uses nucleus sampling, while MAS-BIT uses greedy decoding.
MoDAl-1 uses an area 6v encoder, whereas MoDAl-Full combines three neural encoders and includes area 44 signals~\citep{chen2026modal}.
Their \WER[1]{modalone}{t12}\% and \WER[1]{modal}{t12}\% results come from the 1,200-sentence competition holdout, with 95\% confidence intervals across five seeds; results on the 880-sentence model-selection set are listed separately in Table~\ref{tab:author_evaluations}.
Both configurations use CTC pretraining followed by direct autoregressive text decoding.
BIT End-to-End + Ensemble combines independently trained decoders through an LLM merger and is retained as a separate ensemble reference.
The T12 system merges ten models with a fine-tuned GPT-3.5, while the T15 system merges 29 models with a fine-tuned GPT-4; both mergers receive sentence hypotheses and phoneme outputs~\citep{zhang2026bit}.

\paragraph{Additional published end-to-end variants.}
The RNN, BIT-TFS, and BIT-Human end-to-end configurations in Table~\ref{tab:published_comparison} vary the encoder within BIT~\citep{zhang2026bit}.
We retain the one-decimal T12 holdout and T15 public holdout WER values printed in their Figure 2A.
The RNN configuration pairs its encoder with an end-to-end LLM decoder.
BrainWhisperer has both a direct Whisper decoding path and a separate phoneme/WFST path~\citep{boccato2026brainwhisperer}; Table~\ref{tab:published_comparison} includes its direct-decoding competition results.
BrainWhisperer-cross jointly trains on the Card, Willett, and Kunz datasets.
\red{BEST fuses hierarchical phoneme and character representations before a Whisper-based decoder~\citep{feng2026best}. Table~\ref{tab:published_comparison} uses its no-external-LM row from Table~1 of that paper: \WER[2]{bestnolm}{t12}\% on T12 holdout and \ReportedBestFifteenWER\% on T15 holdout, whose public/private subset is unspecified.}

\paragraph{Decoding on other evaluation sets.}
\label{app:other_evaluations}
Table~\ref{tab:author_evaluations} broadens the coverage of neural-to-text decoders while retaining each study's evaluation set and inference procedure.
These values are reported separately from the matched holdout comparison because the partitions, training data, and selection procedures differ.
\citet{fiedler2025wav2vec2} reserve the first block of each training day for a custom 880-sentence test, leaving 7,920 training sentences; the original labeled test serves as validation.
Both listed models are fully fine-tuned; we report the best run in each setting from the WER column without external language-model decoding.

\citet{olak2026decoding} exclude competition holdout data and report native greedy and beam-search decoding from a Transformer with a BART word decoder.
MFCC targets provide auxiliary supervision during training rather than serving as an additional input at inference.
Their paper calls the 880-sentence labeled data both test and validation, so we retain the neutral labeled-set designation.
The internal phoneme selectors score word-decoder candidates using neural-conditioned phoneme likelihood or agreement with predicted phonemes after grapheme-to-phoneme conversion (G2P).
These configurations use no independent language-model score and are listed separately from native word decoding; candidates are generated using beam search and nucleus sampling.

BrainWhisperer's Card Test results are distinct from those in the competition column; the Test size and split identifiers are not specified~\citep{boccato2026brainwhisperer}.
Its cross-dataset model is a jointly trained decoder, not an ensemble.
For \citet{feng2024towards}, we distinguish the full 880-sentence test from its 720-sentence vocal subset.
We list the additional MoDAl values reported on the 880-sentence set used to select the models and group them accordingly~\citep{chen2026modal}.

\begin{table}[ht]
\centering
\caption{\footnotesize \emph{Reported neural-to-text decoding on author-defined evaluation sets.} Values are WER (\%) as reported. \red{\citeauthor{fiedler2025wav2vec2} report} best runs with full fine-tuning. \red{\citeauthor{olak2026decoding} report} mean $\pm$ SD over five seeds; MoDAl values are mean $\pm$ 95\% CI over five seeds.}
\label{tab:author_evaluations}
\small
\setlength{\tabcolsep}{3pt}
\begin{tabular}{@{}>{\raggedright\arraybackslash}p{0.37\linewidth}>{\raggedright\arraybackslash}p{0.21\linewidth}>{\raggedright\arraybackslash}p{0.23\linewidth}r@{}}
\toprule
\textbf{Model} & \textbf{Evaluation set} & \textbf{Inference} & \textbf{WER (\%)} \\
\midrule
\multicolumn{4}{l}{\textit{Direct decoding}} \\
Wav2Vec2Conformer~\citep{fiedler2025wav2vec2} & T12 custom test (880) & Character CTC; no external LM & \ReportedFiedlerConformerWER \\[0.3em]
Wav2Vec2~\citep{fiedler2025wav2vec2} & T12 custom test (880) & Character CTC; no external LM & \ReportedFiedlerWavWER \\[0.3em]
Transformer + MFCC + BART~\citep{olak2026decoding} & T12 labeled set (880) & Native beam search & \ReportedOlakBeamWER\,$\pm$\,\ReportedOlakBeamSD \\[0.3em]
Transformer + MFCC + BART~\citep{olak2026decoding} & T12 labeled set (880) & Greedy & \ReportedOlakGreedyWER\,$\pm$\,\ReportedOlakGreedySD \\[0.3em]
BrainWhisperer~\citep{boccato2026brainwhisperer} & T15 Card Test & Native Whisper decoder & \ReportedBrainWhispererTestWER \\[0.3em]
BrainWhisperer-cross~\citep{boccato2026brainwhisperer} & T15 Card Test & Native Whisper decoder & \ReportedBrainWhispererCrossTestWER \\[0.3em]
GRU-BPE + Llama2-7B~\citep{feng2024towards} & T12 labeled test (880; all) & Autoregressive & \ReportedFengBpeTestWER \\[0.3em]
GRU-Phone + Llama2-7B~\citep{feng2024towards} & T12 labeled test (880; all) & Autoregressive & \ReportedFengPhoneTestWER \\[0.3em]
BGRU-Phone + Llama2-7B~\citep{feng2024towards} & T12 labeled test (880; all) & Autoregressive & \ReportedFengBiGruTestWER \\[0.3em]
BGRU-Phone + Llama2-7B~\citep{feng2024towards} & T12 vocal subset (720) & Autoregressive & \ReportedFengVocalTestWER \\
\midrule
\multicolumn{4}{l}{\textit{Word decoding with internal phoneme selection}} \\
Transformer + MFCC + BART~\citep{olak2026decoding} & T12 labeled set (880) & Phoneme-head likelihood; G2P & \ReportedOlakPhoneWER\,$\pm$\,\ReportedOlakPhoneSD \\[0.3em]
Transformer + MFCC + BART~\citep{olak2026decoding} & T12 labeled set (880) & Predicted-phoneme agreement; G2P & \ReportedOlakAgreementWER\,$\pm$\,\ReportedOlakAgreementSD \\
\midrule
\multicolumn{4}{l}{\textit{Labeled evaluation used for model selection}} \\
MoDAl-1~\citep{chen2026modal} & T12 selection set (880) & Autoregressive & \ReportedMoDAlOneTestWER\,$\pm$\,\ReportedMoDAlOneTestCI \\[0.3em]
MoDAl-Full~\citep{chen2026modal} & T12 selection set (880) & Autoregressive & \ReportedMoDAlFullTestWER\,$\pm$\,\ReportedMoDAlFullTestCI \\[0.3em]
MoDAl, two area-6v encoders~\citep{chen2026modal} & T12 selection set (880) & Autoregressive & \ReportedMoDAlTwoTestWER\,$\pm$\,\ReportedMoDAlTwoTestCI \\
\bottomrule
\end{tabular}
\end{table}

\paragraph{Other decoding protocols.}
\citet{khanday2026endtoend} report validation CER without WER, so their character-CTC model has no numerical entry here.
The multi-user system of \citet{Fogg2026.07.23.739430} and the neural ensembles of \citet{benster2024cross,yoon2026ensembles} use phoneme-to-word language-model decoding and remain cascaded systems.
LightBeam uses lexicon constraints and external language-model fusion~\citep{feghhi2026lightbeam}; removing a WFST does not make its deployment end-to-end under our definition\red{, and Table~\ref{tab:cascade_results} reports our reproduced BIT + LightBeam holdout WER}.

\begin{table}[ht]
\centering
\caption{\footnotesize \emph{Published cascaded references.} Values are corpus holdout WER (\%) reported in the benchmark tables of \citet{zhang2026bit}. These systems use external language-model decoding and are not single-model end-to-end baselines. The published BIT cascade uses $K=100$; T15 private scores were not reported.}
\label{tab:published_cascaded}
\small
\setlength{\tabcolsep}{6pt}
\begin{tabular}{lccc}
\toprule
\textbf{Method} & \textbf{T12 holdout} & \textbf{T15 public} & \textbf{T15 private} \\
\midrule
RNN baseline~\citep{willett2023high,card2024accurate} & \WER[2]{rnn}{t12} & \WER[2]{rnn}{t15public} & -- \\
Time-masked Transformer~\citep{feghhi2025time} & \WER[2]{timemasked}{t12} & -- & -- \\
Published BIT cascade~\citep{zhang2026bit} & \WER[2]{bitcascade}{t12} & \WER[2]{bitcascade}{t15public} & -- \\
\bottomrule
\end{tabular}
\end{table}

\begin{table}[ht]
\centering
\caption{\footnotesize \emph{Candidate-sentence counts and holdout decoding.} Values are WER (\%); T15 public and private scores refer to separate subsets. $K$ is the candidate-sentence limit in our reproduced cascade \red{using either the released BIT encoder or MAS-BIT's fine-tuned participant-specific encoder branch and phoneme readout; BIT + LightBeam uses the released BIT encoders}. Blue shading marks MAS-BIT.}
\label{tab:cascade_results}
\small
\setlength{\tabcolsep}{7pt}
\begin{tabular}{lrccc}
\toprule
& & T12 & \multicolumn{2}{c}{T15} \\
\cmidrule(lr){3-3}\cmidrule(lr){4-5}
\textbf{Method} & $K$ & \textbf{Holdout} & \textbf{Public} & \textbf{Private} \\
\midrule
Reproduced cascade + OPT~\citep{zhang2026bit} & 100 & \WER{cascade100}{t12} & \WER{cascade100}{t15public} & \WER{cascade100}{t15private} \\
 & 75 & \WER{cascade75}{t12} & \WER{cascade75}{t15public} & \WER{cascade75}{t15private} \\
 & 50 & \WER{cascade50}{t12} & \WER{cascade50}{t15public} & \WER{cascade50}{t15private} \\
 & 37 & \WER{cascade37}{t12} & \WER{cascade37}{t15public} & \WER{cascade37}{t15private} \\
 & 25 & \WER{cascade25}{t12} & \WER{cascade25}{t15public} & \WER{cascade25}{t15private} \\
 & 18 & \WER{cascade18}{t12} & \WER{cascade18}{t15public} & \WER{cascade18}{t15private} \\
 & 12 & \WER{cascade12}{t12} & \WER{cascade12}{t15public} & \WER{cascade12}{t15private} \\
 & 9 & \WER{cascade9}{t12} & \WER{cascade9}{t15public} & \WER{cascade9}{t15private} \\
 & 6 & \WER{cascade6}{t12} & \WER{cascade6}{t15public} & \WER{cascade6}{t15private} \\
 & 4 & \WER{cascade4}{t12} & \WER{cascade4}{t15public} & \WER{cascade4}{t15private} \\
 & 3 & \WER{cascade3}{t12} & \WER{cascade3}{t15public} & \WER{cascade3}{t15private} \\
 & 2 & \WER{cascade2}{t12} & \WER{cascade2}{t15public} & \WER{cascade2}{t15private} \\
 & 1 & \WER{cascade1}{t12} & \WER{cascade1}{t15public} & \WER{cascade1}{t15private} \\
\midrule
\red{Fine-tuned-encoder cascade + OPT} & \red{100} & \red{\WER{cm1cascade100}{t12}} & \red{\WER{cm1cascade100}{t15public}} & \red{\WER{cm1cascade100}{t15private}} \\
 & \red{75} & \red{\WER{cm1cascade75}{t12}} & \red{\WER{cm1cascade75}{t15public}} & \red{\WER{cm1cascade75}{t15private}} \\
 & \red{50} & \red{\WER{cm1cascade50}{t12}} & \red{\WER{cm1cascade50}{t15public}} & \red{\WER{cm1cascade50}{t15private}} \\
 & \red{37} & \red{\WER{cm1cascade37}{t12}} & \red{\WER{cm1cascade37}{t15public}} & \red{\WER{cm1cascade37}{t15private}} \\
 & \red{25} & \red{\WER{cm1cascade25}{t12}} & \red{\WER{cm1cascade25}{t15public}} & \red{\WER{cm1cascade25}{t15private}} \\
 & \red{18} & \red{\WER{cm1cascade18}{t12}} & \red{\WER{cm1cascade18}{t15public}} & \red{\WER{cm1cascade18}{t15private}} \\
 & \red{12} & \red{\WER{cm1cascade12}{t12}} & \red{\WER{cm1cascade12}{t15public}} & \red{\WER{cm1cascade12}{t15private}} \\
 & \red{9} & \red{\WER{cm1cascade9}{t12}} & \red{\WER{cm1cascade9}{t15public}} & \red{\WER{cm1cascade9}{t15private}} \\
 & \red{6} & \red{\WER{cm1cascade6}{t12}} & \red{\WER{cm1cascade6}{t15public}} & \red{\WER{cm1cascade6}{t15private}} \\
 & \red{4} & \red{\WER{cm1cascade4}{t12}} & \red{\WER{cm1cascade4}{t15public}} & \red{\WER{cm1cascade4}{t15private}} \\
 & \red{3} & \red{\WER{cm1cascade3}{t12}} & \red{\WER{cm1cascade3}{t15public}} & \red{\WER{cm1cascade3}{t15private}} \\
 & \red{2} & \red{\WER{cm1cascade2}{t12}} & \red{\WER{cm1cascade2}{t15public}} & \red{\WER{cm1cascade2}{t15private}} \\
 & \red{1} & \red{\WER{cm1cascade1}{t12}} & \red{\WER{cm1cascade1}{t15public}} & \red{\WER{cm1cascade1}{t15private}} \\
\midrule
\red{BIT + LightBeam~\citep{feghhi2026lightbeam}} & \red{--} & \red{\WER{lightbeam}{t12}} & \red{\WER{lightbeam}{t15public}} & \red{\WER{lightbeam}{t15private}} \\
\midrule
\rowcolor{blue!10} \textbf{MAS-BIT (ours)} & -- & \textbf{\WER{mas}{t12}} & \textbf{\WER{mas}{t15public}} & \textbf{\WER{mas}{t15private}} \\
\bottomrule
\end{tabular}
\end{table}

\begin{table}[ht]
\centering
\caption{\footnotesize \emph{Holdout performance of MAS-BIT and its ablations.} Values are WER (\%) on the same hidden holdouts. The five core controls, the rank-24 single-agent variant\red{, the sequential dual-view refinement control, the detached-latent-message control, and the message-only arbiter control} use the shared backbone, two-encoder frontend, and checkpoint-selection protocol\red{; the Borealis-5B single-agent variant replaces the 1.7B decoder with a 4B audio-trained decoder and rank-13 adapter}.}
\label{tab:ablation_results}
\small
\setlength{\tabcolsep}{9pt}
\begin{tabular}{lccc}
\toprule
& T12 & \multicolumn{2}{c}{T15} \\
\cmidrule(lr){2-2}\cmidrule(lr){3-4}
\textbf{Method} & \textbf{Holdout} & \textbf{Public} & \textbf{Private} \\
\midrule
\rowcolor{blue!10} \textbf{MAS-BIT} & \WER{mas}{t12} & \WER{mas}{t15public} & \WER{mas}{t15private} \\
Single agent, two views & \WER{single}{t12} & \WER{single}{t15public} & \WER{single}{t15private} \\
\quad Rank-24 adapter & \WER{single_r24}{t12} & \WER{single_r24}{t15public} & \WER{single_r24}{t15private} \\
\quad \red{Borealis-5B backbone (rank-13 adapter)} & \red{\WER{single_borealis}{t12}} & \red{\WER{single_borealis}{t15public}} & \red{\WER{single_borealis}{t15private}} \\
\red{Sequential dual-view refinement} & \red{\WER{refinement}{t12}} & \red{\WER{refinement}{t15public}} & \red{\WER{refinement}{t15private}} \\
Hard-text communication & \WER{hard}{t12} & \WER{hard}{t15public} & \WER{hard}{t15private} \\
\red{Detached latent messages} & \red{\WER{softsg}{t12}} & \red{\WER{softsg}{t15public}} & \red{\WER{softsg}{t15private}} \\
\red{Message-only arbiter} & \red{\WER{message_only}{t12}} & \red{\WER{message_only}{t15public}} & \red{\WER{message_only}{t15private}} \\
Two neural experts & \WER{nna}{t12} & \WER{nna}{t15public} & \WER{nna}{t15private} \\
Separate participant training & \WER{separate}{t12} & \WER{separate}{t15public} & \WER{separate}{t15private} \\
No staged training & \WER{joint}{t12} & \WER{joint}{t15public} & \WER{joint}{t15private} \\
\bottomrule
\end{tabular}
\end{table}

\clearpage
\section{Representational Similarity Analysis}
\label{app:rsa}
\providecommand{\RSAValue}[1]{\csname rsavalue@#1\endcsname}
\expandafter\def\csname rsavalue@t12z_naudiobefore\endcsname{0.191}
\expandafter\def\csname rsavalue@t12z_naudioafter\endcsname{0.343}
\expandafter\def\csname rsavalue@t12z_naudiochange\endcsname{+0.152 [+0.123, +0.179]}
\expandafter\def\csname rsavalue@t12z_ntextbefore\endcsname{0.141}
\expandafter\def\csname rsavalue@t12z_ntextafter\endcsname{0.196}
\expandafter\def\csname rsavalue@t12z_ntextchange\endcsname{+0.055 [+0.035, +0.074]}
\expandafter\def\csname rsavalue@t12second_viewaudiobefore\endcsname{0.199}
\expandafter\def\csname rsavalue@t12second_viewaudioafter\endcsname{0.233}
\expandafter\def\csname rsavalue@t12second_viewaudiochange\endcsname{+0.034 [+0.017, +0.051]}
\expandafter\def\csname rsavalue@t12views\endcsname{0.582}
\expandafter\def\csname rsavalue@t12viewsci\endcsname{[0.562, 0.603]}
\expandafter\def\csname rsavalue@t12neuralviews\endcsname{0.950}
\expandafter\def\csname rsavalue@t12neuralviewsci\endcsname{[0.946, 0.954]}
\expandafter\def\csname rsavalue@t12hard\endcsname{0.322}
\expandafter\def\csname rsavalue@t12hardchange\endcsname{+0.021 [+0.011, +0.030]}
\expandafter\def\csname rsavalue@t15z_naudiobefore\endcsname{0.097}
\expandafter\def\csname rsavalue@t15z_naudioafter\endcsname{0.239}
\expandafter\def\csname rsavalue@t15z_naudiochange\endcsname{+0.142 [+0.123, +0.163]}
\expandafter\def\csname rsavalue@t15z_ntextbefore\endcsname{0.067}
\expandafter\def\csname rsavalue@t15z_ntextafter\endcsname{0.165}
\expandafter\def\csname rsavalue@t15z_ntextchange\endcsname{+0.098 [+0.083, +0.113]}
\expandafter\def\csname rsavalue@t15second_viewaudiobefore\endcsname{0.050}
\expandafter\def\csname rsavalue@t15second_viewaudioafter\endcsname{0.133}
\expandafter\def\csname rsavalue@t15second_viewaudiochange\endcsname{+0.083 [+0.066, +0.101]}
\expandafter\def\csname rsavalue@t15views\endcsname{0.639}
\expandafter\def\csname rsavalue@t15viewsci\endcsname{[0.616, 0.660]}
\expandafter\def\csname rsavalue@t15neuralviews\endcsname{0.949}
\expandafter\def\csname rsavalue@t15neuralviewsci\endcsname{[0.945, 0.951]}
\expandafter\def\csname rsavalue@t15hard\endcsname{0.223}
\expandafter\def\csname rsavalue@t15hardchange\endcsname{+0.016 [+0.010, +0.022]}
\newcommand{\RSAInterfaceRows}{%
T12 & $Z_N$--audio & 0.191 & 0.343 & +0.152 [+0.123, +0.179] \\
T12 & $Z_N$--text & 0.141 & 0.196 & +0.055 [+0.035, +0.074] \\
T12 & $Z_P$--audio & 0.199 & 0.233 & +0.034 [+0.017, +0.051] \\
T15 & $Z_N$--audio & 0.097 & 0.239 & +0.142 [+0.123, +0.163] \\
T15 & $Z_N$--text & 0.067 & 0.165 & +0.098 [+0.083, +0.113] \\
T15 & $Z_P$--audio & 0.050 & 0.133 & +0.083 [+0.066, +0.101] \\
}
\newcommand{\RSABranchRows}{%
T12 & Native branch--audio & 0.277 & 0.287 & +0.010 [+0.005, +0.014] \\
T12 & Cross-participant branch--audio & 0.014 & 0.198 & +0.184 [+0.150, +0.219] \\
T12 & $H_1$--$H_2$ & 0.047 & 0.374 & +0.326 [+0.299, +0.354] \\
T15 & Native branch--audio & 0.262 & 0.260 & -0.002 [-0.006, +0.003] \\
T15 & Cross-participant branch--audio & 0.036 & 0.156 & +0.121 [+0.098, +0.142] \\
T15 & $H_1$--$H_2$ & 0.061 & 0.448 & +0.387 [+0.369, +0.405] \\
}
\newcommand{\RSALengthRows}{%
T12 & Training change & +0.152 & +0.226 & +0.196 & +0.196 & +0.189 \\
T12 & MAS-BIT $-$ Hard text & +0.021 & +0.012 & +0.014 & +0.015 & +0.015 \\
T15 & Training change & +0.142 & +0.132 & +0.125 & +0.138 & +0.129 \\
T15 & MAS-BIT $-$ Hard text & +0.016 & +0.008 & +0.012 & +0.010 & +0.012 \\
}

We use sentence-level representational similarity analysis (RSA)~\citep{kriegeskorte2008representational} to examine how the learned interfaces organize neural evidence.
The analysis follows three parts of the architecture: alignment of the fused interface, adaptation of the two encoder branches, and the geometry of the expert input views.
All comparisons below use sentence-pooled representations from the selected checkpoints and their initialization.

\begin{figure}[ht]
\centering
\includegraphics[width=\linewidth]{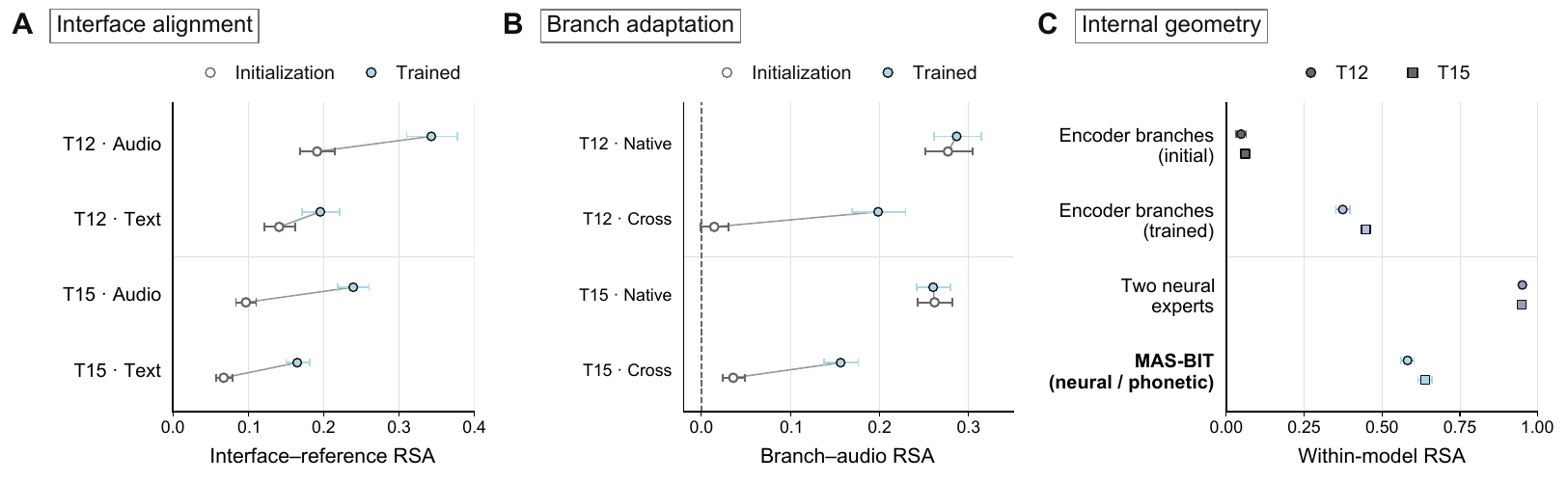}
\caption{\footnotesize \emph{Joint training aligns the neural interface while preserving distinct expert views.} \textbf{(A)} Pooled $Z_N$--audio/text RSA before and after training. \textbf{(B)} Encoder-branch alignment with the audio reference; native and cross-participant branches are defined by the input participant and encoder origin. \textbf{(C)} Within-model RSA between encoder branches before/after training, and between the two expert input views in the two-neural-expert control and MAS-BIT. Points and intervals are word-count-stratified estimates and pointwise 95\% sentence-bootstrap intervals. Audio references use the 0.6B Qwen audio tower on fixed-voice TTS; text references use the 1.7B text decoder.}
\label{fig:rsa_alignment}
\end{figure}

\subsection{Joint training aligns the fused neural interface}

Training increases the alignment of $Z_N$ with both audio and text sentence geometry (Figure~\ref{fig:rsa_alignment}A; Table~\ref{tab:rsa_interfaces}).
Audio-reference RSA rises from \RSAValue{t12z_naudiobefore} to \RSAValue{t12z_naudioafter} on T12 and from \RSAValue{t15z_naudiobefore} to \RSAValue{t15z_naudioafter} on T15.
The paired changes are \RSAValue{t12z_naudiochange} and \RSAValue{t15z_naudiochange}, respectively; all 2,000 bootstrap changes are positive for each participant.
Text-reference alignment increases by 0.055/0.098 on T12/T15, and the phonetic interface $Z_P$ also becomes more aligned with the audio reference.
These results locate the training-associated changes at the continuous interfaces read by the experts and arbiter.

\begin{table}[ht]
\centering
\caption{\footnotesize \emph{Training aligns neural and phonetic interfaces with reference sentence geometry.} Values are pooled, word-count-stratified RSA. Changes use paired bootstrap draws; brackets give pointwise 95\% intervals.}
\label{tab:rsa_interfaces}
\small
\setlength{\tabcolsep}{5pt}
\begin{tabular}{llrrr}
\toprule
Participant & Interface--reference & Initialization & Trained & Change [95\% interval] \\
\midrule
\RSAInterfaceRows
\bottomrule
\end{tabular}
\end{table}

The increase in $Z_N$--audio RSA is positive under all five length treatments: word-count grouping, ungrouped RSA, and partial correlations controlling for word-count difference, valid neural-position-count difference, or both (Table~\ref{tab:rsa_sensitivity}).
The estimates range from $+0.152$ to $+0.226$ on T12 and from $+0.125$ to $+0.142$ on T15.

\subsection{Cross-participant branches acquire shared sentence structure}

Each recording is processed by both pretrained encoders through participant-specific input projections.
For T12, the T12-initialized encoder is the native branch and the T15-initialized encoder is the cross-participant branch; these roles reverse for T15.
The cross-participant input projections are newly initialized, while the native projections retain the released initialization.
The fused interface combines these differently initialized pathways and is learned jointly for transcription; its training-associated alignment gain is measured relative to the initialized fused interface.

The native branches retain similar audio-reference alignment before and after training: 0.277 to 0.287 on T12 and 0.262 to 0.260 on T15.
The cross-participant branches show larger changes, from 0.014 to 0.198 on T12 and from 0.036 to 0.156 on T15 (Figure~\ref{fig:rsa_alignment}B).
Their paired increases are $+0.184$ [0.150, 0.219] and $+0.121$ [0.098, 0.142], respectively.
The two branches also develop more shared sentence structure: $H_1$--$H_2$ RSA rises from 0.047/0.061 to 0.374/0.448 on T12/T15 (Figure~\ref{fig:rsa_alignment}C).
Thus training adapts the cross-participant pathway alongside the pretrained native pathway, and the nonlinear fusion presents their joint evidence through $Z_N$.

\begin{table}[ht]
\centering
\caption{\footnotesize \emph{Adaptation of the encoder branches and their shared sentence geometry.} Native/cross-participant denotes encoder origin relative to the input participant. The final row for each participant measures similarity between the two branches. Brackets give pointwise 95\% paired-bootstrap intervals.}
\label{tab:rsa_branches}
\small
\setlength{\tabcolsep}{5pt}
\begin{tabular}{llrrr}
\toprule
Participant & Comparison & Initialization & Trained & Change [95\% interval] \\
\midrule
\RSABranchRows
\bottomrule
\end{tabular}
\end{table}

\subsection{Neural and phonetic views share structure and remain distinct}

The two-neural-expert control provides each expert with an independent fusion projector and role adapter, but both inputs are transformations of the same continuous encoder states.
Its two input views have closely matched sentence geometry: RSA is \RSAValue{t12neuralviews} on T12 and \RSAValue{t15neuralviews} on T15.
In MAS-BIT, $Z_N$--$Z_P$ RSA is \RSAValue{t12views} and \RSAValue{t15views}, respectively (Figure~\ref{fig:rsa_alignment}C).
The neural and phonetic interfaces therefore preserve shared sentence structure while exposing more differentiated geometry to the experts.
Together with the phonetic-expert holdout comparison in Section~\ref{sec:ablation_results}, this provides converging evidence for complementary neural and phonetic views.

The communication interface is also associated with a change in the learned neural geometry.
MAS-BIT's $Z_N$--audio RSA is 0.343/0.239, compared with \RSAValue{t12hard}/\RSAValue{t15hard} for hard-text communication.
The paired differences are \RSAValue{t12hardchange} on T12 and \RSAValue{t15hardchange} on T15, and the difference remains positive under all five length treatments (Table~\ref{tab:rsa_sensitivity}).
This comparison links MAS-BIT with latent messages to a more aligned sentence-level neural interface than the hard-text control.

\begin{table}[ht]
\centering
\caption{\footnotesize \emph{Neural-interface alignment changes across five length treatments.} Entries are differences in pooled $Z_N$--audio RSA. Training change compares MAS-BIT with initialization; the second contrast compares MAS-BIT with hard-text communication. Partial correlations control for differences in reference word count, valid neural-position count, or both.}
\label{tab:rsa_sensitivity}
\small
\setlength{\tabcolsep}{4pt}
\begin{tabular}{llrrrrr}
\toprule
& & & & \multicolumn{3}{c}{Partial correlation} \\
\cmidrule(lr){5-7}
Participant & Contrast & Grouped & Ungrouped & Words & Positions & Both \\
\midrule
\RSALengthRows
\bottomrule
\end{tabular}
\end{table}

\subsection{Representation extraction and statistical analysis}

\paragraph{Representations and references.}
We extract every provided validation utterance (880 T12 and 1,426 T15) using the frozen checkpoint's FP32 inference path at batch size one.
For each valid time position, we retain the encoder states $H_1,H_2$, the fused neural interface $Z_N$, and the phonetic interface $Z_P$ defined in Equation~\ref{eq:evidence_views}.
Each transformation, including the nonlinear fusion projector, is applied before temporal averaging.
In the two-neural-expert control, the second input view is the output of its separate neural projector.
The audio reference is the mean-pooled Qwen3-ASR-0.6B audio-tower output on fixed-voice TTS of the transcript; the text reference is the mean-pooled final-layer Qwen3-ASR-1.7B text-decoder state on the transcript.
Comparisons across training states use the same reference features and utterance identities.

\paragraph{Word-count-stratified RSA.}
We z-score each feature across the full participant-specific sentence set and form representational dissimilarity matrices using one minus Pearson correlation between sentence vectors.
We retain groups with identical reference word counts and at least 20 utterances, then compute the Spearman correlation $\rho_g$ between distinct upper-triangle dissimilarities in each group.
The overall score is
\begin{equation}
\rho_{\mathrm{RSA}}=\frac{\sum_g\binom{n_g}{2}\rho_g}{\sum_g\binom{n_g}{2}},
\end{equation}
where $n_g$ is the original group size.
T12 contributes 848 sentences and 48,531 within-group pairs (2--10 words); T15 contributes 1,386 sentences and 134,018 pairs (3--11 words).
The ungrouped and partial estimators use all provided validation sentences.
For partial RSA, we regress each ranked dissimilarity vector on the ranked absolute length differences with an intercept, then correlate the residuals.

\paragraph{Paired uncertainty estimates.}
We draw 2,000 bootstrap samples with replacement within each word-count group, sharing draws across models, representations, and references within a participant (seed 0).
For distinct original sentences $i,j$, the pair weight is $c_i c_j$, where $c_i$ is the sentence's multiplicity in the draw; pairs between copies of the same original sentence are excluded.
Weighted midranks retain this repeated-pair distribution.
Full-data standardization and original group weights remain fixed.
We report pointwise 95\% sentence-bootstrap intervals from the 2.5th and 97.5th percentiles of the estimates or paired differences, holding the checkpoint weights fixed.

\end{document}